\documentclass{aa}
\usepackage{graphicx}
\usepackage[intlimits]{amsmath}
\usepackage{float}
\usepackage[below]{placeins}
\usepackage{hyperref}
   \hypersetup{
     colorlinks=true,
     citecolor=blue,
     linkcolor=blue,
    }
\usepackage{txfonts}
\usepackage{natbib,txfonts}
\usepackage{subfig}
\usepackage{longtable,lscape}

\begin{document}
   \title{The parsec-scale jet of PKS\,1749+096}
   \author{R.-S. Lu
          \inst{1,2,3,4}
          \and
          Z.-Q. Shen
          \inst{1,5}
          \and
          T. P. Krichbaum\inst{3}
          \and
          S. Iguchi\inst{6}  
          \and
          S.-S. Lee\inst{7}
	  \and
          J. A. Zensus\inst{3}
          }

   \institute{
Key Laboratory for Research in Galaxies and Cosmology, Shanghai Astronomical Observatory, Chinese Academy of Sciences, 80 Nandan Rd, Shanghai 200030, China\\
            \email{rslu@shao.ac.cn}
             \and
Graduate School of the Chinese Academy of Sciences, Beijing 100039, China
             \and
             Max-Planck-Institut f\"ur Radioastronomie, Auf dem H\"ugel 69, D-53121 Bonn, Germany
             \and
             Massachusetts Institute of Technology, Haystack Observatory, Route 40, Westford, MA 01886, USA
             \and
             Key Laboratory of Radio Astronomy, Chinese Academy of Sciences, China
             \and
National Astronomical Observatory of Japan, 2-2-1 Osawa, Mitaka, Tokyo 181-8588, Japan
             \and
Korean VLBI Network, Korea Astronomy and Space Science Institute, P. O. Box 88, Yonsei University, Seongsanro 262, Seodaemun, Seoul 120-749, Republic of Korea
             }

   \date{Received 20 January 2012; Accepted 6 April 2012}


\abstract
{PKS\,1749+096 is a BL Lac object showing weak extended jet emission to the northeast of the compact VLBI core on parsec scales.}
{We aim at better understanding the jet kinematics and variability of this source and finding clues that may applicable to other BL Lac objects.}
{The jet was studied with multi-epoch multi-frequency high-resolution VLBI observations.}
{The jet is characterized by a one-sided curved morphology at all epochs and all frequencies. The VLBI core, located at the southern end of the jet, was identified based on its spectral properties. The equipartition magnetic field of the core was investigated, through which we derived a Doppler factor of 5, largely consistent with that derived from kinematics (component C5).
The study of the detailed jet kinematics at 22 and 15\,GHz, spanning a period of more than10 years, indicates the possible existence of a bimodal distribution of the jet apparent speed. Ballistic and non-ballistic components are found to coexist in the jet. Superluminal motions in the range of 5--21\,$c$ were measured in 11 distinct components. We estimated the physical jet parameters with the minimum Lorentz factor of 10.2 and Doppler factors in the range of 10.2--20.4 (component C5). The coincidence in time of the component's ejection and flares supports the idea that, at least in PKS\,1749+096, ejection of new jet components is connected with major outbursts in flux density. For the best-traced component (C5) we found that the flux density decays rapidly as it travels downstream the jet, accompanied by a steepening of its spectra, which argues in favor of a contribution of inverse Compton cooling. These properties make PKS\,1749+096 a suitable target for an intensive monitoring to decipher the variability phenomenon of BL Lac objects.}
{}


   \maketitle
%
\section{Introduction}
\label{sect:Introduction}
PKS\,1749$+$096 (also known as OT\,081 and 4C +09.57) is an
ultra-luminous BL Lac object with an optical polarization of up to
32 $\%$ \citep{1999ApJS..121..131F}. \citet{2000A&A...363..887D} classified this source as a high-frequency
peaker (HFP), while \cite{2005A&A...435..839T} suggested that it is
a flat-spectrum source with an inverted spectrum during flares.
Strong variability is often seen from the radio to X-ray
regime. The variability at high radio frequencies is
outstanding: the quiescent flux density is less than 1\,Jy and
during the outbursts the source reached up to more than 10\,Jy with the
fractional variability index, defined as Var$_{\Delta
S}$=(S$_{max}$-S$_{min}$)/$S_{min}$, about 13 and 18 at 37 and
90\,GHz, respectively \citep{2005A&A...435..839T}.
In gamma-ray band, it was recently detected with Fermi \citep{2009ApJ...700..597A}
though no counterpart was found by EGRET according to \citet{2001ApJS..135..155M}.

This compact BL Lac object was unresolved on arc-second scales by
VLA observations \citep{2003AJ....125.2447R}.
On mas scales, it was dominated by the compact, strong core, which contains over 90\% of its total flux \citep[][and references therein]{2001ApJ...549..840H}, and showed a curved structure with faint jet features extending to northeast \citep[e.g.,][]{1999MNRAS.307..725G,2005AJ....130.1389L}. Superluminal motion of jet features in PKS 1749+096 has been reported with apparent speeds in the range of $\sim$ 1--14 $c$ \citep{2000PASJ...52.1037I,2001ApJ...549..840H,2004ApJ...609..539K,2009AJ....137.3718L}.  PKS\,1749$+$096 is also among the most compact sources investigated with 215\,GHz VLBI \citep{1997A&A...323L..17K}. Therefore it is suitable for VLBI studies at short millimeter wavelengths.

In this paper, we present results from high-resolution VLBI studies of PKS\,1749$+$096. In $\S$~\ref{sect:Observations}, we describe the
observations and data analysis. In $\S$~\ref{sect:Results}, we present
results followed by discussion in $\S$~\ref{sect:Discussion}. Finally, a brief summary
is given in $\S$~\ref{sect:Summary}. At a redshift of 0.32 \citep{1988A&A...191L..16S}, the luminosity distance to PKS\,1749$+$096 is D$_L$ = 1674\,Mpc and 1\,mas of angular separation corresponds to 4.64 pc, a proper motion of 0.1\,mas yr$^{-1}$ corresponds to a speed of $\beta_{\rm app}$ = 2.0\,c (adopting H$_0$ = 71\,km\,s$^{-1}$ Mpc$^{-1}$, $\Omega_M$ = 0.27, $\Omega_\Lambda$ = 0.73). Throughout this paper, the spectral index $\alpha$ is defined by S $_{\nu} \propto \nu^{\alpha}$.

\section{Observations and data analysis}
\label{sect:Observations}
VLBI observations of PKS\,1749+096 at 22\,GHz were made for projects to investigate OVV\,1633+382 after a major flare (code: BK090, BK092, and BK107), where PKS\,1749+096 was observed as a fringe finder. These observations were performed with VLBA (BK090 and BK092) and VLBA plus the Effelsberg telescope (BK107) roughly every 2--3 months from December 2001 to February 2005 (14 epochs in total). Data were recorded with four IFs in left and right circular polarization with an IF bandwidth of 8\,MHz and two-bit sampling, giving a data rate of 256\,Mbps (except in 2001.937 with a data rate of 128\,Mbps, two IFs in each polarization). The typical on-source integration time was about 20 minutes.

To study the jet kinematics for PKS\,1749+096 in more detail, 42 epoch data at 15 GHz from the ``2\,cm Survey'' and the follow up MOJAVE\footnote{http://www.physics.purdue.edu/MOJAVE/.} program between 1995 and 2005, spanning 11 years, were analyzed. Part of these data have been used to investigate the statistical properties of a sample of active galaxies~\citep{2001ApJ...549..840H,2009AJ....138.1874L}. Moreover, \citet{2000PASJ...52.1037I} made observations of this source at four frequencies including 15\,GHz with the VLBA in 1998, which we re-calibrated and re-mapped. In addition to these 15\,GHz data, two simultaneous multi-frequency (8, 15, 22, and 43\,GHz) data sets for PKS\,1749+096 from the VLBA archive on May 07, 1999 (code: BI012) and on September 13, 2001 (code: BI018) were analyzed. These observations were made in frequency-switch mode and the data were recorded in VLBA format in left circular polarization (LCP) (1999) and in dual circular polarization (2001) with four IFs of 8\,MHz each. Two-bit sampling was used in the recording. In each frequency, the total integration time was $\sim$100 (1999) and $\sim$40 (2001) minutes.

The archival 15\,GHz data from the 2\,cm/MOJAVE observing program were provided as fringe-fitted and calibrated \textit{uv} FITS files, which we re-mapped and self-calibrated before model-fitting. The post-correlation data reduction for the other observations was performed within the AIPS software in the usual manner. Opacity corrections were made at frequencies $\geq$15\,GHz by solving for receiver temperature and zenith opacity for each antenna. The data were then exported into DIFMAP for self-calibration imaging. We have fitted circular Gaussian components to the calibrated visibility data to obtain a quantitative description of the structure and to simplify and unify the comparison. The final fitting of the jet components was stopped when no significant improvement of the reduced $\chi_\nu^2$ values was obtained and no significant features were left in the residual map after subtraction of the modeled components. We tried to use as few as possible model parameters to quantify the data. We noted that although the model fitting results are generally not unique, like any other results from model-fitting of VLBI data, the differences are very small.

The formal errors of the fit parameters were estimated by using the formulae from \citet{1999ASPC..180..301F}. The position errors were estimated by $\sigma_{\mathrm r} = \frac{\sigma_{\rm{rms}}\ d}{2\ I_{\rm{peak}}}$, where $\sigma_{\rm{rms}}$ is the post-fit rms, and $d$ and $I_{\rm{peak}}$ are the size and the peak intensity of the component. In case of very compact components and high peak flux-density, this tends to underestimate the error~\citep[e.g.,][]{2010AA...511A..57B}. We therefore included an additional minimum error according to the map grid size, which roughly matches 1/5 of the beam size at each frequency and corresponds to 0.1\,mas at 8\,GHz and 0.05\,mas at 15 and 22\,GHz, and 0.03\,mas at 43\,GHz. The final flux density uncertainties were obtained by combining the uncertainties in the fitting process with a 5\,\%--10\,\% error (5\,\% at 8, 15, and 22\,GHz and 10\,\% at 43\,GHz) related to the absolute amplitude calibration.

The component identification across epochs was made based on the consistency in the evolution of flux density, core separation, position angle (PA), and full width at half-maximum (FWHM). The compact core is denoted by the letter ``D'' and the main jet components are labeled by the letter ``C'' followed by a number indicating the order of appearance. Components that appeared in only one epoch are labeled as ``x''. In some cases, complicated circumstances arose when components appeared to split or two components appeared to merge from one epoch to another.

\section{Results}
\label{sect:Results}

\begin{figure*}%
\centering
\subfloat[][2001.701, 8\,GHz]{%
\label{fig:example-1}%
\includegraphics[width=0.3\textwidth,clip]{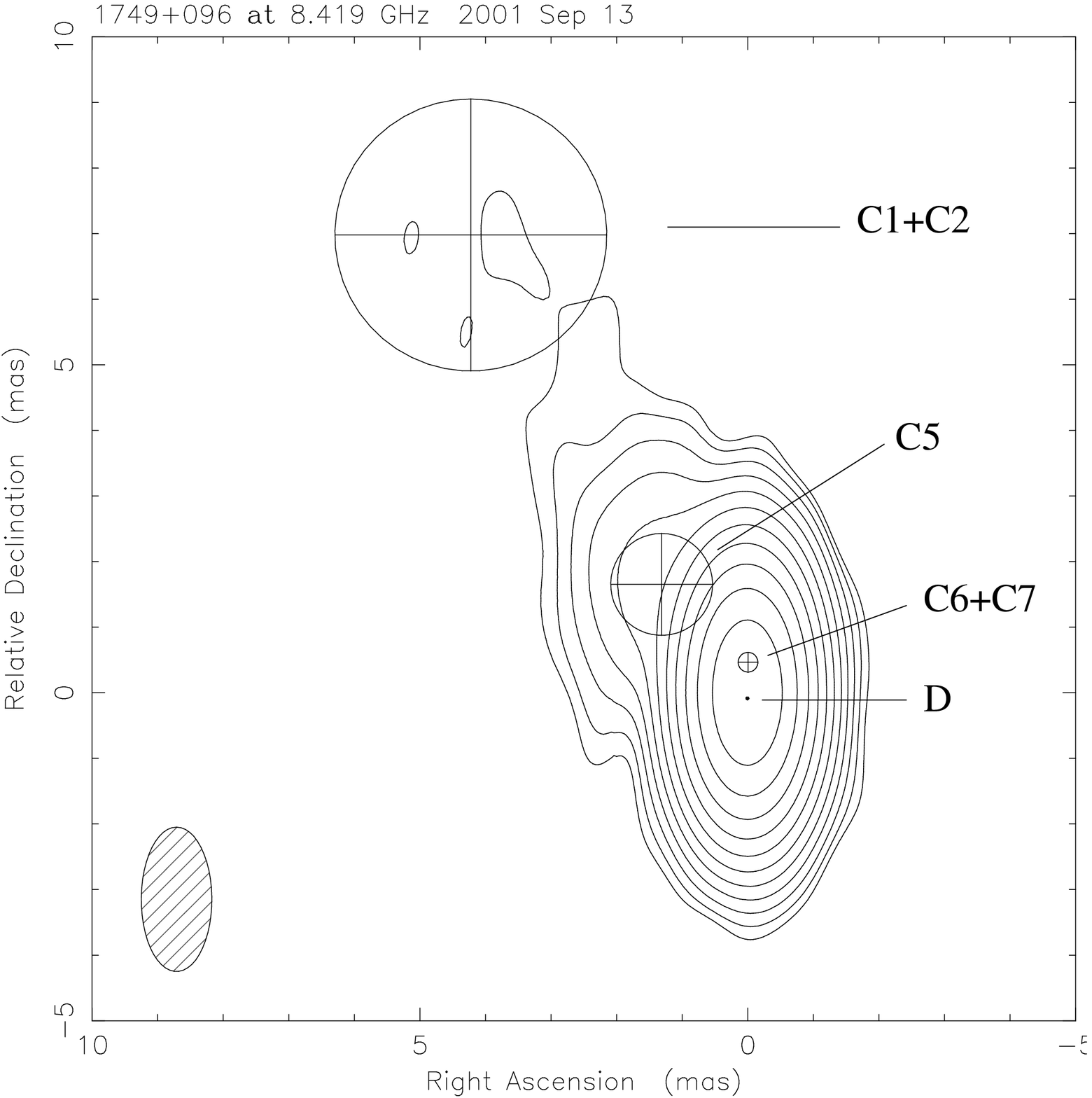}}
\hspace{3pt}%
\subfloat[][2001.701, 15\,GHz]{%
\label{fig:example-2}%
\includegraphics[width=0.3\textwidth,clip]{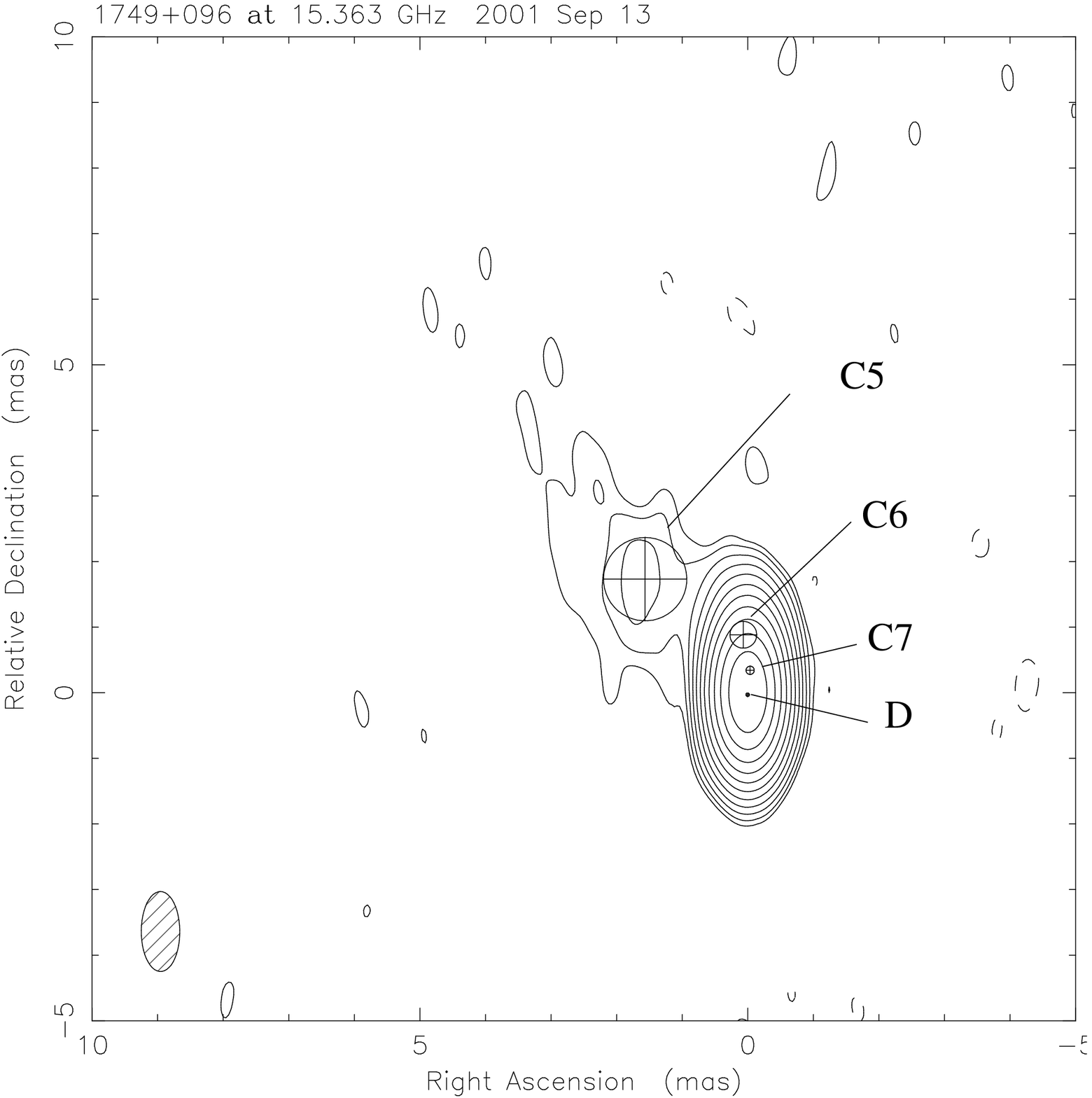}}\\
\hspace{3pt}%
\subfloat[][2001.701, 22\,GHz]{%
\label{fig:example-3}%
\includegraphics[width=0.3\textwidth,clip]{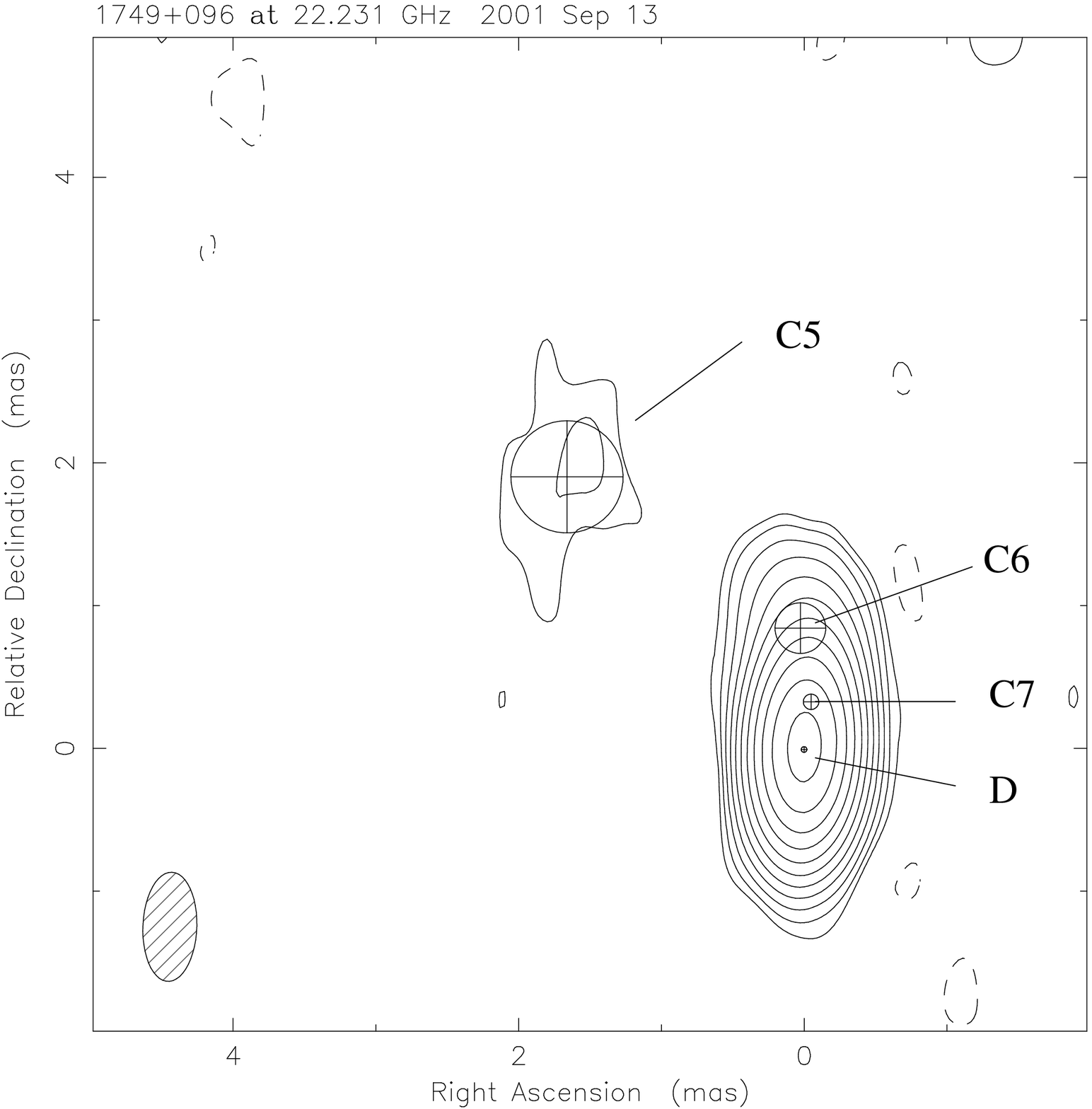}}
\hspace{3pt}%
\subfloat[][2001.701, 43\,GHz]{%
\label{fig:example-4}%
\includegraphics[width=0.3\textwidth,clip]{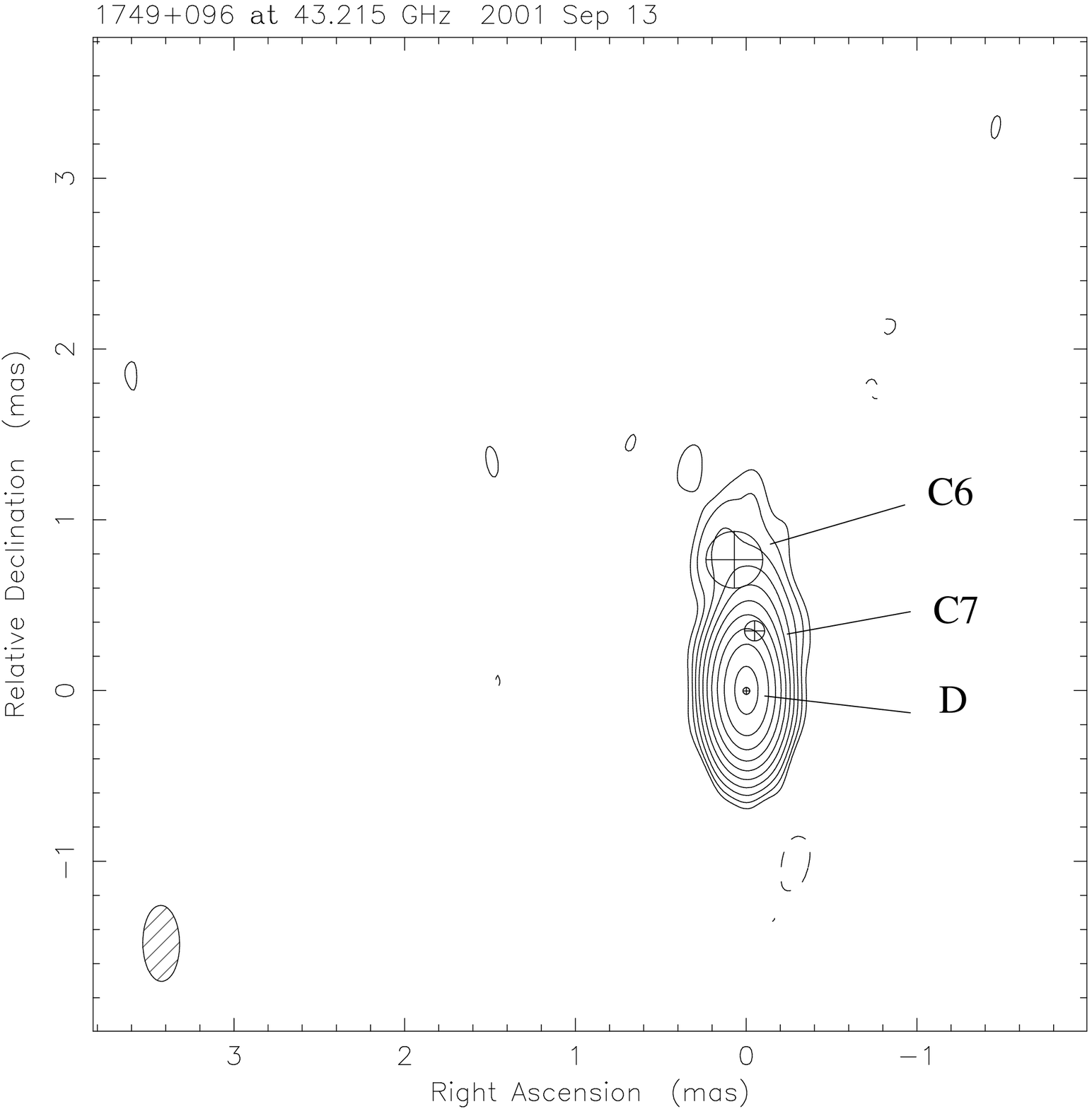}}
\caption[Clean images of PKS\,1749+096 at 8, 15, 22, and 43\,GHz]{Example of clean images of PKS\,1749+096 at 8, 15, 22, and 43\,GHz at epoch 2001.701 (2001/09/13). Parameters of the images are given in Table~\ref{tab:para_example}. Fitting parameters of the circular Gaussian component are given in Table~\ref{tab:online_model} of the online material.}
\label{fig:fig_example}
\end{figure*}

As an example, the final CLEAN images with fitted Gaussian components at 8, 15, 22, and 43\,GHz (epoch:2001.701) are shown in Fig.~\ref{fig:fig_example} with detailed parameters of these images presented in Table~\ref{tab:para_example}. All resulting CLEAN images and their parameters are presented as online material in Fig.~\ref{fig:online_maps} and Table~\ref{tab:online_para}. In Table~\ref{tab:online_model} of the online material section, the observation epoch and model-fitting results (including component identification, and model parameters of jet components, such as the flux density, the core separation, the PA, and the size) are summarized.

\begin{table*}
\centering
\caption{Description of clean images of PKS\,1749+096 shown in Fig.~\ref{fig:fig_example}. The columns list the observing frequency, the peak flux density, the parameters of the restoring elliptical Gaussian beam: the full width at half-maximum (FWHM) of the major and minor axes and the PA of the major axis, the rms noise level (3\,$\sigma$) and the contour levels of the image, expressed in percentage of the peak intensity.}
\label{tab:para_example}
\begin{tabular}{lllllrr}
\hline
&&\multicolumn{3}{c}{Restoring Beam}&\\[10pt]
\cline{3-5}
$\nu$&$S_{\rm peak}$&Major&Minor&PA&$3\times\sigma$&Contours\\
\hline
 GHz&Jy/beam&mas&mas&deg&mJy/beam&\\
(1)&(2)&(3)&(4)&(5)&(6)&(7)\\
\hline
8&2.83&2.20&1.08&0.0&1.3&0.05, 0.1, 0.2, ..., 51.2\\
15&3.60&1.22&0.59&0.1&2.0&-0.05, 0.05, 0.1, ..., 51.2\\
22&3.70&0.76&0.38&-2.2&3.1&-0.075, 0.075, 0.15, ..., 76.8\\
43&3.69&0.44&0.22&0.8&3.8&-0.15, 0.15, 0.3, ..., 76.8\\
\hline
\end{tabular}
\end{table*}

\subsection{Parsec-scale morphology}
\label{sect:3.1}
On parsec scales, PKS\,1749+096 showed one-sided extended jet emission to the northeast of the compact VLBI core~\citep{1996AJ....112.1877G,2000A&A...364..391L,2007ASPC..373..237L}.
As shown in Fig.~\ref{fig:fig_example}, there exist bending jet structures extending to the northeast of the core (Section \ref{sect:3.2}) that display similar morphology at each frequency at the time of our observations. The extended emission seen at 8\,GHz can be detected up to $\sim$8\,mas from the core, but it becomes weaker at higher frequencies and is gradually resolved out with increasing resolution. The jet structure is highly core-dominated and, on average, the ratio of core flux to total VLBI flux ($\frac{S_{\rm core}}{S_{\rm {total}}}$) is 66.7, 83.0, 88.0, 77.9\,\% at 8, 15, 22 and 43\,GHz, respectively.

\subsection{Core identification}
\label{sect:3.2}
\begin{figure*}
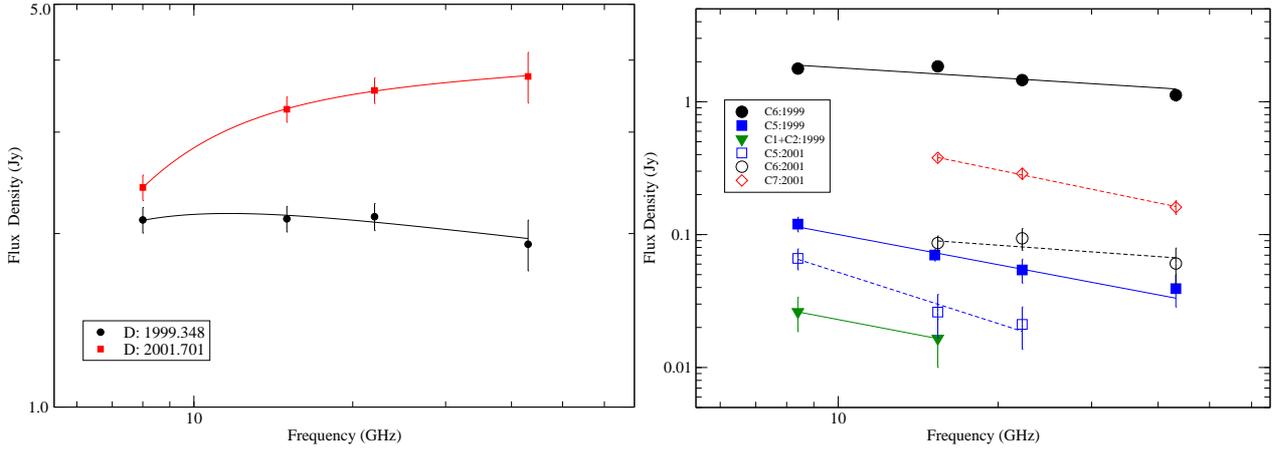

\centering
\includegraphics[width=0.45\textwidth,clip]{./fgs/fg2_1.eps}
\includegraphics[width=0.45\textwidth,clip]{./fgs/fg2_2.eps}
\caption[Component spectral of PKS\,1749+096]{Components' spectra of PKS\,1749+096. Left: Spectrum of the core. SSA fits are shown as solid lines. Right: Spectra of jet components.}
\label{fig:spectra}
\end{figure*}

The two-epoch (1999.348, 2001.701) multi-frequency observations enable us to
study the spectra of these VLBI components and then to identify the VLBI core.
In Fig.~\ref{fig:spectra} (left), we show the spectra of component D located at the south end of the jet.
The spectra of other components are shown in Fig.~\ref{fig:spectra} (right). At 1999.348, component D had a flat spectrum, while at 2001.701, it showed an obvious turnover below 43\,GHz, with its optically thin spectrum not measured at this epoch. In addition, component D is also the most compact component based on the modeled size (see Table~2 of the online material). Therefore, we identify component D as the compact core of PKS~1749+096. We used the synchrotron self-absorption (SSA) model~\citep{1974ApJ...192..261J} to fit the spectrum of each component,
\begin{equation}
S_{\nu}=S_{0}\nu^{2.5}[1-\rm exp(-\tau_{s}\nu^{\alpha-2.5})],
\end{equation}
where, $\nu$ is the observing frequency in GHz, $S_{0}$ the intrinsic flux density in Jy at 1\,GHz, $\tau_{\rm s}$ the SSA opacity at 1\,GHz, and $\alpha$ the optically thin spectral index. The best-fit results are $\alpha_{\rm D}=-0.11\pm0.07, S_0=58.4\pm42.2\,\rm mJy, \tau_{s}=50.3\pm46.5$ (1999.348), and $\alpha_{\rm D}=0.05\pm0.01, S_0=24.6\pm0.6\,\rm mJy, \tau_{s}=125.8\pm7.1$~(2001.701). For the first epoch, the accuracy of the fitted parameters is limited by the accuracy of the individual flux density measurements at each frequency and by the limited frequency range of our observations. Jet components show a typically steep spectrum ($\alpha \leq-0.75$) with the exception of C6 ($\alpha=-0.3$--$-0.2$), which may be due to its interaction with the surrounding interstellar medium. The fitted spectral indices for all these components are summarized in Table \ref{tab:spectra}.

\begin{table}
\caption[Results of spectral fitting for jet components of PKS\,1749+096]{Results of spectral fitting for jet components of PKS\,1749+096\label{tab:spectra}}
\centering
\begin{tabular}{ccc}
  \hline
  ID& 1999.348 &2001.701\\
  \hline
  D &$-0.11\pm0.07$ &$~~~0.05\pm0.01$\\
  C7& ... & $-0.82\pm0.03$\\
  C6& $-0.25\pm0.11$ &$-0.28\pm0.25$\\
  C5&$-0.75\pm0.10$ &$-1.28\pm0.21$\\
  C1+C2& $-0.76\pm0.81$ & ... \\
  \hline
\end{tabular}
\end{table}

The magnetic field $B$ of a homogeneous synchrotron self-absorbed source component can be calculated via~\citep[i.e.,][]{1983ApJ...264..296M}
\begin{equation}
\label{eq:eq1}
B_{\mathrm syn}=10^{-5}b(\alpha)\nu_{\mathrm max}^5\theta^{4}S_{\mathrm max}^{-2}\delta/(1+z),
\end{equation}
here $\nu_{\rm max}$ is the peak frequency in GHz, $\theta$ the source angular size in mas, $S_{\rm max}$ the peak flux density in Jy, $\delta$ the Doppler factor, and $b(\alpha)$ a tabulated function of the spectral index $\alpha$. By using $S_{\mathrm max}$ = 2.2\,Jy and $\nu_{\mathrm max}=11.6$\,GHz from the above SSA fitting for 1999.348 data and adopting $b (\alpha) \simeq1.8$ and $\theta=\theta_{G}\times1.8$\,mas, where $\theta_{G}$ is the fitted-Gaussian FWHM and the factor of 1.8 accounts for the Gaussian approximation to an optically thin uniform sphere, we obtained $B_{\rm syn}=3.9\times10^{-2}\delta$\,mG.

Assuming the energy equipartition between the radiative particles and magnetic field, one
can also estimate the magnetic field using the synchrotron luminosity $L$.
The energy of the particles can be written as
\begin{equation}
E_{e}=f(\alpha,\nu_{1},\nu_{2})LB^{-\frac{3}{2}},
\end{equation}
where, $L=4\pi D^{2}_{\rm L} \int^{\nu_2}_{\nu_1}S\,d\nu$, $f(\alpha,\nu_{1},\nu_{2})$ is a tabulated function,
$\nu_{1}$ and $\nu_{2}$ are the low and high cutoff frequency of normally $10^7$ and $10^{11}$\,Hz, respectively~\citep{1970ranp.book.....P}. In the case of spherical symmetry, the energy density of the magnetic field is $\frac{B^2}{8\pi}$, and
the magnetic field energy of a source (with radius $R$) is
\begin{equation}
E_{B}=\frac{B^{2}}{8\pi}\cdot\frac{4}{3}\pi R^{3}=\frac{B^{2}R^{3}}{6}.
\end{equation}

If the energy ratio of heavy particles to electrons is $\eta$, then the total energy can be expressed as
\begin{equation}\label{eq:energy}
\begin{split}
E_{\mathrm{total}}&=(1+\eta)E_{e}+E_{B}\\
&=(1+\eta)f(\alpha, \nu_{1},\nu_{2})LB^{-\frac{3}{2}}+\frac{B^{2}R^{3}}{6}.
\end{split}
\end{equation}

$\eta$ is determined by the matter content of the jet, and in an electron-positron pair plasma, $\eta \approx 1$, while in a ``normal'' electron-proton plasma, $\eta \approx 2000$. It seems to be reasonable to adopt $\eta=100$~\citep{1970ranp.book.....P}. Note that $E_{e}\varpropto B^{-\frac{3}{2}}$ and $E_{B}\varpropto B^{2}$, so the total energy $E_{\mathrm{total}}$ has a minimum, and in this case, the magnetic filed is
\begin{equation}\label{eq:equipartition}
\begin{split}
B_{\mathrm{eq}} &=\left(\frac{9}{2}(1+\eta)f(\alpha,\nu_{1}\nu_{2})LR^{-3}\right)^{2/7}\\
&\simeq4.7\times10^{-2}(S_{\mathrm{max}}\nu_{\mathrm{max}}D_{\rm L}^{-1}\theta^{-3})^{2/7},
\end{split}
\end{equation}
where $f(\alpha,\nu_1,\nu_2)=1.06\times10^{12}\left(\frac{2\alpha+2}{2\alpha+1}\right)\frac{\nu_{1}^{(1+2\alpha)/2}-\nu_{2}^{(1+2\alpha)/2}}{\nu_{1}^{1+\alpha}-\nu_{2}^{1+\alpha}}$, $f(-0.11,10^7,10^{11})=0.74\times10^7$, and $D_{\rm L}$ is the luminosity distance in Gpc. Using the same parameters for deriving $B_{syn}$, we obtain $B_{eq}=0.8\,\rm G$. This is much larger than that derived from SSA, indicating that either the radiation is strongly Doppler-boosted or that the source is particle-dominated. In the former case, since $B_{\mathrm{syn}}$ and $B_{\mathrm{eq}}$ depend differently on the Doppler factor $\delta$ (cf. Equations \ref{eq:eq1} and \ref{eq:equipartition}), one obtains $\frac{B_{\mathrm{eq}}}{B_{\mathrm {syn}}}\simeq(\frac{\delta}{1+z})^{6-\frac{16\alpha}{7}}$~\citep[e.g.,][]{2006A&A...456...97F}. Correspondingly, the derived Doppler factor is the equipartition Doppler factor $\delta_{\rm eq}$, and for the core component D, we obtain~$\delta^{\mathrm D}_{\mathrm{eq}}=5.0$. In this case, we get $B_{\mathrm syn} =  0.2$\,mG, which seems to be a low value for an AGN jet. In the latter case, the ratio of energy density in particles ($u_{p}$) to magnetic field ($u_{m}$) can be written as $\frac{u_{p}}{u_{m}}=(\frac{B_{\mathrm eq}}{B_{\mathrm syn}})^{17/4}$, according to~\citet{1994ApJ...426...51R}. Thus, we obtain $\frac{u_p}{u_m} \simeq 2.1 \times10^{18}$ (ignoring relativistically beaming effect), indicating that the core is strongly particle-dominated.

\subsection{Jet kinematics}
\label{sect:kinematics}
\subsubsection{22\,GHz}
\begin{figure*}
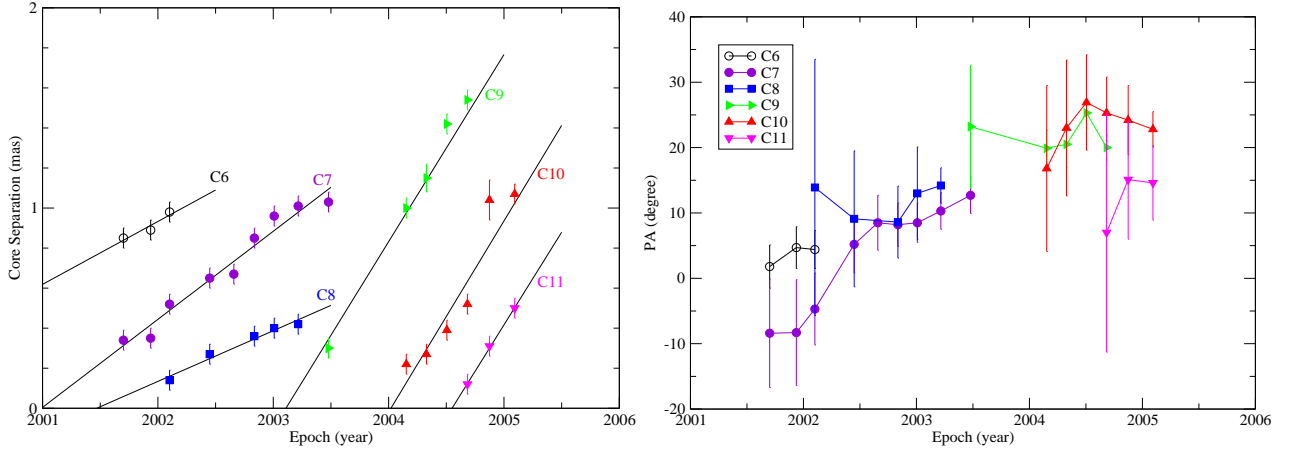

\centering
\includegraphics[width=0.45\textwidth,clip]{./fgs/fg3_1.eps}
\includegraphics[width=0.45\textwidth,clip]{./fgs/fg3_2.eps}
\caption{Core separation (left) and PA (right) of jet components plotted as a function of time at 22\,GHz. The solid lines represent linear fits (left, see Table~\ref{tab:linear_fit_22GHz}). C5 is not shown to obtain a clear plot for the inner jet components.}
\label{fig:linear_fit_22GHz}
\end{figure*}

To study the jet kinematics, we show in Fig.~\ref{fig:linear_fit_22GHz} (left)
the core separation as a function of time along with the linear regression fits through which the apparent speeds of the individual jet components are measured. The fitted results are summarized in Table~\ref{tab:linear_fit_22GHz}. Figure~\ref{fig:linear_fit_22GHz} (right) plots the time evolution of the PA of jet components. As a result seven components (C5--C11) are identified at 22\,GHz. However, the identification of C9 is possibly subject to some uncertainties because of the time gaps during 2003--2004. All these components move away from the core, with proper motion in the range of 0.25--1.05\,mas/yr, corresponding to apparent speeds of 5--21\,$c$. Interestingly, the apparent speeds seem to follow a bimodal distribution.
C9, C10, and C11 move 2--3 times faster than the other components (C5, C6, C7, and C8).
It can be seen from Fig.~\ref{fig:linear_fit_22GHz} (right) that the PAs for most jet components show no significant variations. The averages of the PAs for these components are summarized in Table~\ref{tab:linear_fit_22GHz}, too. One exception is for component C7, whose PA systematically varies with time. A discussion of this result is deferred to the following section.

\begin{table}
\begin{minipage}[ht!]{\hsize}
\caption{\label{tab:linear_fit_22GHz}Results of the linear fits to the component motion at 22\,GHz.
Listed are the component Id., number of epochs for the fit, fitted proper motion $\mu$, apparent velocity $\beta_{\rm app}$, ejection time of each component $t_0$, and average PA.}
\centering
\begin{tabular}{ccccccc}
\hline
Id. &\#  & $\mu$& $\beta_{\rm app}$ & $t_{0}$&PA\\
\hline
&& [mas/year] &[$c$] & [year] &[degree]\\
\hline
C5 &7&$0.43\pm0.03$&$ 8.6\pm0.6$&$1996.2\pm0.4$&37.2$\pm5.2$\\
C6 &3&$0.31\pm0.10$&$ 6.2\pm2.0$&$1999.0\pm1.0$&$3.7\pm1.5$\\
C7 &9&$0.44\pm0.03$&$ 8.8\pm0.6$&$2001.0\pm0.1$&-\\
C8 &5&$0.25\pm0.03$&$ 5.0\pm0.6$&$2001.5\pm0.1$&$12.9\pm2.5$\\
C9 &5&$1.05\pm0.04$&$21.0\pm0.8$&$2003.2\pm0.1$&$21.9\pm1.2$\\
C10&6&$0.95\pm0.14$&$19.0\pm2.8$&$2004.0\pm0.1$&$23.5\pm3.4$\\
C11&3&$0.93\pm0.04$&$18.6\pm0.8$&$2004.5\pm0.1$&$14.2\pm2.4$\\
\hline
\end{tabular}
\end{minipage}
\end{table}

\subsubsection{15\,GHz}
\label{sect:15GHz}
\begin{figure*}
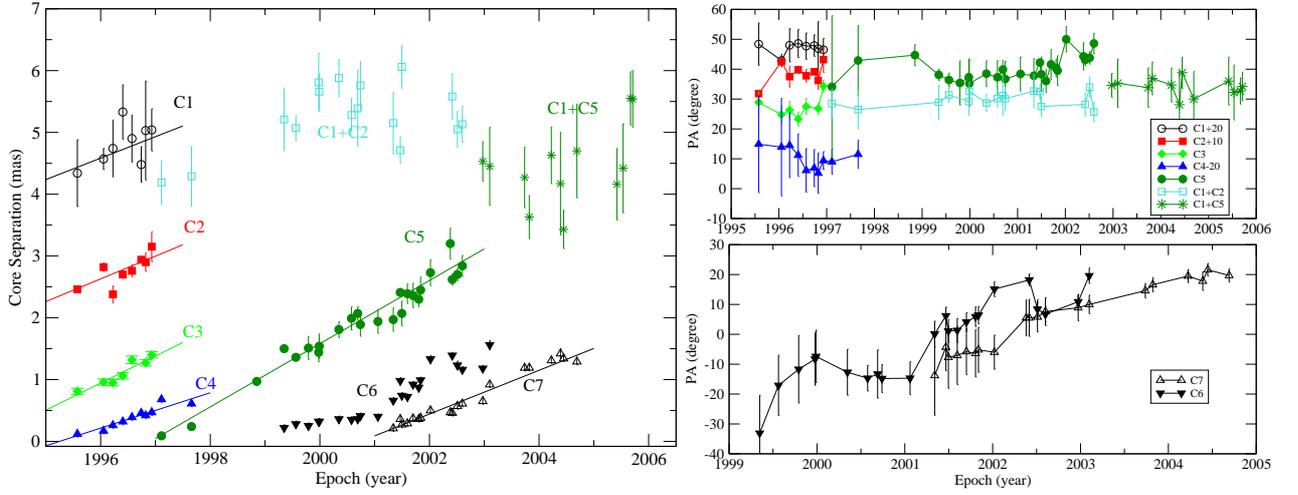

\centering
\includegraphics[width=0.48\textwidth,clip]{./fgs/fg4_1.eps}
\includegraphics[width=0.42\textwidth,clip]{./fgs/fg4_2.eps}
\caption{Core separation (left) and PA (right) of jet components plotted as a function of time at 15\,GHz. The solid lines represent linear fits (see Table~\ref{tab:linear_fit_15GHz}). The PAs of C1, C2, and C4 are shifted for a better visualization (upper right).}
\label{fig:r_t_15GHz}
\end{figure*}

\begin{table}
\begin{minipage}[t]{\hsize}
\caption{\label{tab:linear_fit_15GHz}Results of the linear fits to the component motion at 15\,GHz.
Listed are the component Id., number of epochs for the fit, fitted proper motion $\mu$, apparent velocity $\beta_{\rm app}$, ejection time of each component $t_0$, and average PA.}
\centering
\begin{tabular}{ccccccc}
\hline
Id. &\#  & $\mu$& $\beta_{\rm app}$ & $t_{0}$&PA\\
\hline
&& [mas/year] &[$c$] & [year] &[degree]\\
\hline
C1& 8&$0.35\pm0.25$&$ 7.0\pm5.0$&$1982.8\pm9.5$&$25.6\pm2.6$\\
C2& 8&$0.37\pm0.09$&$ 7.4\pm1.8$&$1988.8\pm1.8$&$28.1\pm3.8$\\
C3& 7&$0.44\pm0.07$&$ 8.8\pm1.4$&$1993.8\pm0.4$&$27.9\pm4.1$\\
C4&10&$0.29\pm0.04$&$ 5.8\pm0.8$&$1995.2\pm0.2$&$29.3\pm2.2$\\
C5&25&$0.51\pm0.02$&$10.2\pm0.4$&$1996.9\pm0.1$&$41.2\pm3.2$\\
C7&20&$0.36\pm0.02$&$ 7.2\pm0.4$&$2000.8\pm0.1$&-\\
\hline
\end{tabular}
\end{minipage}
\end{table}
In view of the bimodal distribution of apparent jet speeds at 22\,GHz and the ``mode'' switch for jet kinematics in some BL Lac objects~\citep{2010A&A...515A.105B}, we then considered the 15\,GHz MOJAVE data with relatively longer time coverage. However, the irregular time sampling (e.g., gaps around 1998--1999 and 2005) and the complexity of the jet kinematics make it nearly impossible to trace all the jet components. We therefore turned to investigate a few features for which we were able to follow the evolution during the studied period to obtain the main characteristics of the jet kinematics. Fig~\ref{fig:r_t_15GHz} (left) shows the core distance evolution of the modeled components with obvious proper motion at 15\,GHz together with the linear regression fit.
Table~\ref{tab:linear_fit_15GHz} summarizes the fitting results. These components show similar speeds, and for component C5 and C7, the fitted speeds and ejection time at 15 and 22\,GHz are consistent (Tables~\ref{tab:linear_fit_22GHz} and \ref{tab:linear_fit_15GHz}), considering the different time coverage and data points at each frequency. For component C6, the core separation does not increase linearly with time and therefore a linear regression was not considered. \citet{2000PASJ...52.1037I} measured the proper motion of component C5 (C1 in their case) $0.68\pm0.14$\,mas yr$^{-1}$. Based on the data of three epochs, the inferred ejection time of May 1997 is slightly later than what we derived for component C5 (1996.9). \citet{2000PASJ...52.1037I} deduced that the core contained an unresolved component extending to the northwest along a PA of $\sim -30^{\circ}$ during their observations in 1998. It turns out that this component can be identified as component C6 from our analysis (Table~\ref{tab:online_model} of the online material). Although bearing relatively large uncertainties because of the limited data (three epochs) at 22\,GHz, the ejection time (around 1999) and PA (along $\sim -30^{\circ}$) of component C6 are identical to the unresolved component. This component was also detected by space-based VLBI at 5\,GHz~\citep[epoch: 2001.24][]{2003ASPC..299...99G}. \citet{2001ApJ...549..840H} reported a proper motion of 0.45\,mas/yr for component C4 (their U3) based on the data of six epochs in 1996, which is slightly faster than our estimated speed of 0.29\,mas/yr. They considered that the position of this component was affected by the strong core in some epochs.
Similarly, this effect at 1997.658 may also explain the relatively low speed. In addition, \citet{2009AJ....138.1874L} reported a maximum apparent speed of 6.8\,$c$ in PKS\,1749+096.

\begin{figure}[ht!]
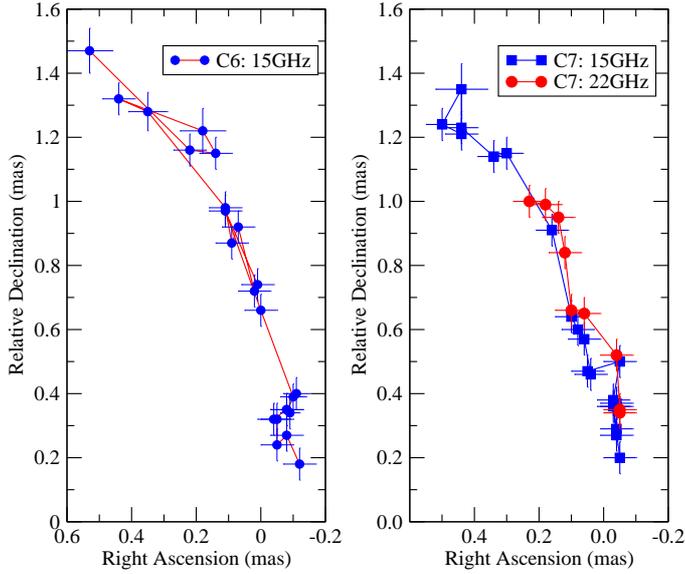

\centering
\includegraphics[scale=0.45,clip]{./fgs/fg5_1.eps}
\includegraphics[scale=0.45,clip]{./fgs/fg5_2.eps}
\caption{Trajectory of the components C6 an C7 in the sky plane.}
\label{fig:C6_7_22_ra_dec}
\end{figure}

In Fig.~\ref{fig:r_t_15GHz}~(right) we show the time evolution of the PA of jet components.
Most components keep nearly constant PA when they move downstream of the jet. The average PAs for these components are given in Table~\ref{tab:linear_fit_15GHz}. Obviously, C6 and C7 are two exceptions with their PAs showing systematic variations with time. In Fig.~\ref{fig:C6_7_22_ra_dec}, the trajectories of component C6 and C7 in the sky plane are displayed, which exhibits a clear non-radial, non-ballistic motion.

\begin{figure}
\centering
\includegraphics[width=0.475\textwidth,clip]{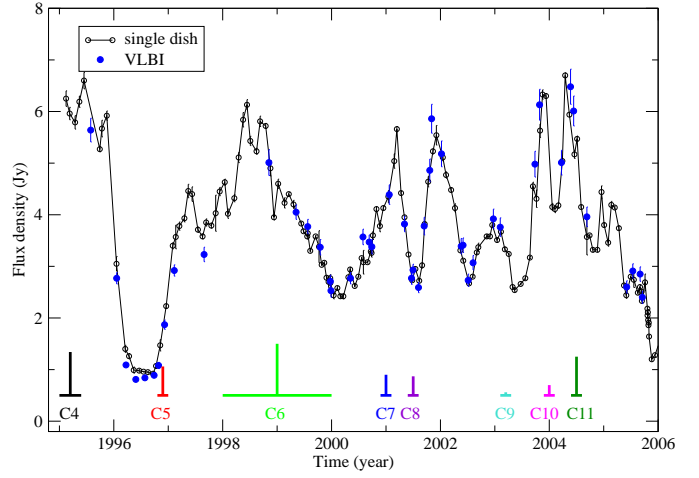}
\caption{Light curves of PKS 1749+096 from total VLBI flux density data (filled circles) and single-dish flux density data (open circles, UMRAO data) at 15\,GHz. The fitted component ejection time is indicated by the vertical lines, whose length is proportional to the flux density of each component (C4--C7: 15\,GHz, C8--C11: 22\,GHz) when first seen.  Horizontal bars along the time axis denote the uncertainty of the ejection time of each labeled component.}
\label{fig:total_flux}
\end{figure}

In many AGN, radio outbursts are believed to be connected with VLBI component ejections~\citep[e.g.,][]{2002A&A...394..851S}.
Fig.~\ref{fig:total_flux} shows the evolution of the UMRAO total flux density (open circles)
and of the VLBI total flux density (filled circles) at 15\,GHz. The agreement of these two quantities demonstrates the reliability of the VLBI amplitude calibration. In Fig.~\ref{fig:total_flux}, the fitted ejection time for components C4--C11 is also marked. 
Evidently, the ejection of jet components is connected with the radio outbursts, even though not all events are linked in a similar manner (e.g., C5 and C8 appear to be ejected early in the rise phase of the flux density), as indicated by the data. Given the time coverage of the available data, and because acceleration/ deceleration could affect the linear back-extrapolation of the ejection time when the component is coincident with the core, it is hard to accurately estimate the time difference between the ejection of a component and the outburst from these data. By considering that there are about 10 years of total flux density measurements and a time window of, on average, about 0.31 yr, within which the flare peaks and component ejections would be considered as related, we estimate the probability that all eight events are connected by chance is $\sim 3.4\times10^{-6}$. Therefore, we conclude that, at least in PKS\,1749+096, there exists correlation between the emergency of VLBI components and radio outbursts. 

 Also shown in Fig.~\ref{fig:total_flux} are the normalized flux density (relative to C6) of each component when they were detected by our observations for the first time. It is interesting to note that the strength of the outburst seems to be correlated with the component brightness (e.g., C9). However, since components are typically not first detected at the same core distance and
they may evolve differently with time, they cannot be readily put into one-to-one correspondence (e.g., C10 and C11). \citet{2009AJ....137.5022N} studied multi-frequency (5, 8, 15, 22, 37, and 90\,GHz) radio light curves of  PKS\,1749+096, and identified five major outbursts, peaked in 1993, 1995, 1998, 2001, and 2002. These outbursts correspond to the ejection time of components C3, C4, C6, C7, and C8\footnote{Note that the position of C8 may be affected by C9 at epoch 2003.216, when C9 was unresolved from the core, we therefore consider that the ejection time of C8 (2001.5) corresponds to the outburst in 2002.}, respectively.

\subsection{The component C5}
\begin{figure}
\centering
\includegraphics[width=0.475\textwidth,clip]{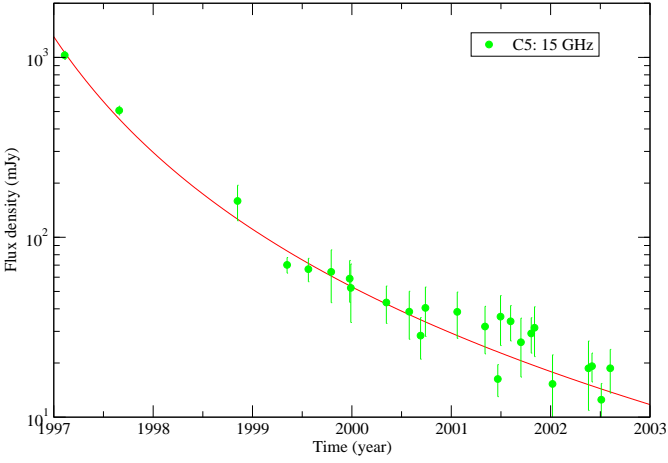}
\caption{Flux density evolution of component C5 at 15\,GHz. The solid line represents a
power-law fit. See text for details.}
\label{fig:C5_flux_t}
\end{figure}

Component C5 is the most reliably determined feature by the data. In Fig.~\ref{fig:C5_flux_t}, we plot its 15 GHz flux density as a function of time. The flux density of C5 shows a rapid decrease after its birth until it reaches a flux density of $\sim $20\,mJy in 2002 at 15\,GHz. In contrast, the total flux density of PKS~1749+096 (see Fig.~\ref{fig:total_flux}) during this period does not show such a monotonic behavior, but instead shows three major flares. The rate of the flux density decay of C5 slows down as it moves outward, and the evolution can be described reasonably well by a power law of the form of $S(t) \sim (t-t_{0})^{\kappa}$ with $t_{0}= 1995.45 \pm 0.01$ and $\kappa=-2.97 \pm0.07$, which is indicated as a solid line in Fig.~\ref{fig:C5_flux_t}.

\begin{figure}
\centering
\includegraphics[width=0.475\textwidth,clip]{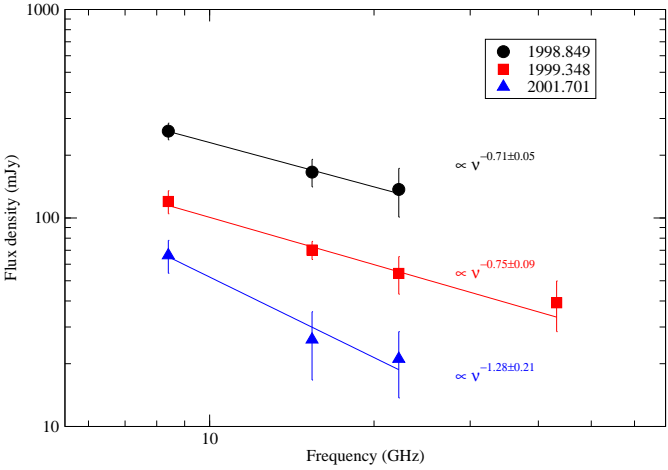}
\caption{Spectral evolution of component C5. The solid lines represent power-law fits.}
 \label{fig:C5_spectra}
\end{figure}

In Fig.~\ref{fig:C5_spectra}, we show the spectrum of C5 determined by simultaneous multi-frequency observations in 1998~\citep{2000PASJ...52.1037I}, 1999, and 2001, respectively.
The spectral indices are $-0.71\pm0.05$\footnote{\citet{2000PASJ...52.1037I} obtained a spectral index of $-0.98\pm0.05$ on the basis of non-simultaneous data.} in 1998.849, $-0.75\pm0.09$ in 1999.348, and $-1.28\pm0.21$ in 2001.701, showing that the spectrum steepens with time at a probability of 97\,\%.  An analysis of multi-frequency (2.3, 8.4, 15.3, 22,2 and 43\,GHz) VLBA data on May 12, 2001 (epoch 2001.362, code: BW055) yields a spectral index of $-1.17\pm0.10$ and additionally confirms the spectral steepening for component C5.

Components showing monotonic decrease in flux density as they move outward from the core have been observed in a number of sources \citep[e.g., 3C\,345,0735+178,][]{1983ApJ...271..536U,1994ApJ...435..140G,1994ApJ...435..128G}. The flux density evolution can be basically explained in terms of geometric Doppler boosting effects and intrinsic effects as the result of radiation or adiabatic expansion loss.

In the former case, the observed flux density ($S$) of a component is related to the intrinsic flux density ($S_{0}$) in the comoving frame by $S=S_{0}\delta^{3-\alpha}$~\citep{1979Natur.277..182S}. This formula indicates that the geometric effect is frequency-independent.
The steepening of the spectra, as shown in Fig.~\ref{fig:C5_spectra}, indicates the frequency-dependent behavior of flux density evolution for component C5. Furthermore, the trajectory needs to be sharply curved to explain the changes in flux density, which is inconsistent with the observed ballistic trajectory (section~\ref{sect:15GHz}).

In the latter case, a simple adiabatic expansion predicts $S(t) \sim (t-t_{0})^{\kappa}$, with
$\kappa=4\alpha-2$~\citep{1962SvA.....6..317K} and would maintain a constant spectral index~\citep{1966Natur.211.1131V}. For the fitted power law index of $\kappa=-2.97\pm0.07$, $\alpha=-0.24\pm0.02$, which is flatter than the measured indices. The fitted $t_0$ is earlier than the extrapolated time of ejection from kinematics, indicating that the expansion may begin with a finite flux density, instead of infinity. A plausible alternative could be adiabatic expansion with a continuous acceleration of the energy index $2\alpha-1$. This would lead to $\kappa=2\alpha-1$~\citep{1962SvA.....6..317K} and $\alpha=-0.98\pm0.04$, which roughly matches the measured spectral indices within errors.

The spectral steepening as seen in Fig.~\ref{fig:C5_spectra} is reminiscent of radiation losses. Following Eq.~\ref{eq:eq1}, we estimate the magnetic field strength to be $B_{\mathrm syn}=268$\,mG adopting $\nu_{\mathrm max}=2.37$\,GHz,  $\theta=2.52$\,mas, $S_{\mathrm max}=0.90$\,Jy, $b(\alpha)=3.6$ and $\delta_{eq}=2.5$~\citep{2000PASJ...52.1037I}.
The cooling time $t_{\frac{1}{2}}$ of synchrotron-emitting electrons follows $t_{\frac{1}{2}}\simeq2.76\times10^{4}B_{\mathrm syn}^{-1.5}\nu_{m}^{-0.5}$, where $t_{\frac{1}{2}}$ is in years, $B_{\mathrm syn}$ in mG, and $\nu_{\mathrm max}$ in GHz. This yields $t_{\frac{1}{2}}\sim$ 4.1 years for component C5, comparable to the time scale of flux density evolution. The estimated equipartition Doppler factor $\delta_{eq}$ of 2.5 indicates $\frac{u_{p}}{u_{m}}\simeq 5.2\times10^9$.  This perhaps means that C5 is particle-dominated and the inverse Compton scattering also contributes to the fast decrease in flux density of C5.

\subsection{Physical parameters}
One can estimate the physical parameters of the parsec-scale jet using the measured apparent speeds.
C5 ($\beta_{\rm app}=10.2\,c$) is the best-determined component in our experiments with as many as 25 data points, and
therefore will be used in the following calculations. The Doppler factor ($\delta$) is determined by the Lorentz factor $\Gamma$, the viewing angle of the jet $\theta$, and the jet speed in the source frame $\beta$:
\begin{equation}\label{eq:doppler}
\delta = \frac{1}{\Gamma (1 - \beta \cos \theta)},
\end{equation}
where $\Gamma=\frac{1}{\sqrt{1-\beta^2}}$.

The maximum allowable viewing angle $\theta_\mathrm{max}$ can be derived via
\begin{equation}
\sin \theta_\mathrm{max} = \frac{2\beta_\mathrm{app}}{(1 + \beta_\mathrm{app}^2)}
\end{equation}
and the minimum Lorentz factor is given by
\begin{equation}
\Gamma_\mathrm{min} = \sqrt{1 + \beta_\mathrm{app}^2}.
\end{equation}

For component C5, $\theta_\mathrm{max}=11.2^{\circ}$, $\Gamma_\mathrm{min}=10.2$. For a given speed in the source frame, $\beta$, the critical viewing angle, which maximizes the apparent speed, is given by $\theta_{\mathrm {cri}}=\mathrm {arcsin}(\frac{1}{\Gamma})$. $\theta_{\mathrm {cri}}=5.6^{\mathrm \circ}$ for component C5. According to equation~\ref{eq:doppler}, the minimum Lorentz factor $\Gamma_{\mathrm{min}}$ and the critical viewing angle $\theta_{\mathrm {cri}}$ lead to $\delta=\Gamma_{\mathrm{min}}=10.2$. This indicates that
the Doppler factor of component C5 is close to that of the core component D ($\delta^{\rm D}_{\rm{eq}}=5.0$). For even smaller viewing angles ($\theta \rightarrow 0$), Doppler factor approaches $\sim2\Gamma=20.4$.

\section{Discussion}
\label{sect:Discussion}
\subsection{Structure variability}
\begin{figure}[ht!]
\centering
\includegraphics[width=0.475\textwidth,clip]{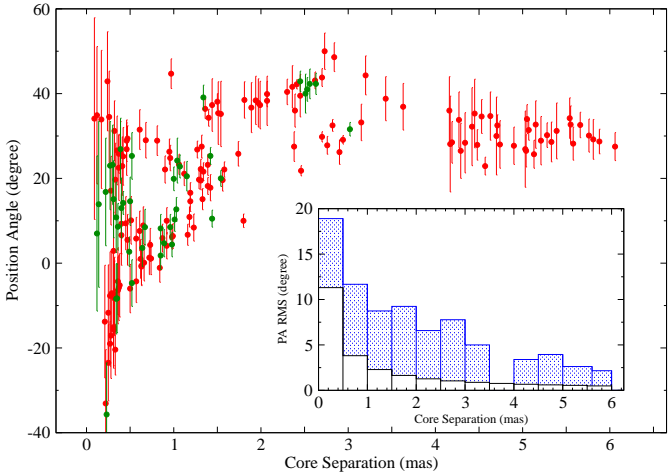}
\caption{Position angle plotted as a function of core separation for all jet components seen at 15 (red) and 22\,GHz (green).
The inset histogram is a plot of the scatter of PAs (shaded) in each bin (0.5\,mas) with comparison to the errors (blank) from the often quoted 1/5 of the beam size (see text for details).}
 \label{fig:pa_r}
\end{figure}

\begin{figure}[ht!]
\centering
\includegraphics[width=0.475\textwidth,clip]{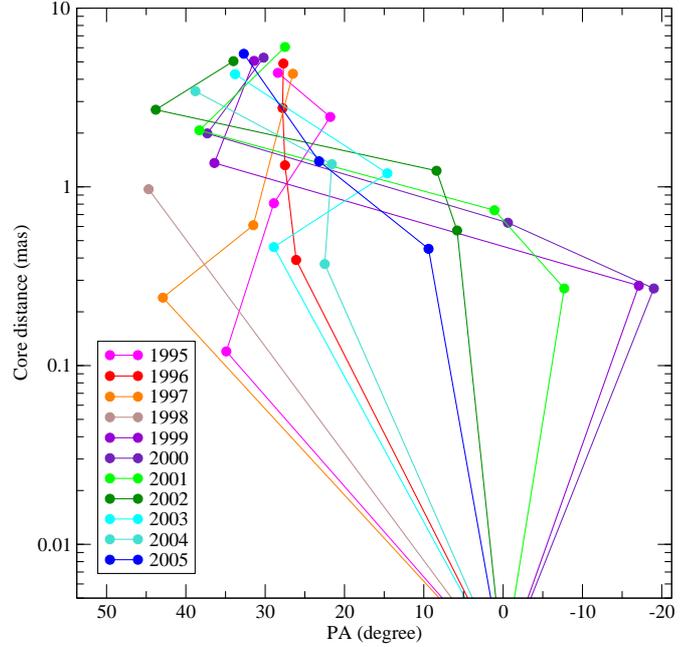}
\caption{Time evolution of jet ridge line at 15\,GHz. Only one epoch is shown for each year for clarity.}
 \label{fig:pa_r2}
\end{figure}

In Fig.~\ref{fig:pa_r}, we plot the PA distribution as a function of core separation at 15 and 22\,GHz.
To distinguish the ambiguity of measurement errors near the core, we also plot within each core separation bin
the scatter of jet PA with comparison to the typical measurement errors derived from the often quoted 1/5 of the beam size, which we took to be 0.5\,mas for these measurements, with most taken at 15\,GHz (Fig.\ref{fig:pa_r}, inset). Clearly, the scatter of the PA in each bin is always larger than the typical errors, and therefore there is no fixed trajectory that all the jet components can follow within the central $\sim 4-6\,\rm{mas}$. It should be noted that the time sampling of these observations is not equal (e.g., there were seven epochs in 1996, but only one epoch in 1998). In view of this, Fig.~\ref{fig:pa_r2} displays the time evolution of the jet ridge line (defined as the line connecting the modeled jet components), which clearly shows the change of the PA.

From Tables~\ref{tab:linear_fit_22GHz} and \ref{tab:linear_fit_15GHz}, it is obvious that the jet components were ejected at various PAs, and this is particularly true for components C6 and C7. Indeed, dramatic jet-ejection-angle variation is not uncommon in many AGNs
\citep[e.g., in PKS\,0048-097,][]{2006A&A...456L...1K}, and a precessing jet scenario is often cited as the explanation. The jet precession can be triggered either by a binary black hole system or by a warped accretion disk. A simple scenario is a precessing ballistic jet where jet components ejected at different times move ballistically to different directions~\citep[e.g.,][]{2003MNRAS.341..405S}. A jet precession would lead to
the change of the inner jet structure and of the component ejection angle, which was indeed observed in PKS\,1749+096. However, these ejection angles are distributed in a random rather than systematic manner, and therefore do not appear to be consistent with a precessing jet scenario.
On the other hand, the variation timescale of the jet ejection angle seems to be much shorter than the precession period~\citep{2005A&A...431..831L}.

\subsection{Kinematics of BL Lac objects}
\citet{2010A&A...515A.105B} recently pointed out that BL Lac objects differ appreciably in the parsec-scale jet kinematics from quasars. According to these authors, the jet kinematics in BL Lac objects cannot always be described by the standard AGN paradigm of apparent superluminal motion. Instead, there exists structural mode change, i.e., transition from ``typical superluminal'' to an unusual ``stationary'' state. For PKS\,1749+096, a similar mode change is not found. The reason could be that the time coverage of the observations is not long enough.
In contrast, we found that there is a bimodal distribution of apparent superluminal
speeds and there coexist ballistic and non-ballistic components, which caused the bending of the jet.
\citet{2011A&A...526A..51K} also detected a component similar to components C6 and C7 in the quasar B0605-085 and attributed this to jet precession. However, the reasons for the existence of a bimodal distribution of apparent speeds and whether this bimodality has any correlation with other properties of the jet, such as the flux density evolution, are unclear. It is not readily obvious that the kinematic behavior of the pc-scale jet is simply a result of a changing viewing angle of the jet. The flux density evolution shows nothing special that could be linked to the sources of each bimodal population, which is complicated by the fact that most, if not all,  of the outbursts are connected with the ejection of jet components.

The apparent jet speeds of BL Lac objects are generally lower than those of quasars, typically $\leq 5\,c$~\citep{2001ApJS..134..181J,2009A&A...494..527H}. Our observations suggest that the jet in PKS\,1749+096 moves faster than in other BL Lac objects with an average apparent jet speed of $\sim9\,c$ and with some components moving as fast as $21\,c$. In this sense, PKS\,1749+096 is an unusual BL Lac object, showing superluminal speeds similar to quasars ($\sim10$--$20\,c$).

\section{Summary}
\label{sect:Summary}
We studied the parsec-scale jet of PKS\,1749+096 by using high-resolution multi-epoch (total of 65 observations in 61 epochs spanning $\sim$10 years) VLBI observations at 8, 15, 22, and 43\,GHz. On parsec scales, PKS\,1749+096 showed a core-jet structure with the jet extending to the northeast. The compact core contains, on average, $\sim$80\,\% of the total VLBI flux density. Using two-epoch (1999.348 and 2001.701) multi-frequency simultaneous observations, we identified the component {\sl D} as the VLBI core based on its flat spectrum and compactness. The derived equipartition Doppler factor for the core was found to be consistent with the Doppler factor of jet component (C5) estimated from kinematics. 
We found that models of adiabatic expansion or Doppler boosting cannot solely account for the flux density decay of component C5. Although the synchrotron cooling time scale is comparable to the decay time scale, this component was probably far away from energy equipartition and was dominated by particles. The fast decay of flux density may be caused partly by inverse Compton scattering effects in addition to the expansion.

The study of the jet kinematics shows that the components exhibited a bimodal distribution of superluminal speeds in PKS\,1749+096, i.e., C9, C10, and C11's proper motions are faster than that of other components. The available evidence shows that ballistic and non-ballistic components coexisted, though the cause for this is still unclear. The superluminal motion of jet components is in the range of 5--21\,$c$. In PKS\,1749+096, the component ejection correlates in time with radio flux density outburst, supporting the commonly assumed connection between the two.

\renewcommand{\thesubfigure}{\thefigure\,(\arabic{subfigure})}
\captionsetup[subfigure]{labelformat=simple,labelsep=colon,
listofformat=subsimple}
\makeatletter
\renewcommand{\p@subfigure}{}
 \onlfig{1}{
  \begin{figure*}
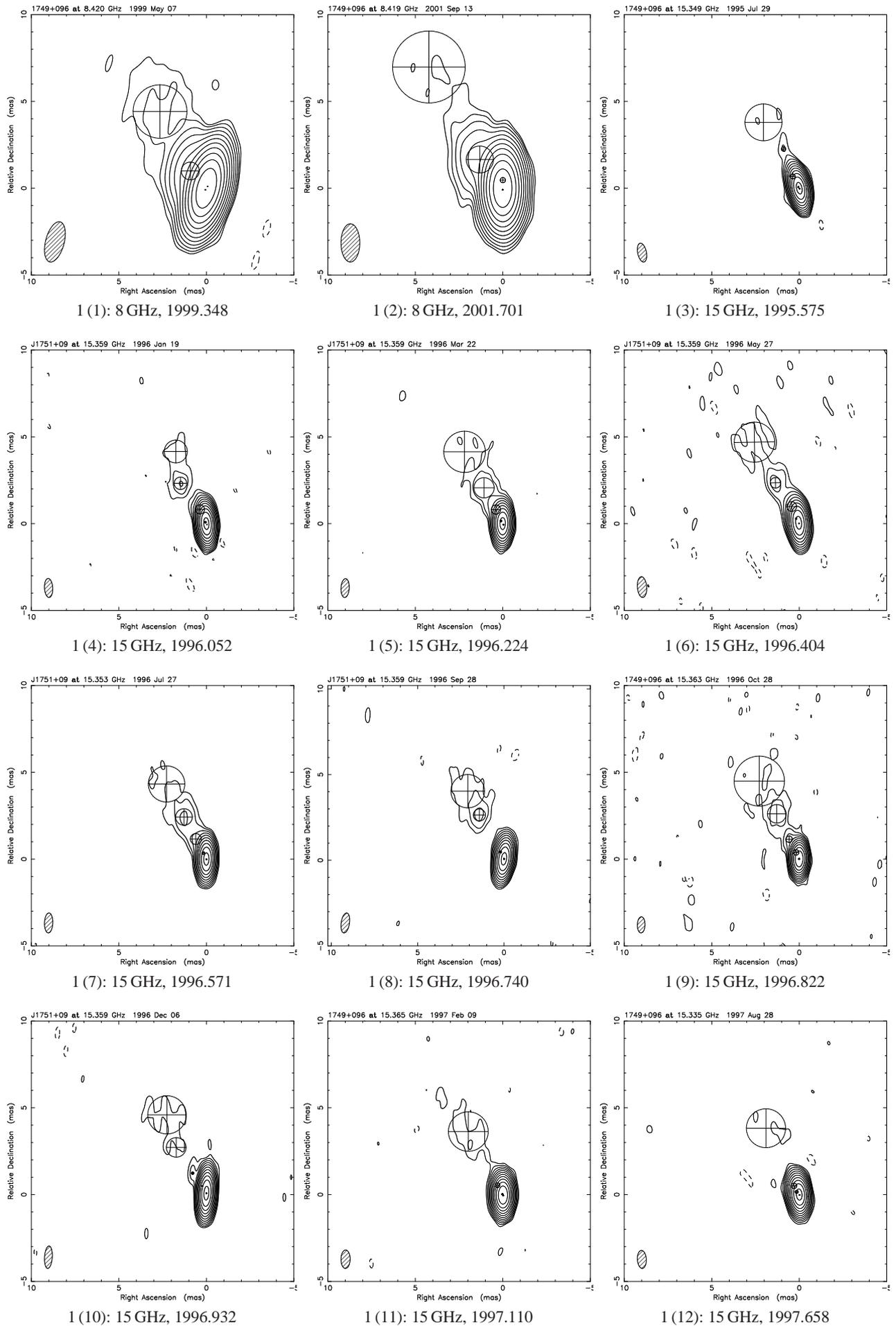
%
  \centering
  \subfloat[][8\,GHz, 1999.348]{%
  \label{fig:x1}%
  \includegraphics[width=0.30\textwidth,clip]{./fgs/X.1999.348.ps}}
  \hspace{3pt}%
  \subfloat[][8\,GHz, 2001.701]{%
  \label{fig:x2}%
  \includegraphics[width=0.30\textwidth,clip]{./fgs/X.2001.701.ps}}
  \hspace{3pt}%
  \subfloat[][15\,GHz, 1995.575]{%
  \label{fig:u-1}%
  \includegraphics[width=0.30\textwidth,clip]{./fgs/U.1995.575.ps}}\\
  \hspace{0pt}%
  \subfloat[][15\,GHz, 1996.052]{%
  \label{fig:u-2}%
  \includegraphics[width=0.30\textwidth,clip]{./fgs/U.1996.052.ps}}
  \hspace{3pt}%
  \subfloat[][15\,GHz, 1996.224]{%
  \label{fig:u-3}%
  \includegraphics[width=0.30\textwidth,clip]{./fgs/U.1996.224.ps}}
  \hspace{3pt}%
  \subfloat[][15\,GHz, 1996.404]{%
  \label{fig:u-4}%
  \includegraphics[width=0.30\textwidth,clip]{./fgs/U.1996.404.ps}}\\
  \hspace{0pt}%
  \subfloat[][15\,GHz, 1996.571]{%
  \label{fig:u-5}%
  \includegraphics[width=0.30\textwidth,clip]{./fgs/U.1996.571.ps}}
  \hspace{3pt}%
  \subfloat[][15\,GHz, 1996.740]{%
  \label{fig:u-6}%
  \includegraphics[width=0.30\textwidth,clip]{./fgs/U.1996.740.ps}}
  \hspace{3pt}%
  \subfloat[][15\,GHz, 1996.822]{%
  \label{fig:u-7}%
  \includegraphics[width=0.30\textwidth,clip]{./fgs/U.1996.822.ps}}\\
  \hspace{0pt}%
  \subfloat[][15\,GHz, 1996.932]{%
  \label{fig:u-8}%
  \includegraphics[width=0.30\textwidth,clip]{./fgs/U.1996.932.ps}}
  \hspace{3pt}%
  \subfloat[][15\,GHz, 1997.110]{%
  \label{fig:u-9}%
  \includegraphics[width=0.30\textwidth,clip]{./fgs/U.1997.110.ps}}
  \hspace{3pt}%
  \subfloat[][15\,GHz, 1997.658]{%
  \label{fig:u-10}%
  \includegraphics[width=0.30\textwidth,clip]{./fgs/U.1997.658.ps}}
  \caption{Clean images of PKS\,1749+096.}%
  \label{fig:online_maps}%
  \end{figure*}
  \begin{figure*}%
  \centering
  \ContinuedFloat
  \hspace{0pt}%
  \subfloat[][15\,GHz, 1998.849]{%
  \label{fig:u-11}%
  \includegraphics[width=0.30\textwidth,clip]{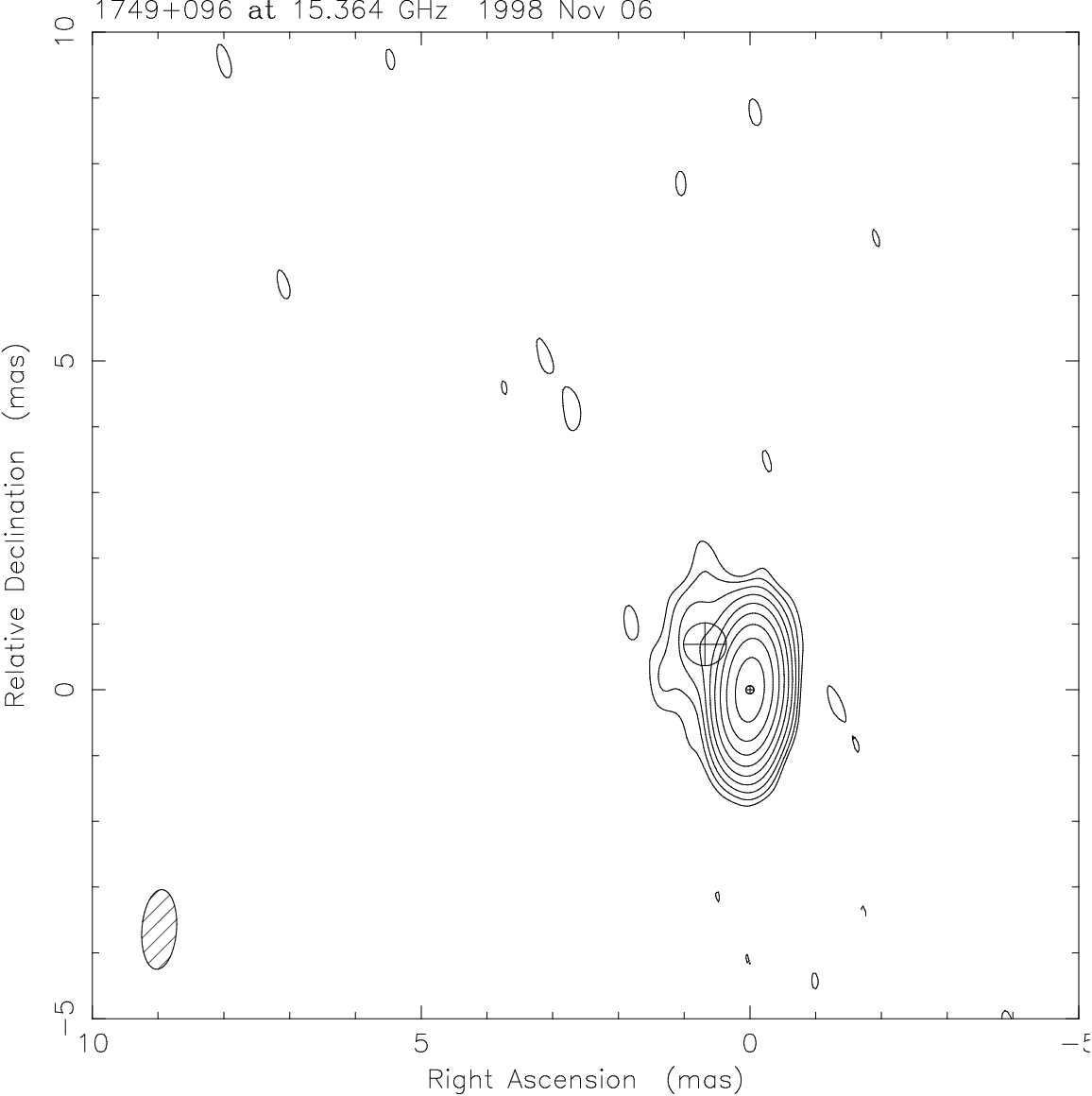}}
  \hspace{3pt}%
  \subfloat[][15\,GHz, 1999.348]{%
  \label{fig:u-12}%
  \includegraphics[width=0.30\textwidth,clip]{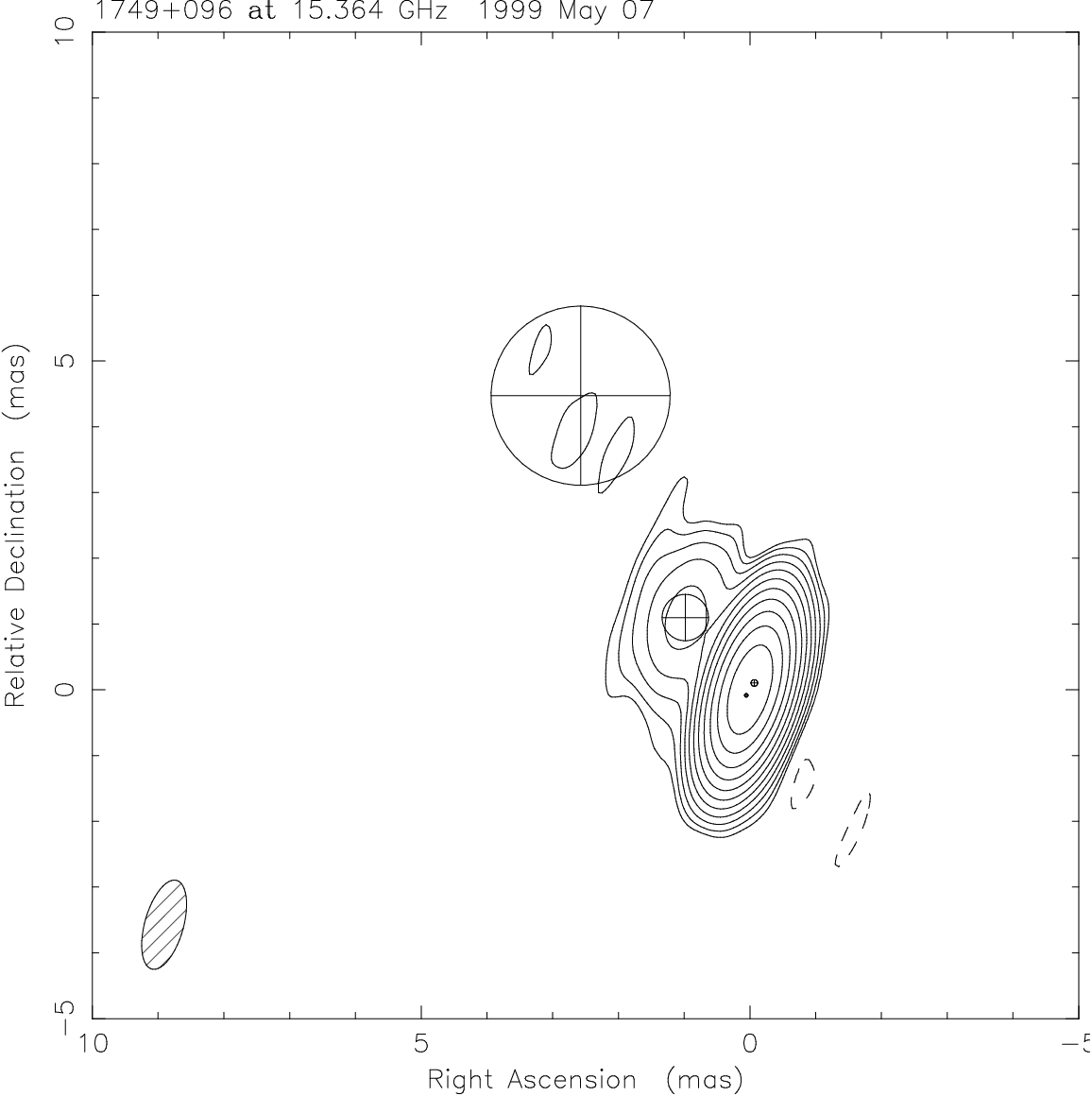}}
  \subfloat[][15\,GHz, 1999.562]{%
  \label{fig:u-13}%
  \includegraphics[width=0.30\textwidth,clip]{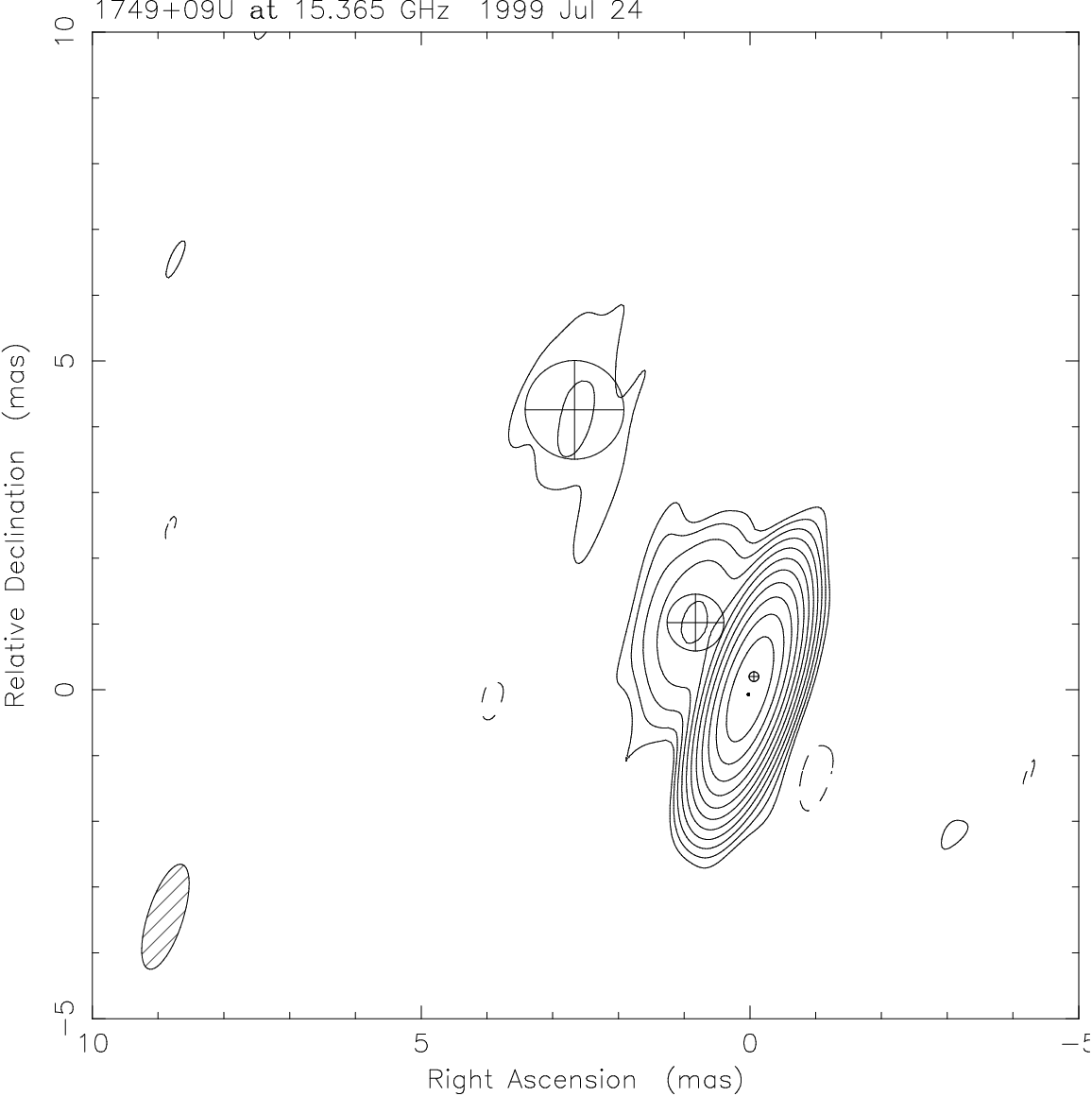}}\\
  \hspace{0pt}%
  \subfloat[][15\,GHz, 1999.792]{%
  \label{fig:u-14}%
  \includegraphics[width=0.30\textwidth,clip]{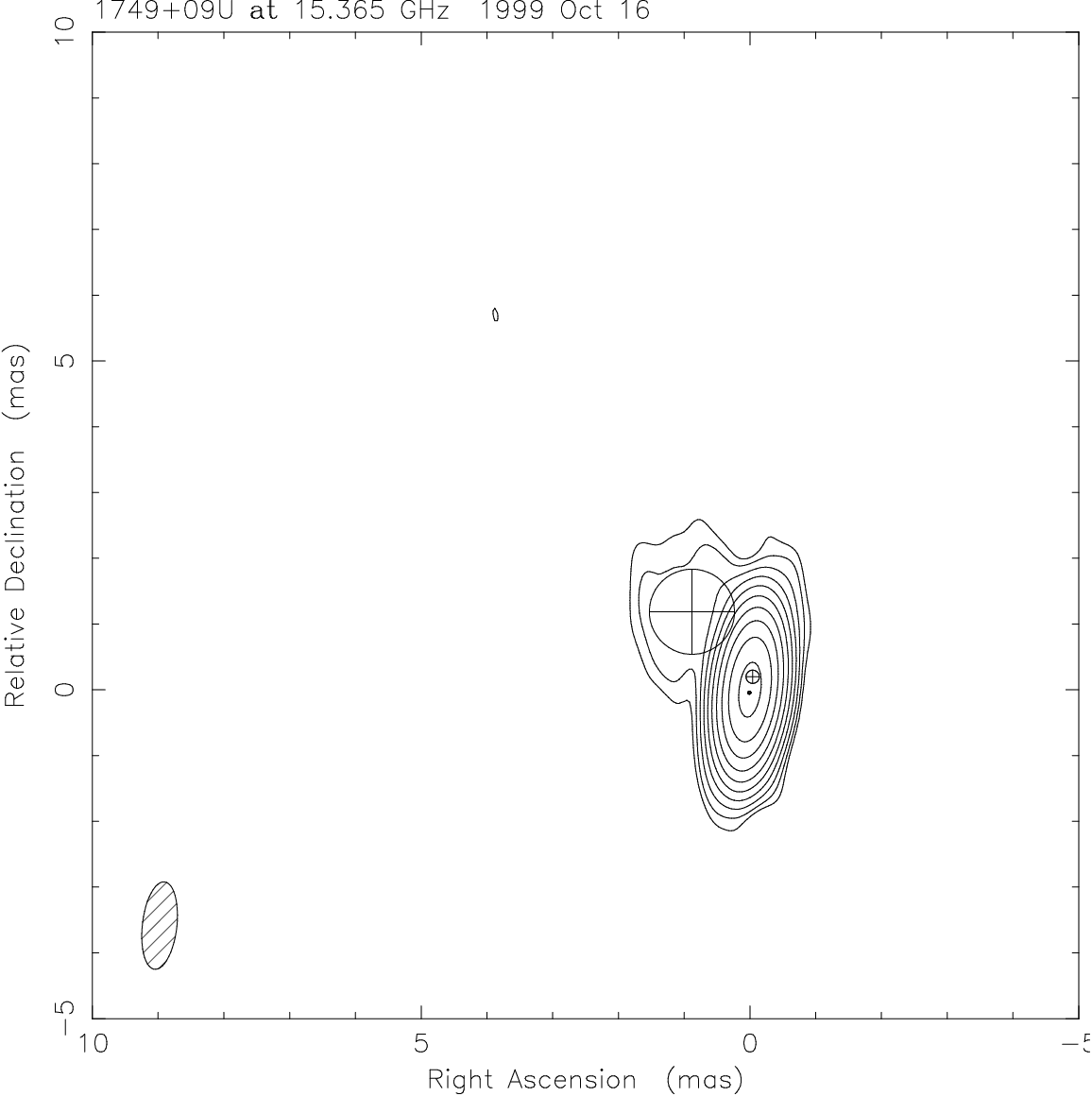}}
  \hspace{3pt}%
  \subfloat[][15\,GHz, 1999.978]{%
  \label{fig:u-15}%
  \includegraphics[width=0.30\textwidth,clip]{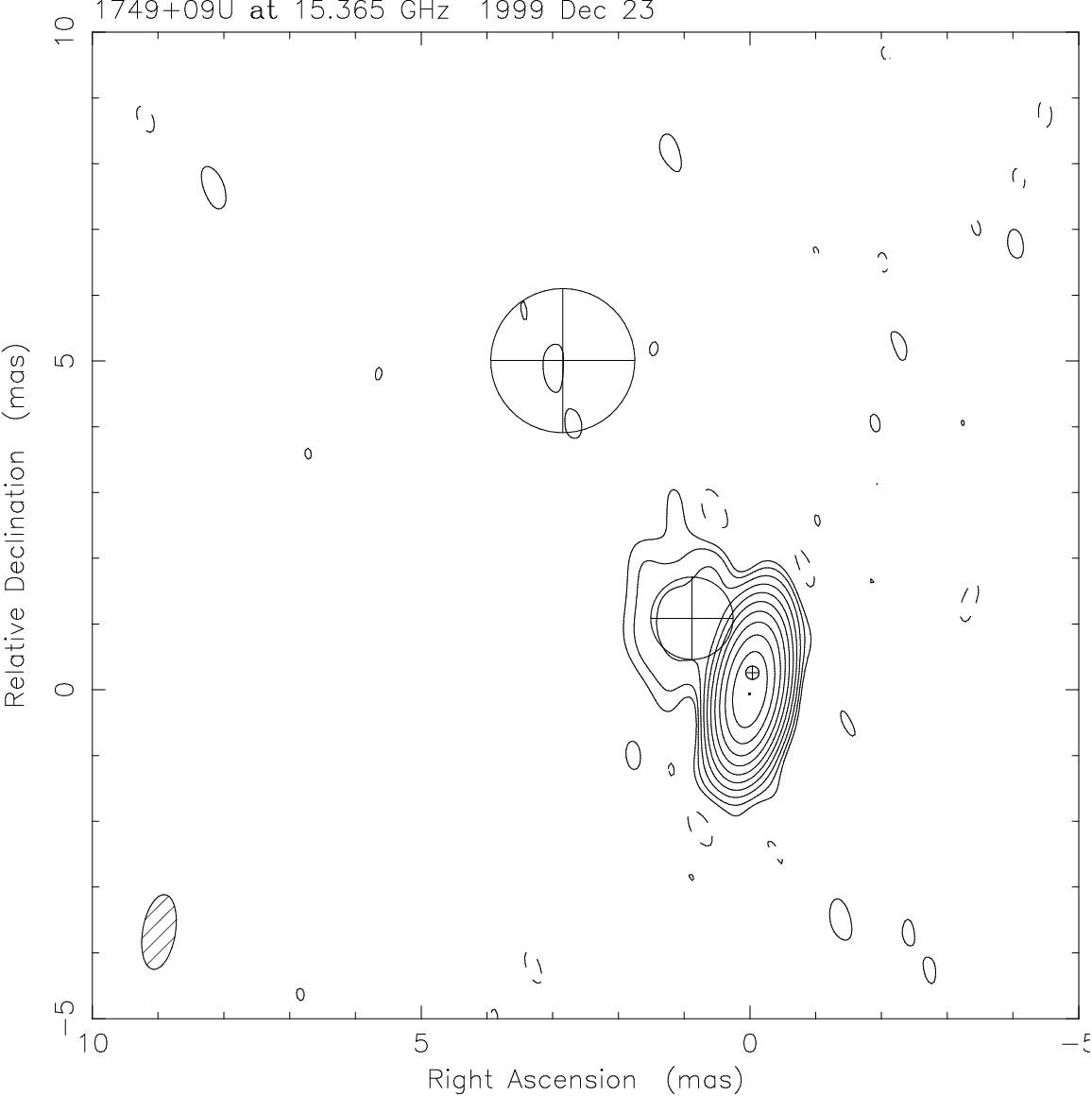}}
  \hspace{3pt}
  \subfloat[][15\,GHz, 1999.989]{%
  \label{fig:u-16}%
 \includegraphics[width=0.30\textwidth,clip]{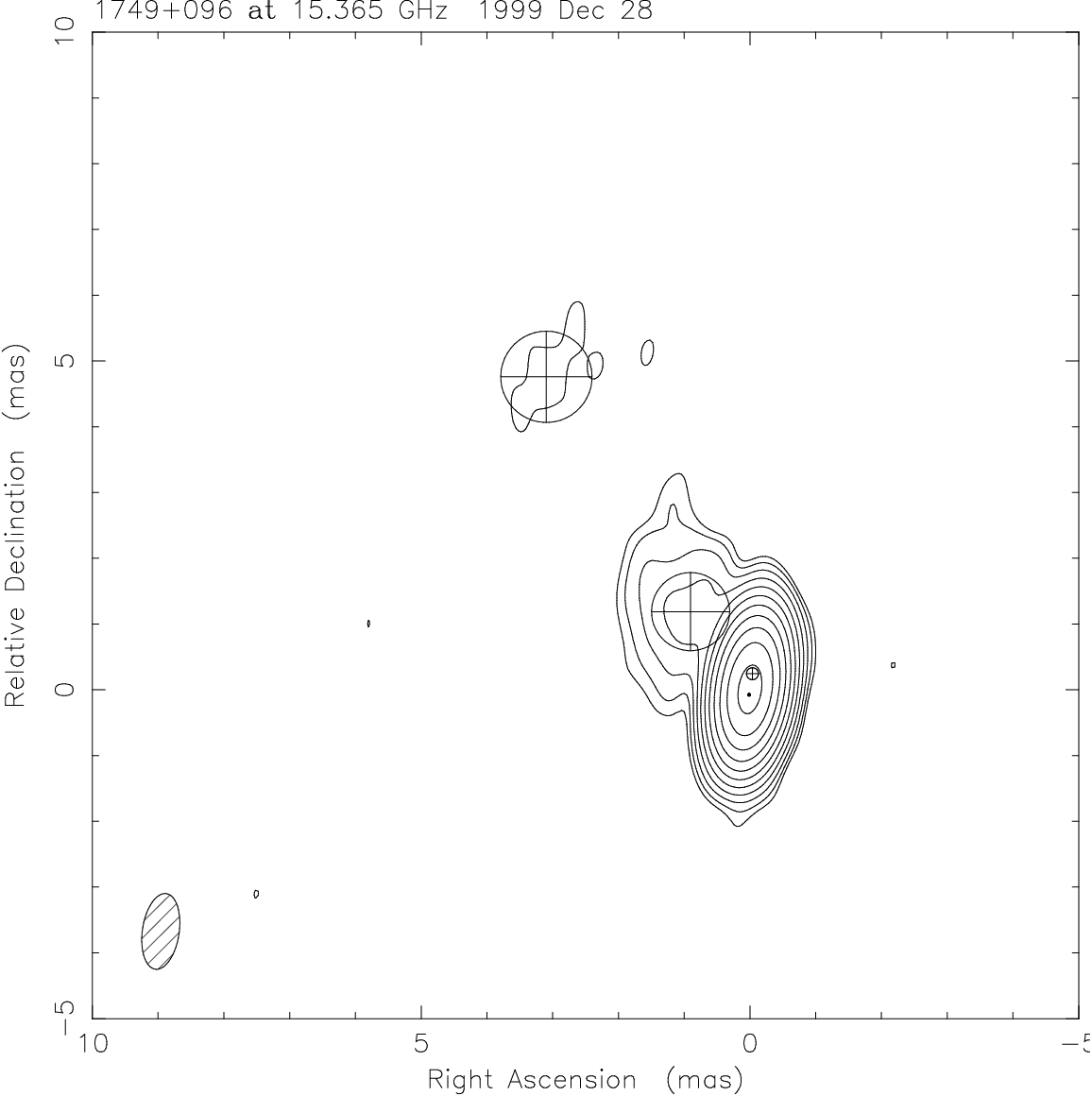}}\\
  \hspace{0pt}%
 \subfloat[][15\,GHz, 2000.347]{%
 \label{fig:u-17}%
 \includegraphics[width=0.30\textwidth,clip]{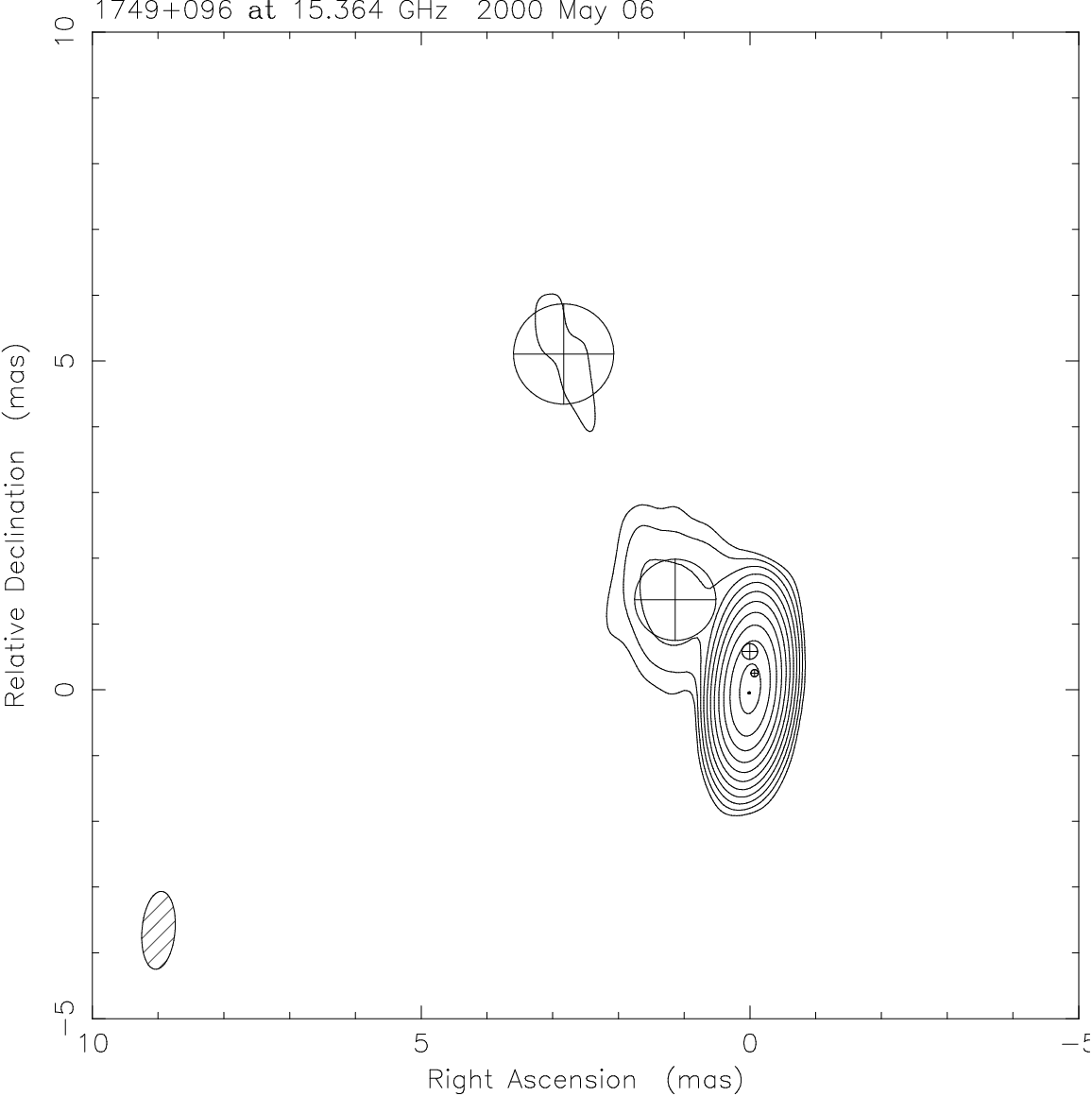}}
 \hspace{3pt}%
 \subfloat[][15\,GHz, 2000.577]{%
 \label{fig:u-18}%
 \includegraphics[width=0.30\textwidth,clip]{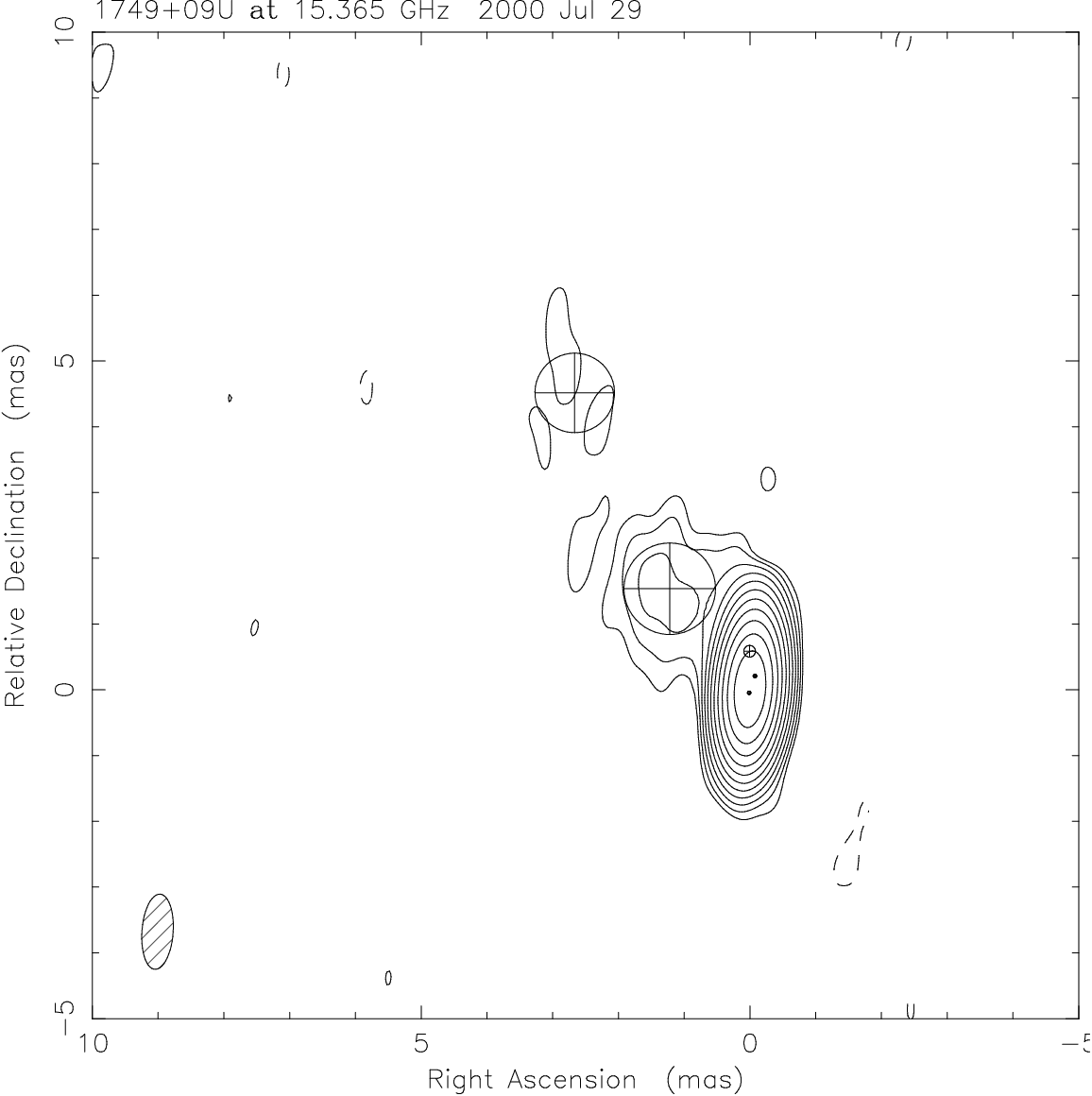}}
  \hspace{3pt}%
 \subfloat[][15\,GHz, 2000.691]{%
 \label{fig:u-19}%
 \includegraphics[width=0.30\textwidth,clip]{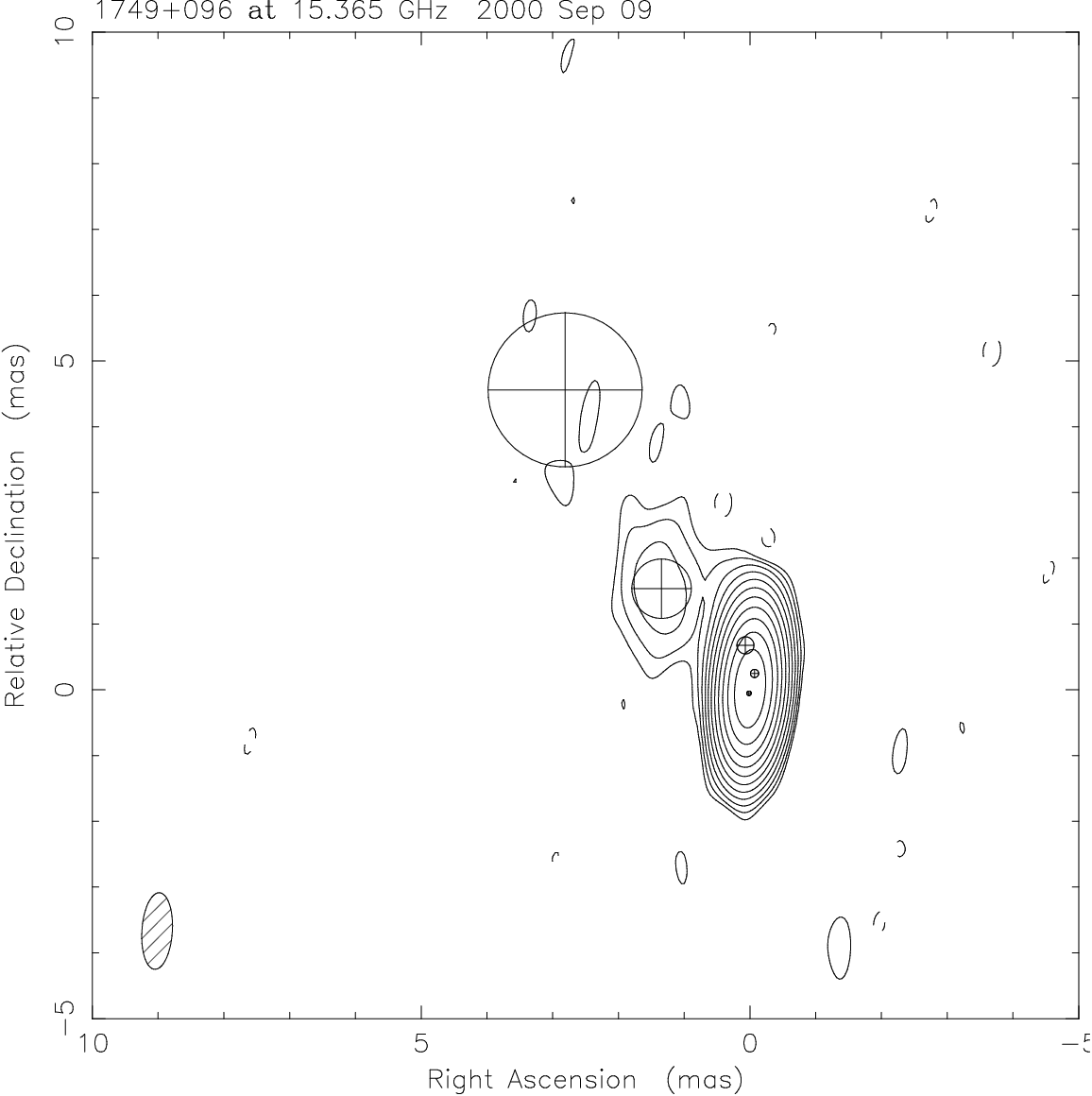}}\\
 \hspace{0pt}%
 \subfloat[][15\,GHz, 2000.740]{%
 \label{fig:u-20}%
 \includegraphics[width=0.30\textwidth,clip]{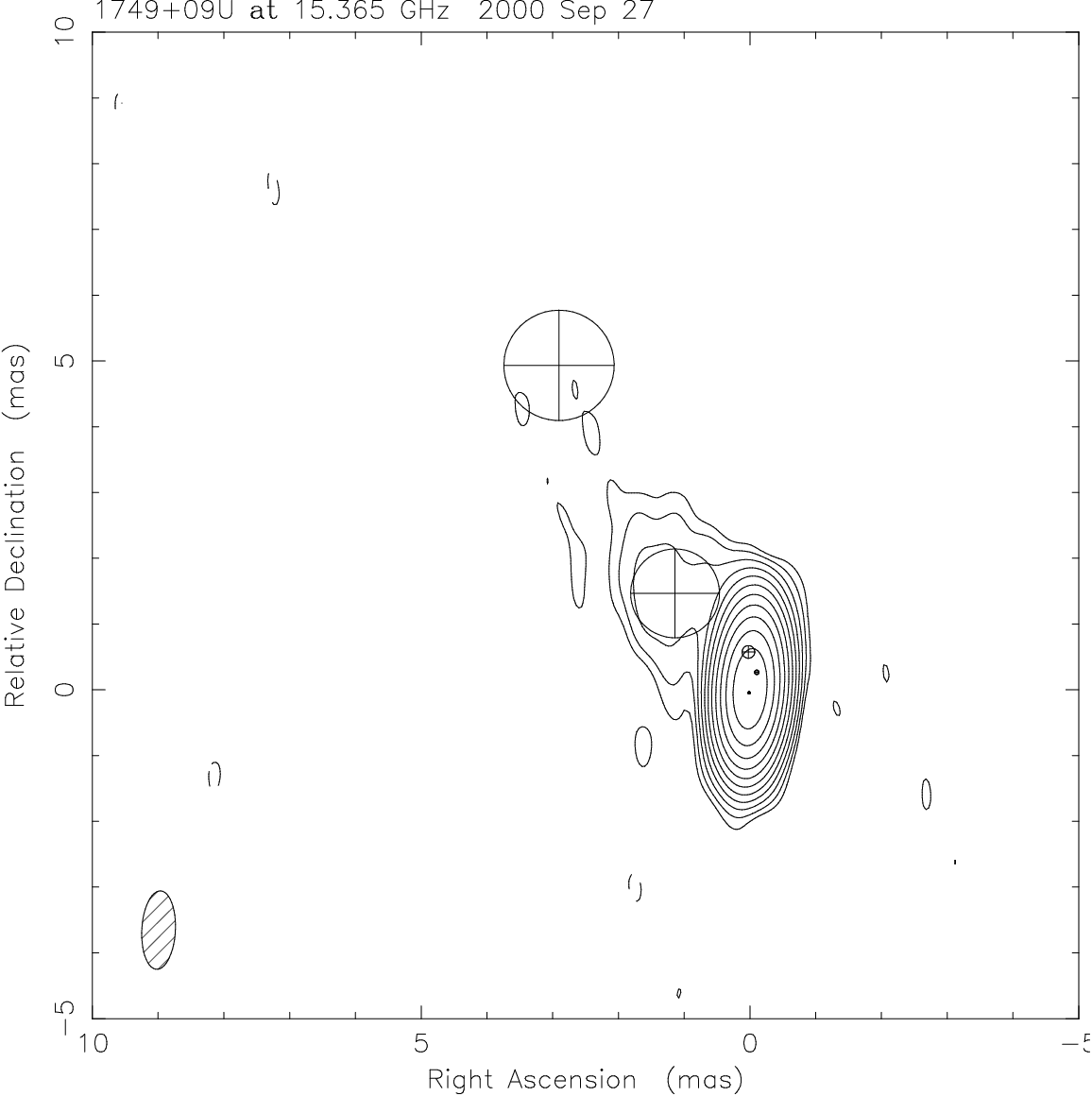}}
 \hspace{3pt}%
 \subfloat[][15\,GHz, 2001.060]{%
 \label{fig:u-21}%
 \includegraphics[width=0.30\textwidth,clip]{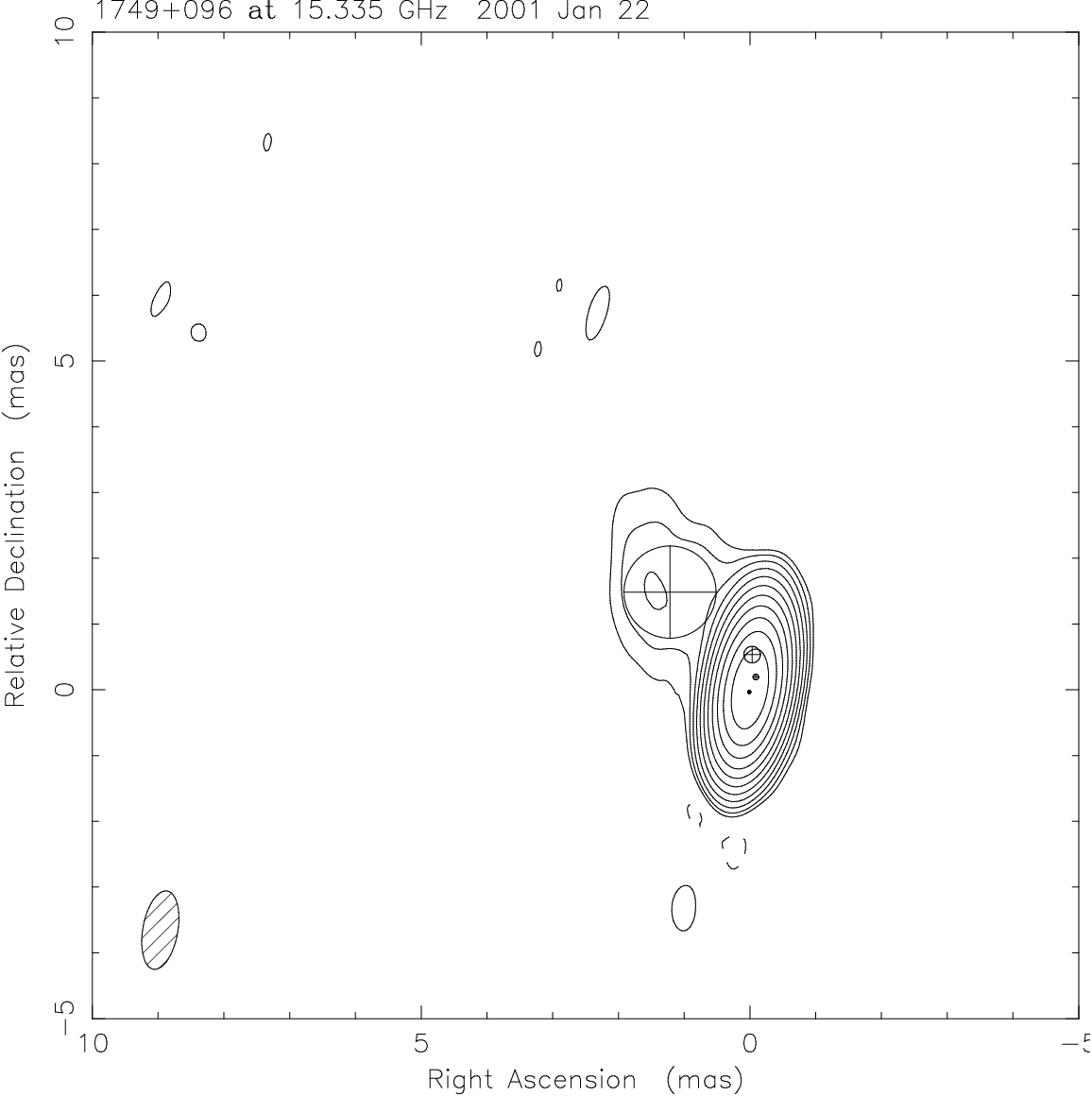}}
 \hspace{3pt}%
 \subfloat[][15\,GHz, 2001.340]{%
 \label{fig:u-22}%
 \includegraphics[width=0.30\textwidth,clip]{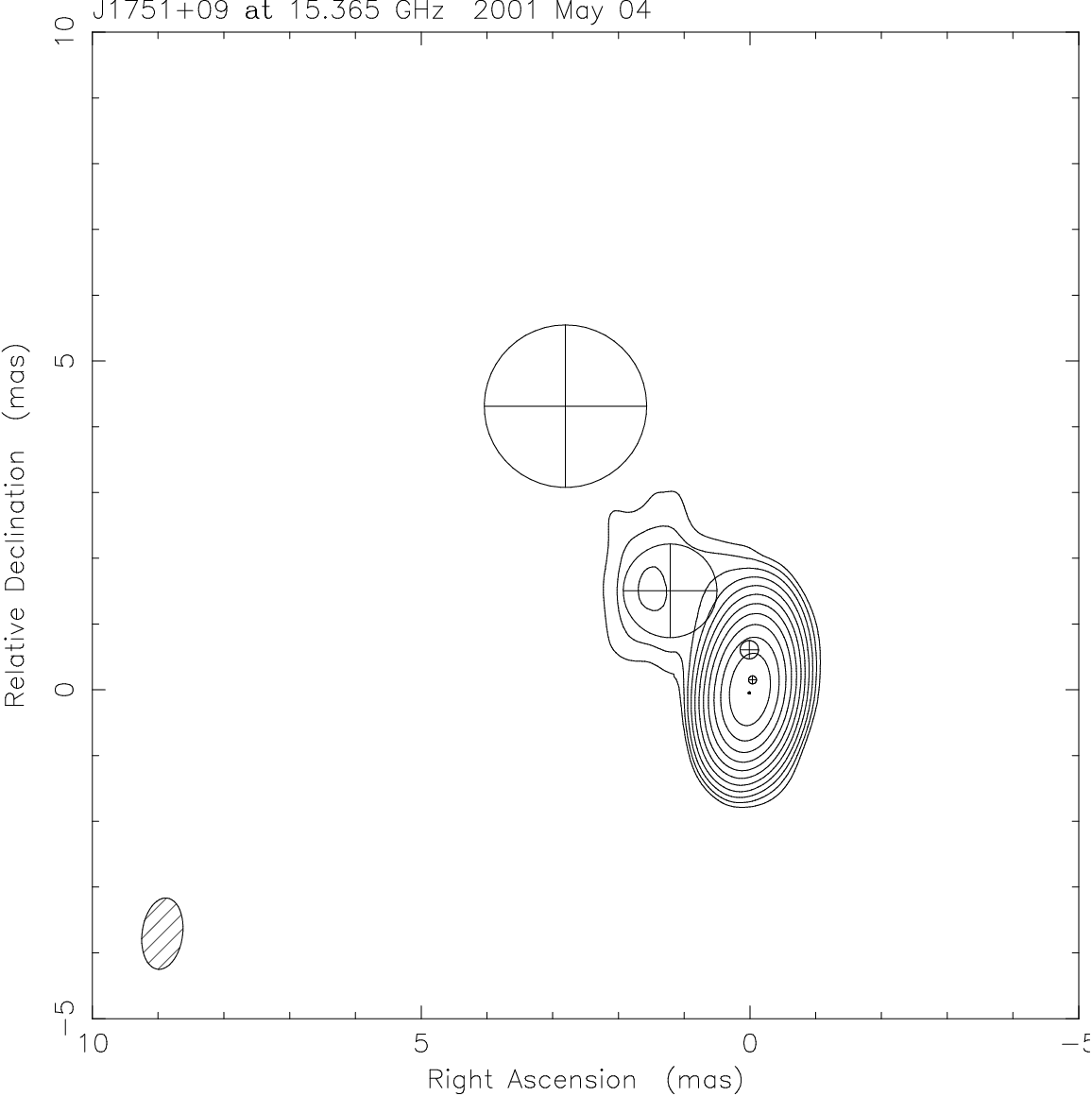}}
 \caption[]{\it{-continued}.}%
 \end{figure*}
 \begin{figure*}%
 \centering
 \ContinuedFloat
 \hspace{0pt}%
 \subfloat[][15\,GHz, 2001.468]{%
 \label{fig:u-23}%
 \includegraphics[width=0.30\textwidth,clip]{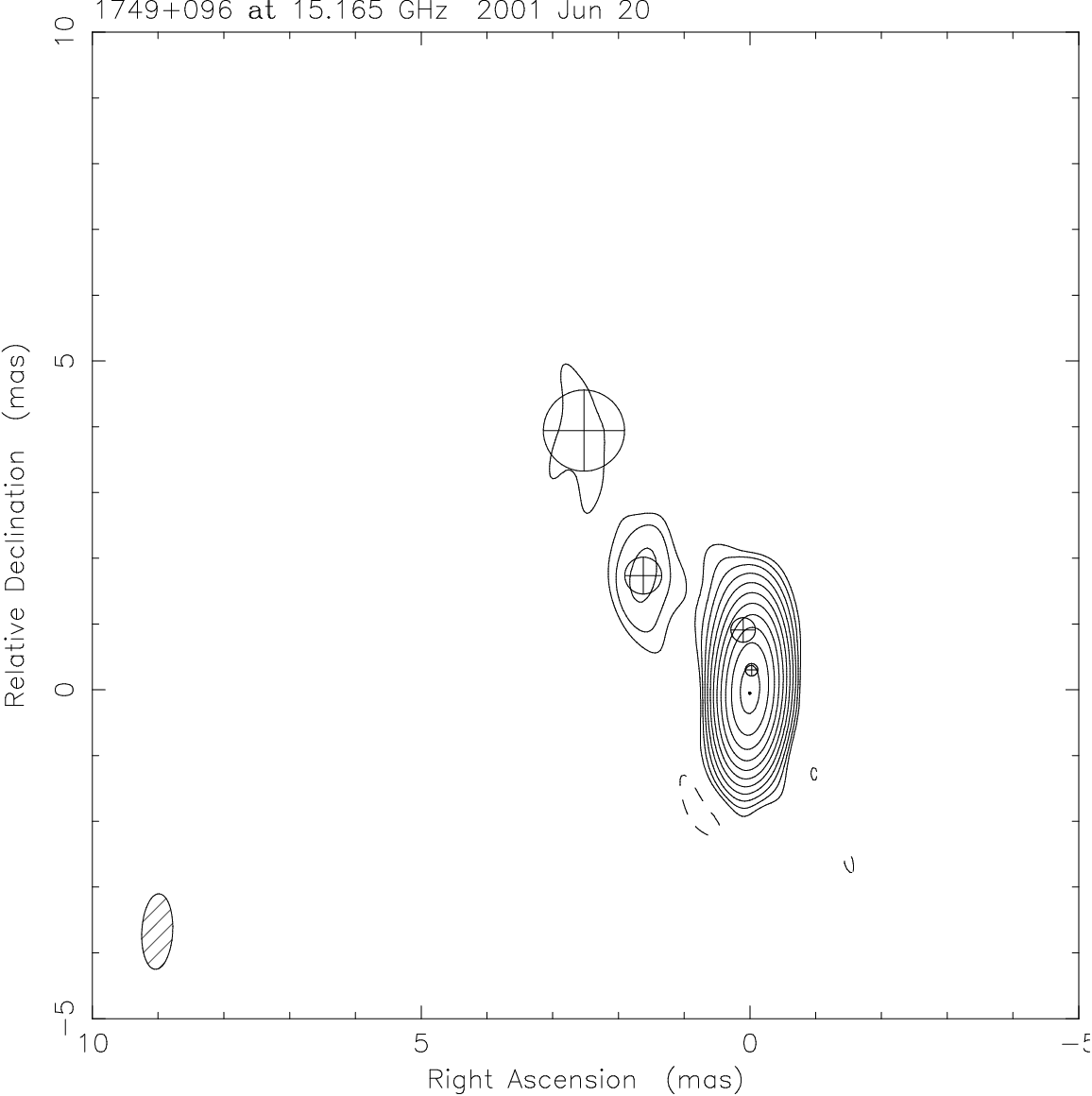}}
 \hspace{3pt}%
 \subfloat[][15\,GHz, 2001.496]{%
 \label{fig:u-24}%
 \includegraphics[width=0.30\textwidth,clip]{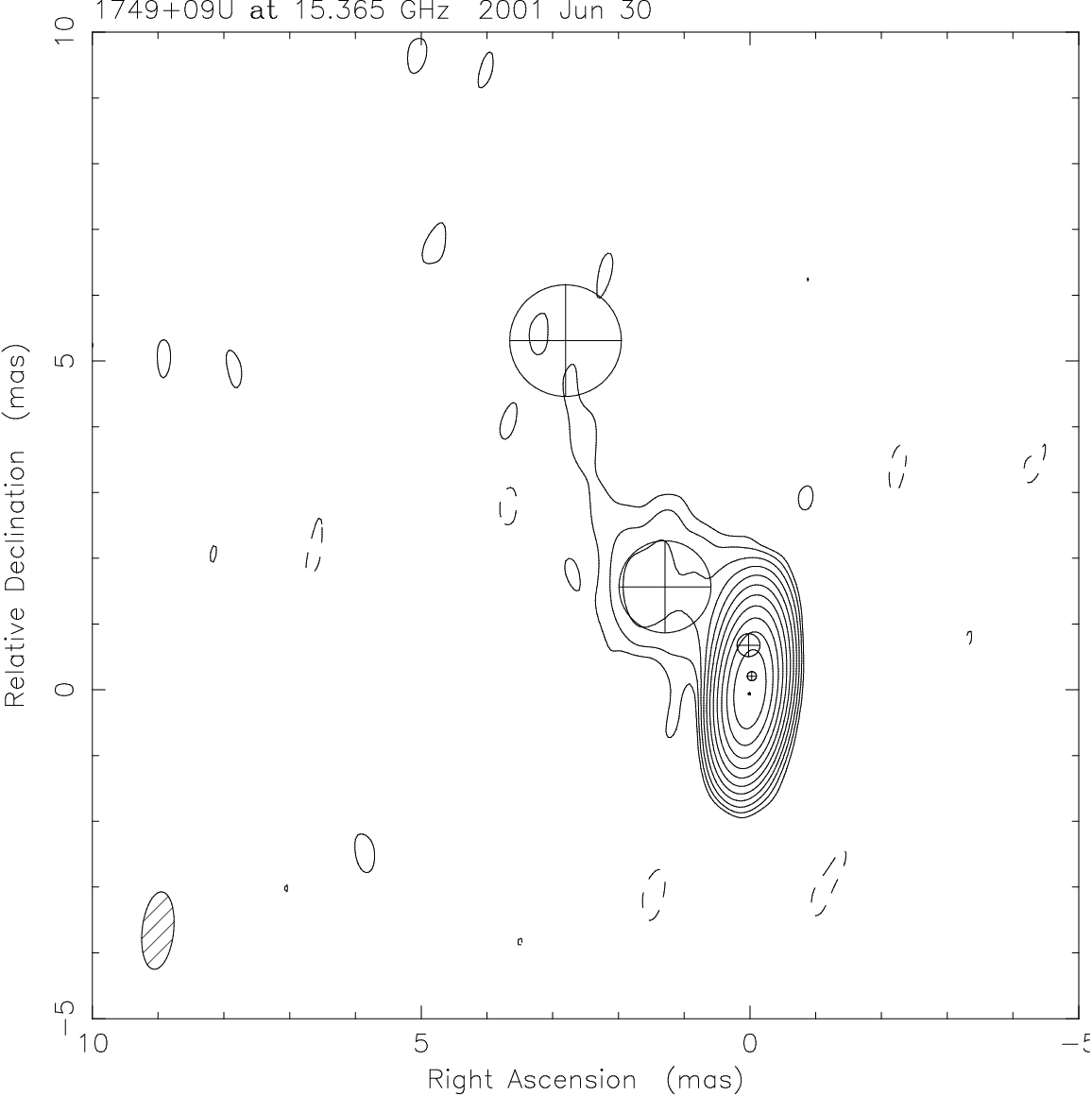}}
 \hspace{3pt}%
 \subfloat[][15\,GHz, 2001.597]{%
 \label{fig:u-25}%
 \includegraphics[width=0.30\textwidth,clip]{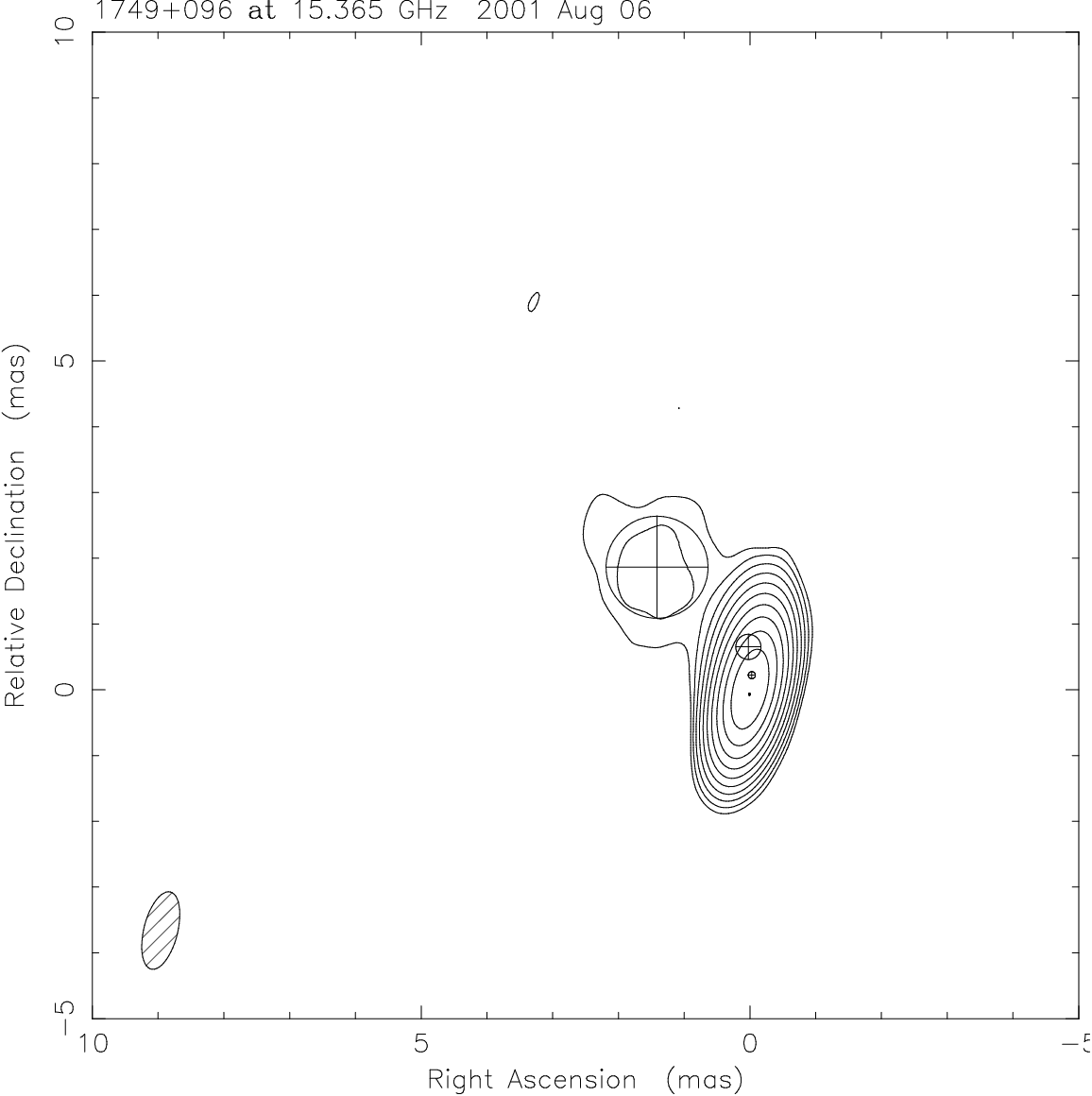}}\\
 \hspace{0pt}%
 \subfloat[][15\,GHz, 2001.701]{%
 \label{fig:u-26}%
 \includegraphics[width=0.30\textwidth,clip]{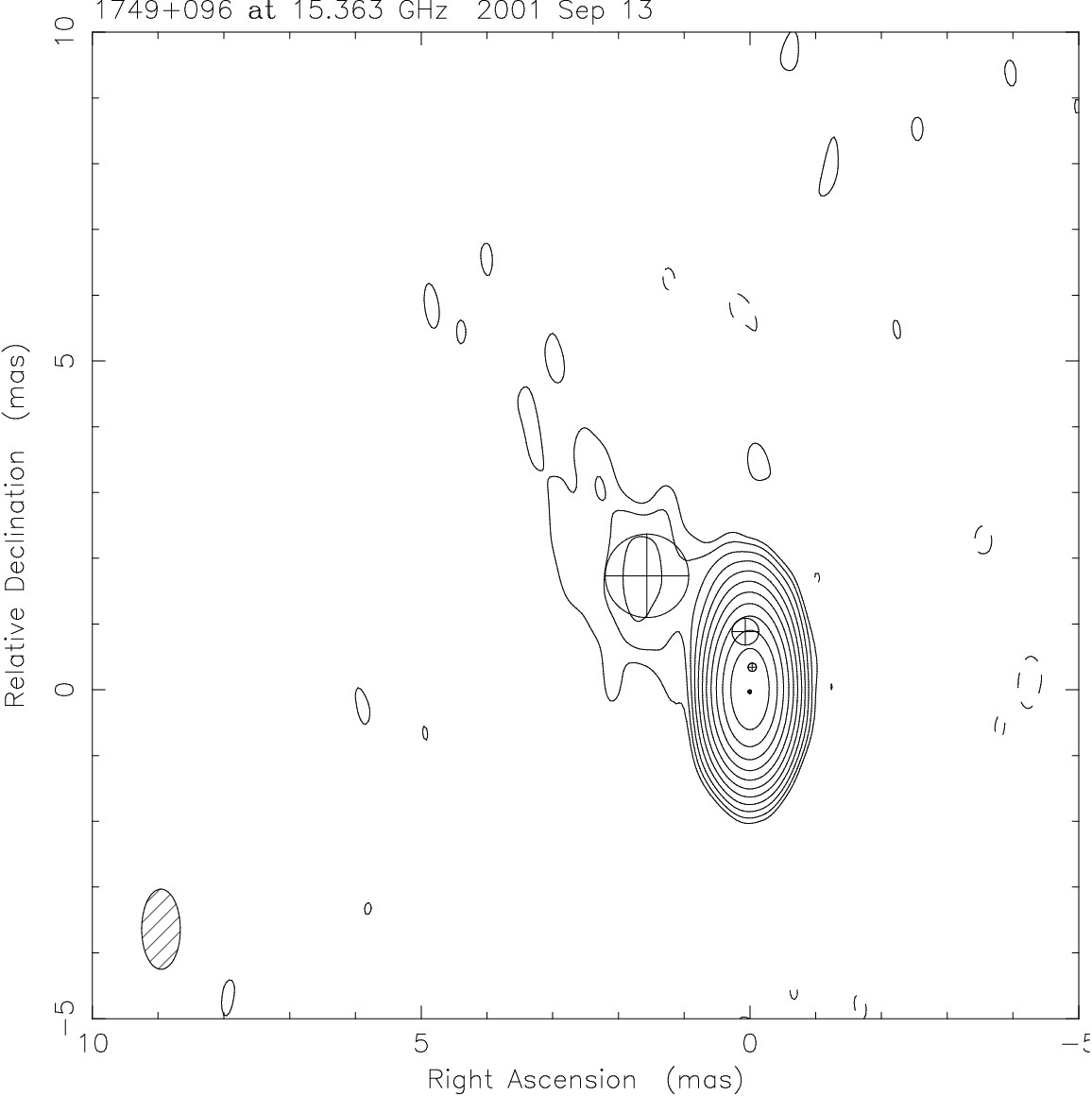}}
 \hspace{3pt}%
 \subfloat[][15\,GHz, 2001.805]{%
 \label{fig:u-27}%
 \includegraphics[width=0.30\textwidth,clip]{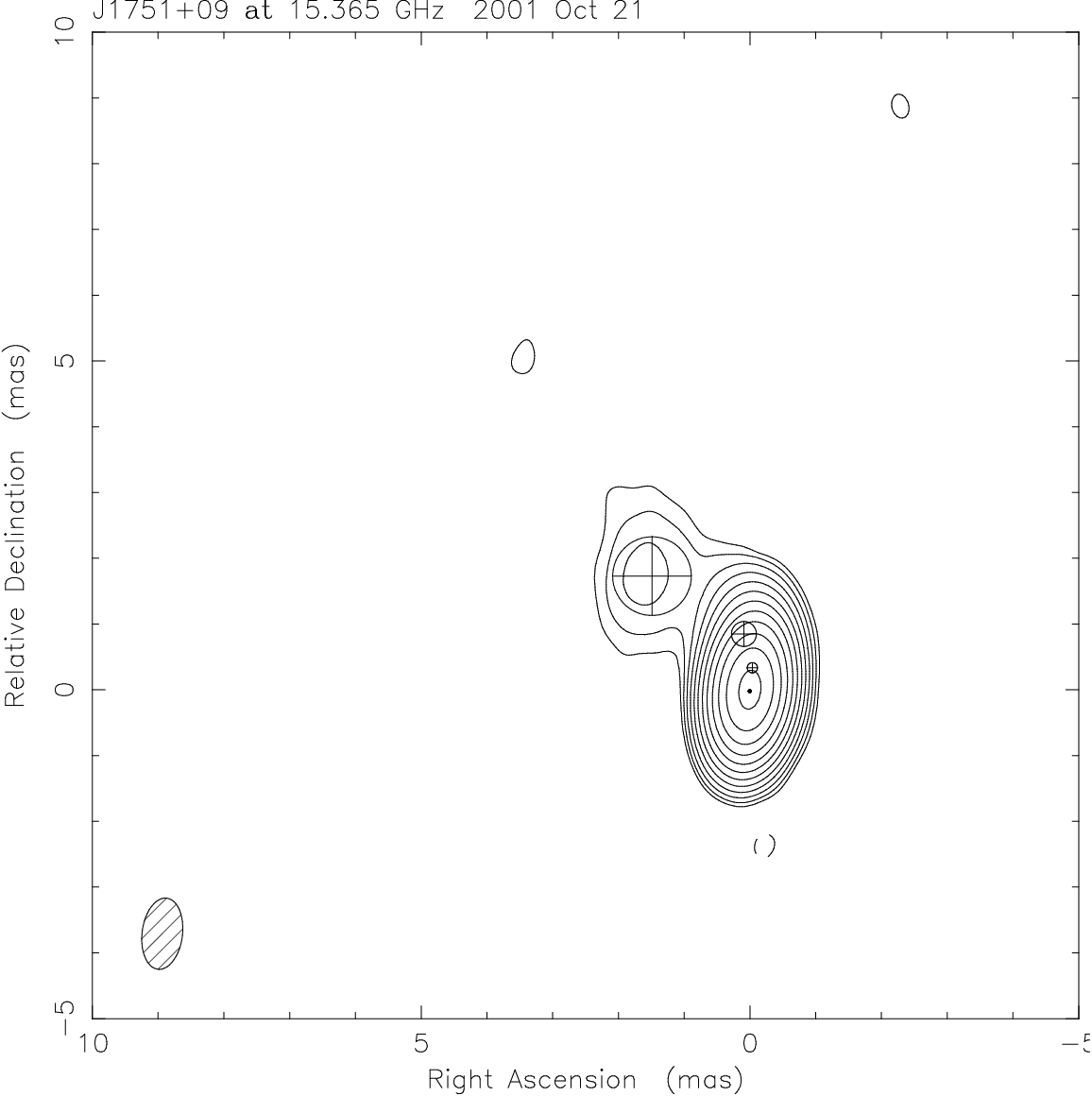}}
 \hspace{0pt}%
 \subfloat[][15\,GHz, 2001.838]{%
 \label{fig:u-28}%
 \includegraphics[width=0.30\textwidth,clip]{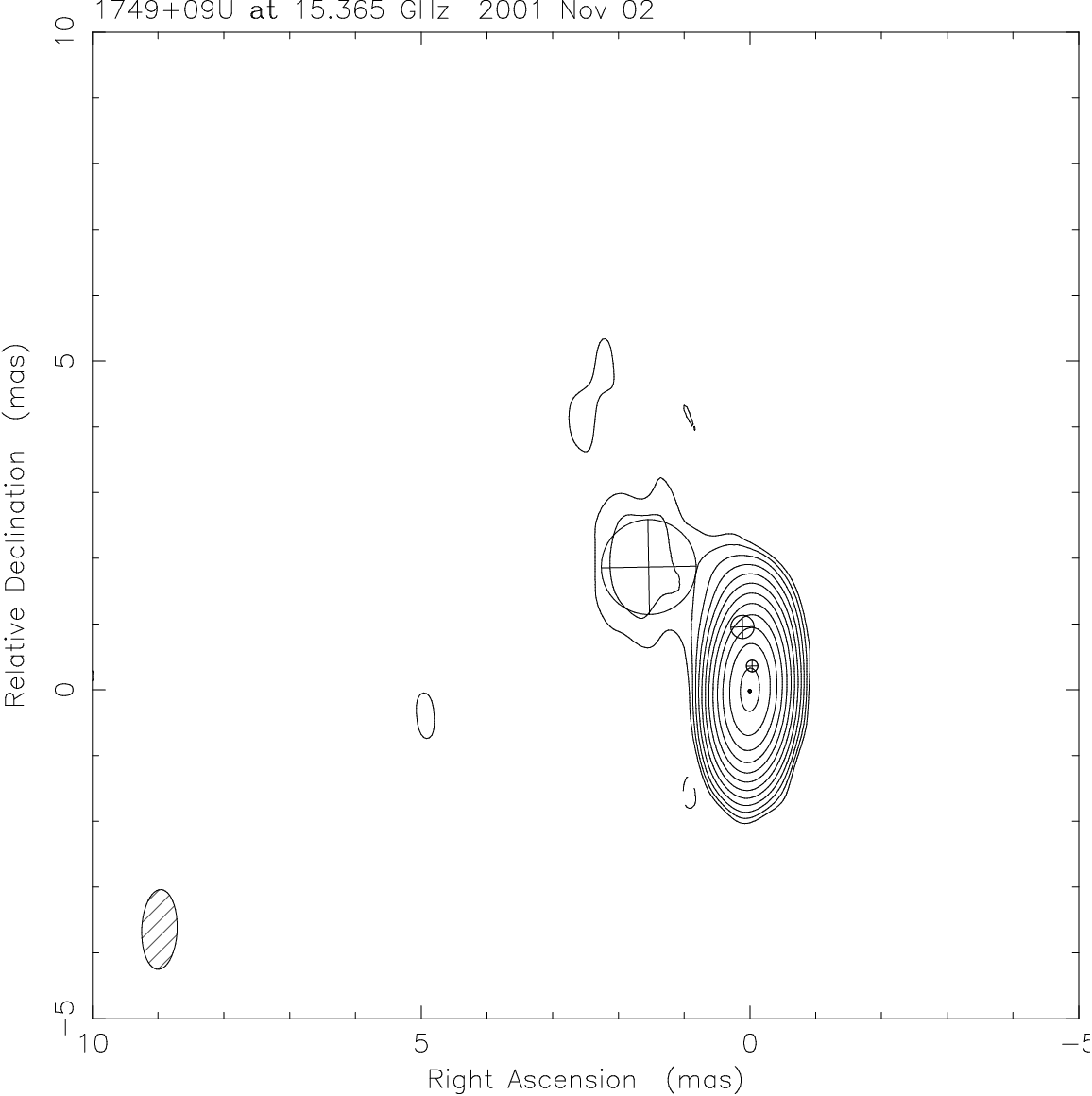}}\\
 \hspace{0pt}%
 \subfloat[][15\,GHz, 2002.019]{%
 \label{fig:u-29}%
 \includegraphics[width=0.30\textwidth,clip]{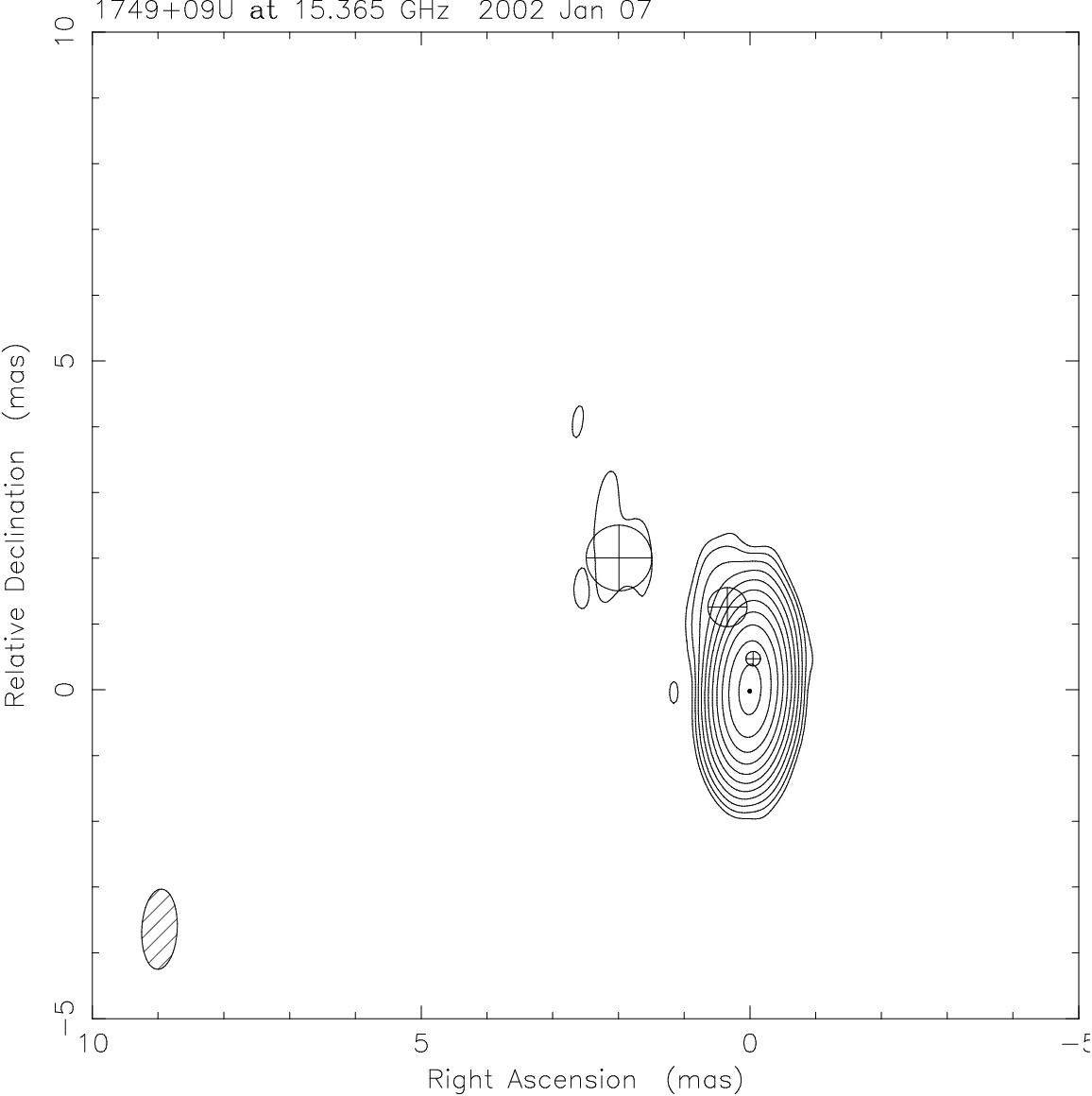}}
 \hspace{3pt}%
 \subfloat[][15\,GHz, 2002.381]{%
 \label{fig:u-30}%
 \includegraphics[width=0.30\textwidth,clip]{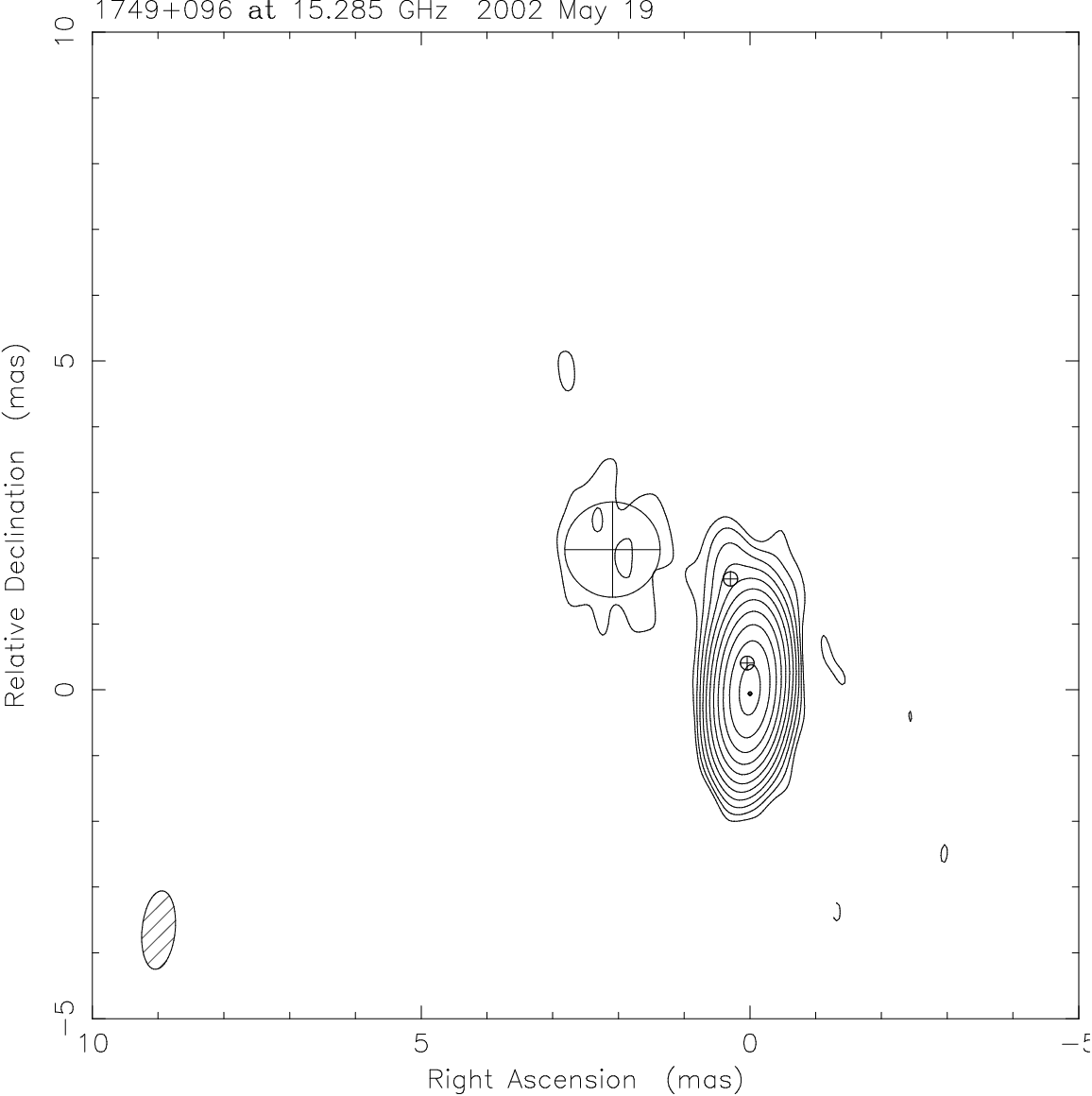}}
 \hspace{0pt}%
 \subfloat[][15\,GHz, 2002.416]{%
 \label{fig:u-31}%
 \includegraphics[width=0.30\textwidth,clip]{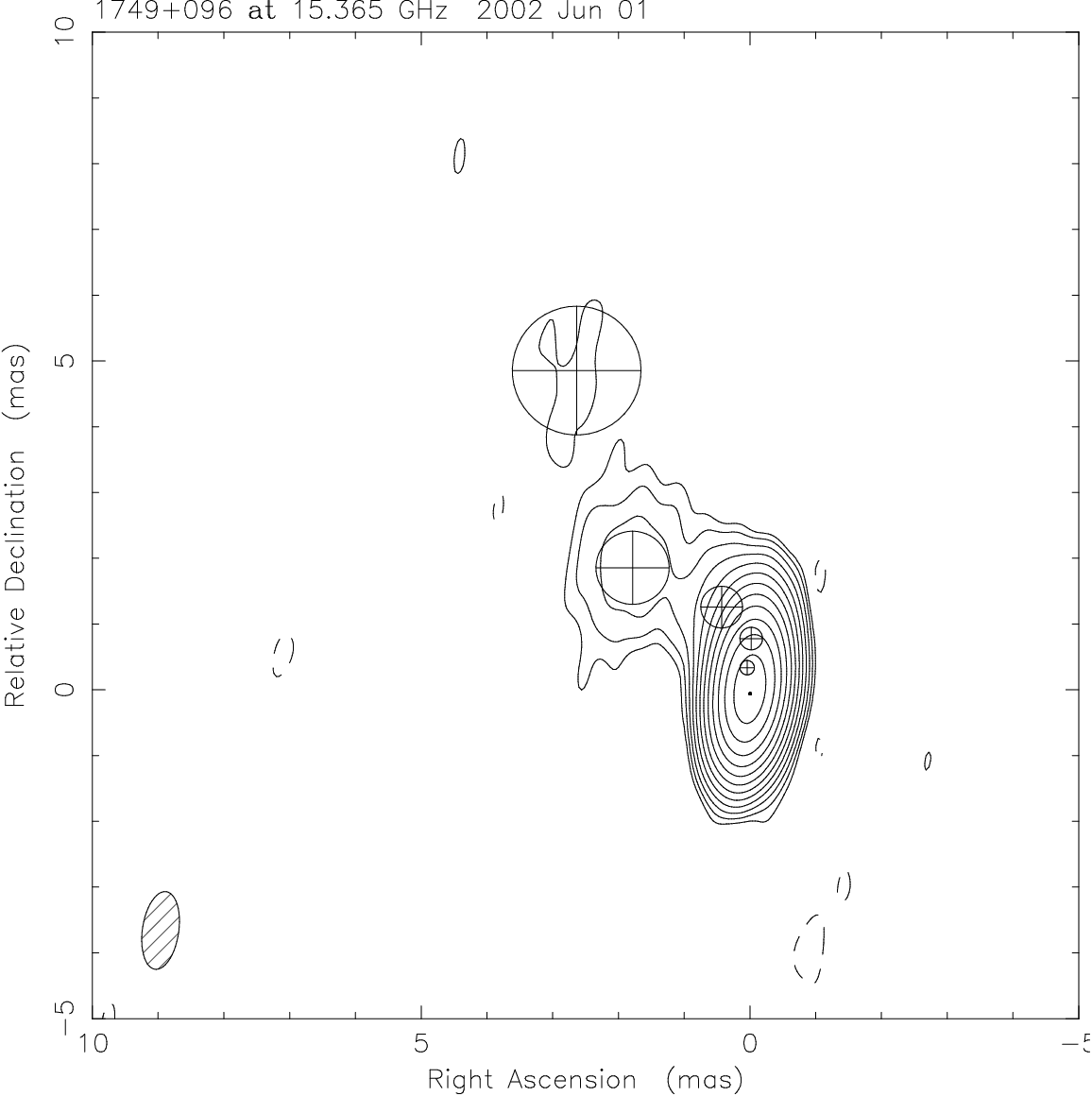}}\\
 \hspace{0pt}%
 \subfloat[][15\,GHz, 2002.510]{%
 \label{fig:u-32}%
 \includegraphics[width=0.30\textwidth,clip]{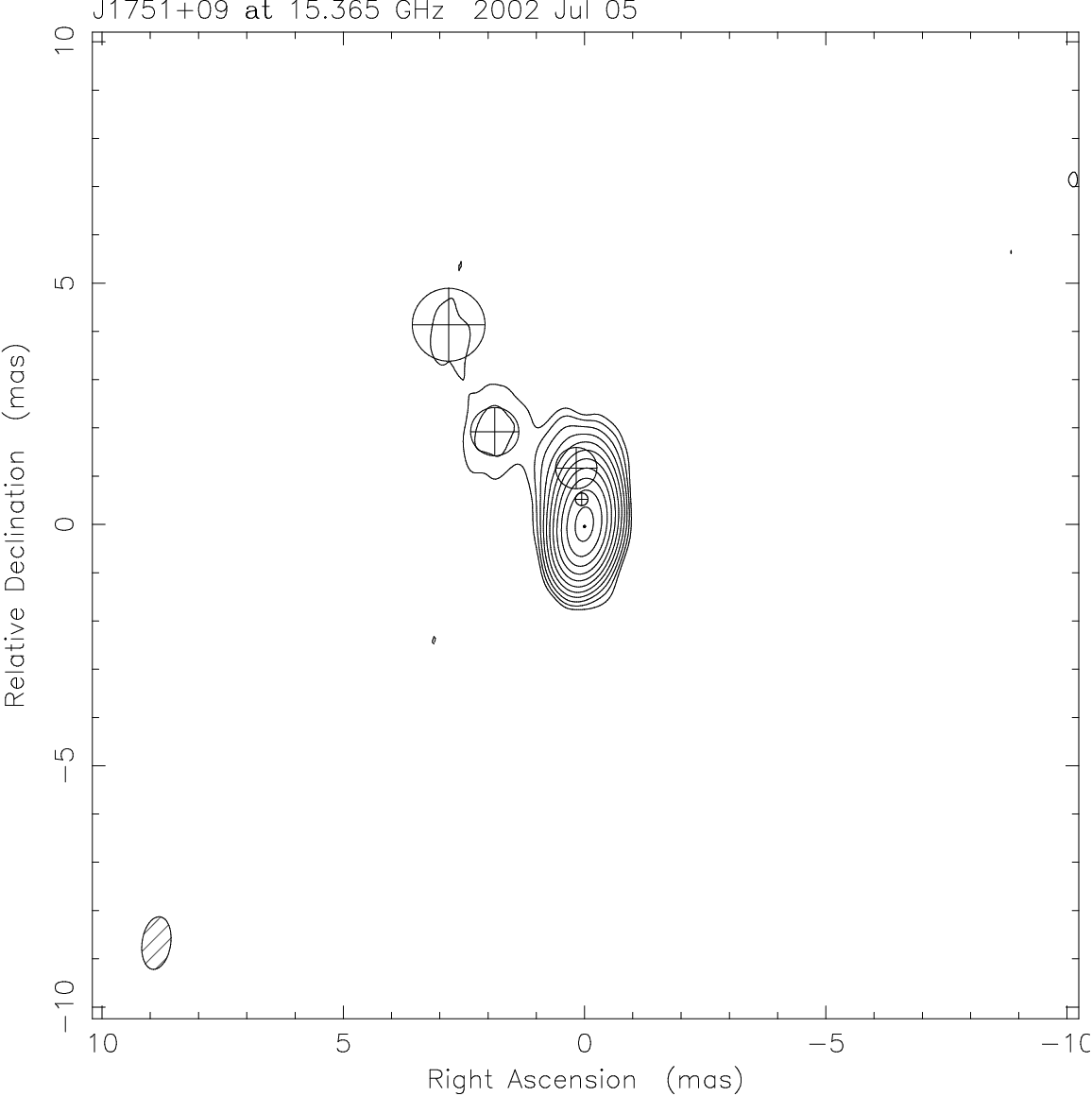}}
 \hspace{3pt}%
 \subfloat[][15\,GHz, 2002.600]{%
 \label{fig:u-33}%
 \includegraphics[width=0.30\textwidth,clip]{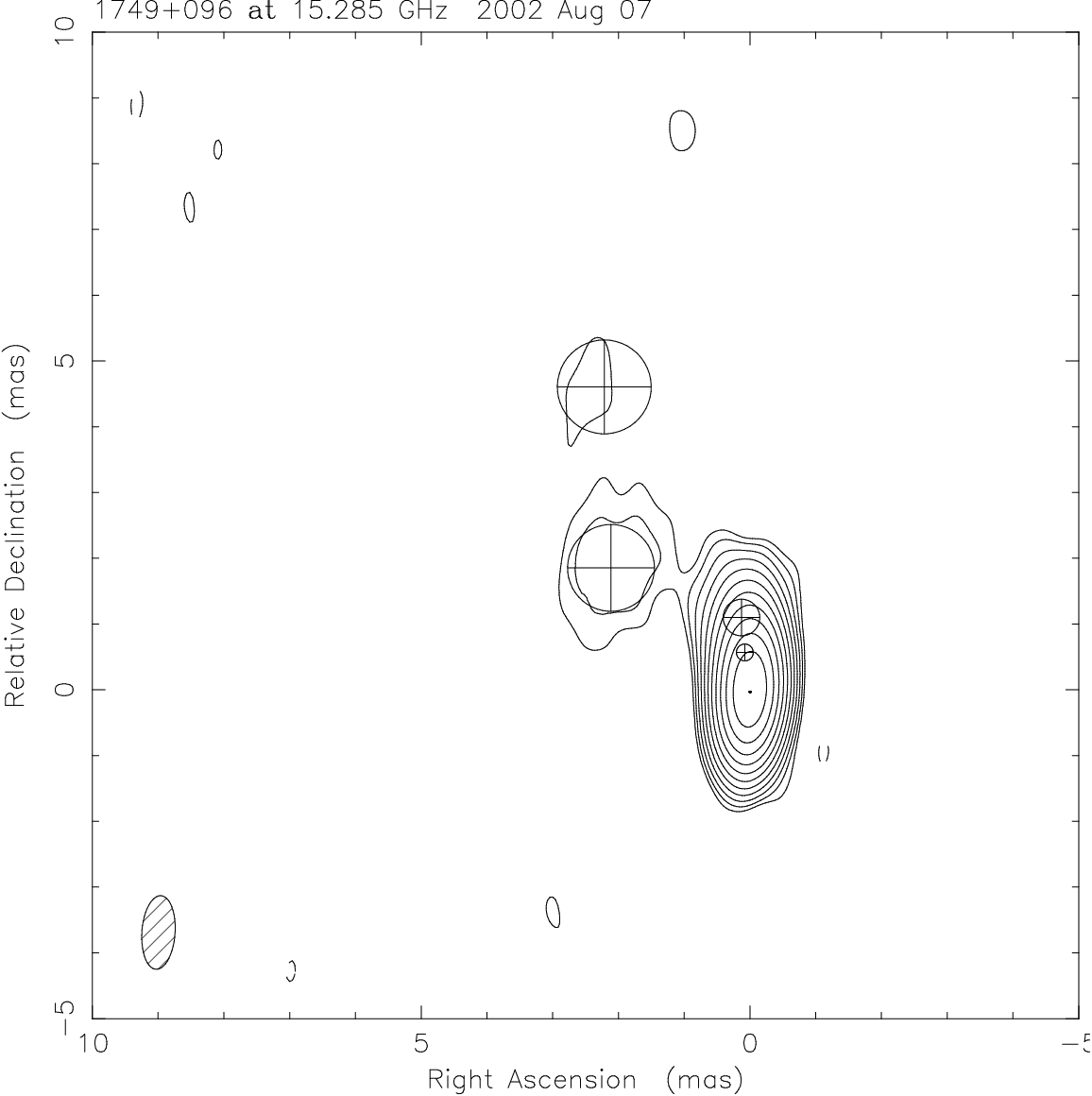}}
 \hspace{3pt}%
 \subfloat[][15\,GHz, 2002.973]{%
 \label{fig:u-34}%
 \includegraphics[width=0.30\textwidth,clip]{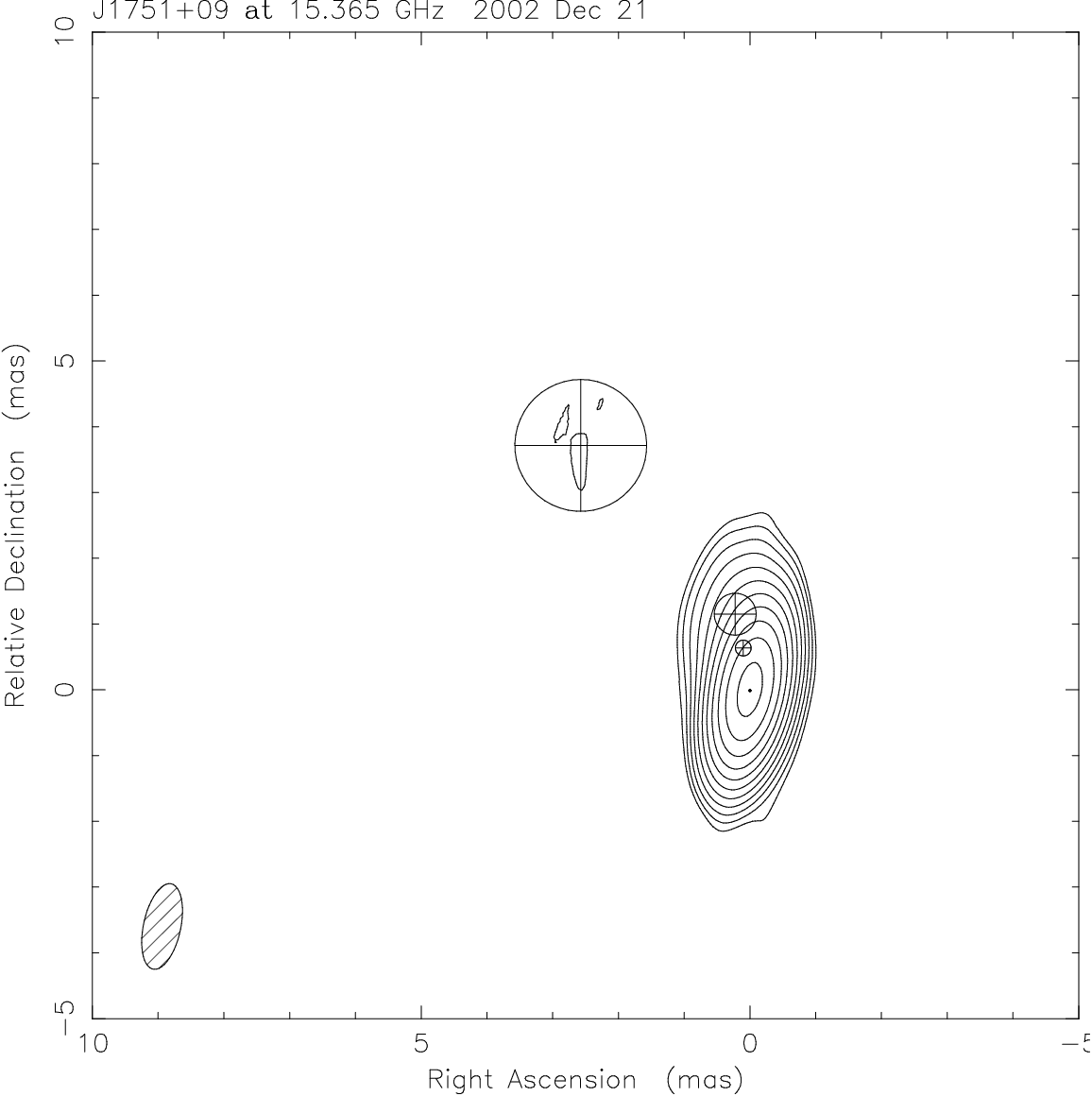}}
 \caption[]{\it{-continued}.}%
  \end{figure*}
  \begin{figure*}%
  \centering
  \ContinuedFloat
 \hspace{0pt}%
 \subfloat[][15\,GHz, 2003.099]{%
 \label{fig:u-35}%
 \includegraphics[width=0.30\textwidth,clip]{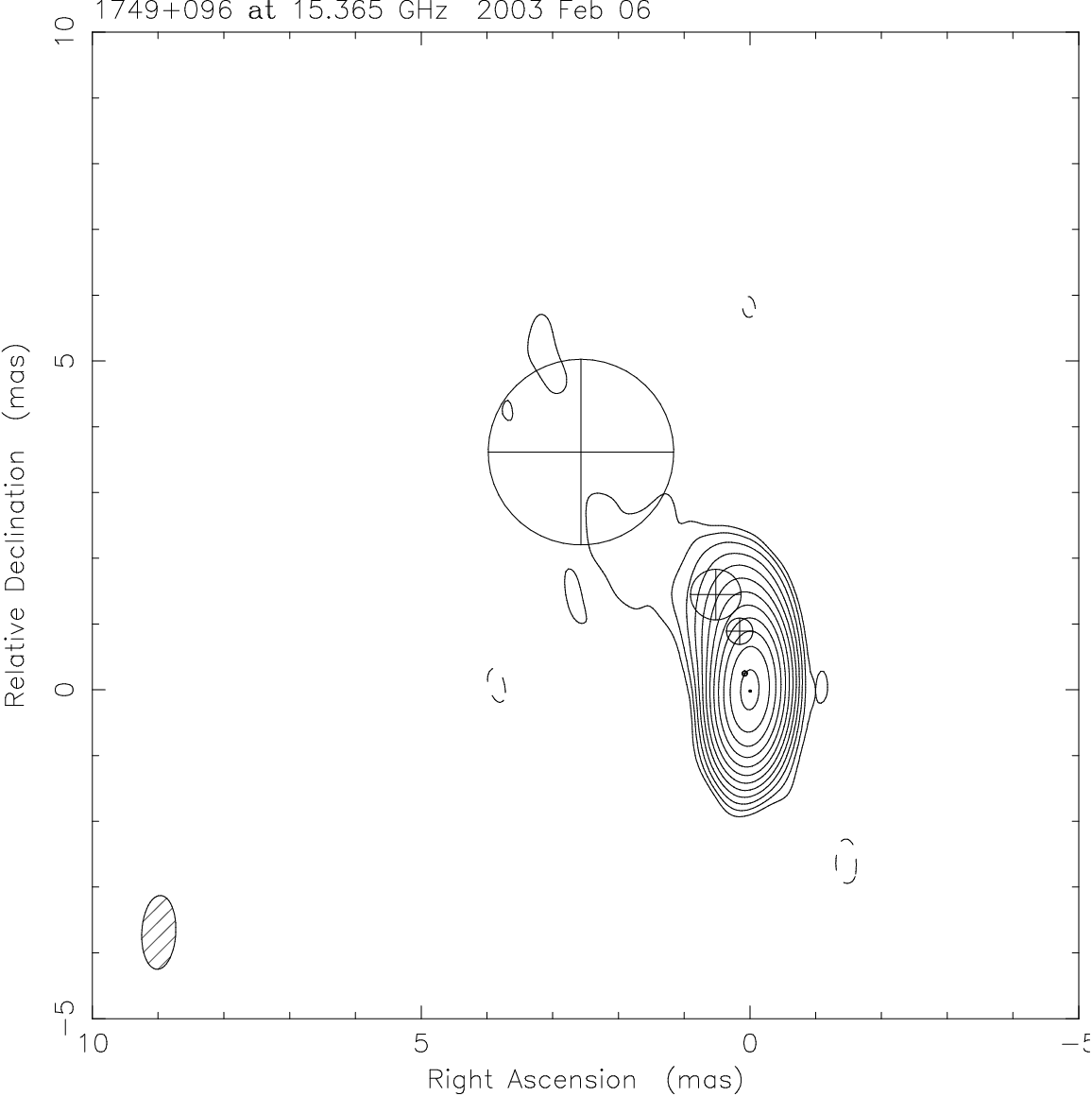}}
 \hspace{3pt}%
 \subfloat[][15\,GHz, 2003.737]{%
 \label{fig:u-36}%
 \includegraphics[width=0.30\textwidth,clip]{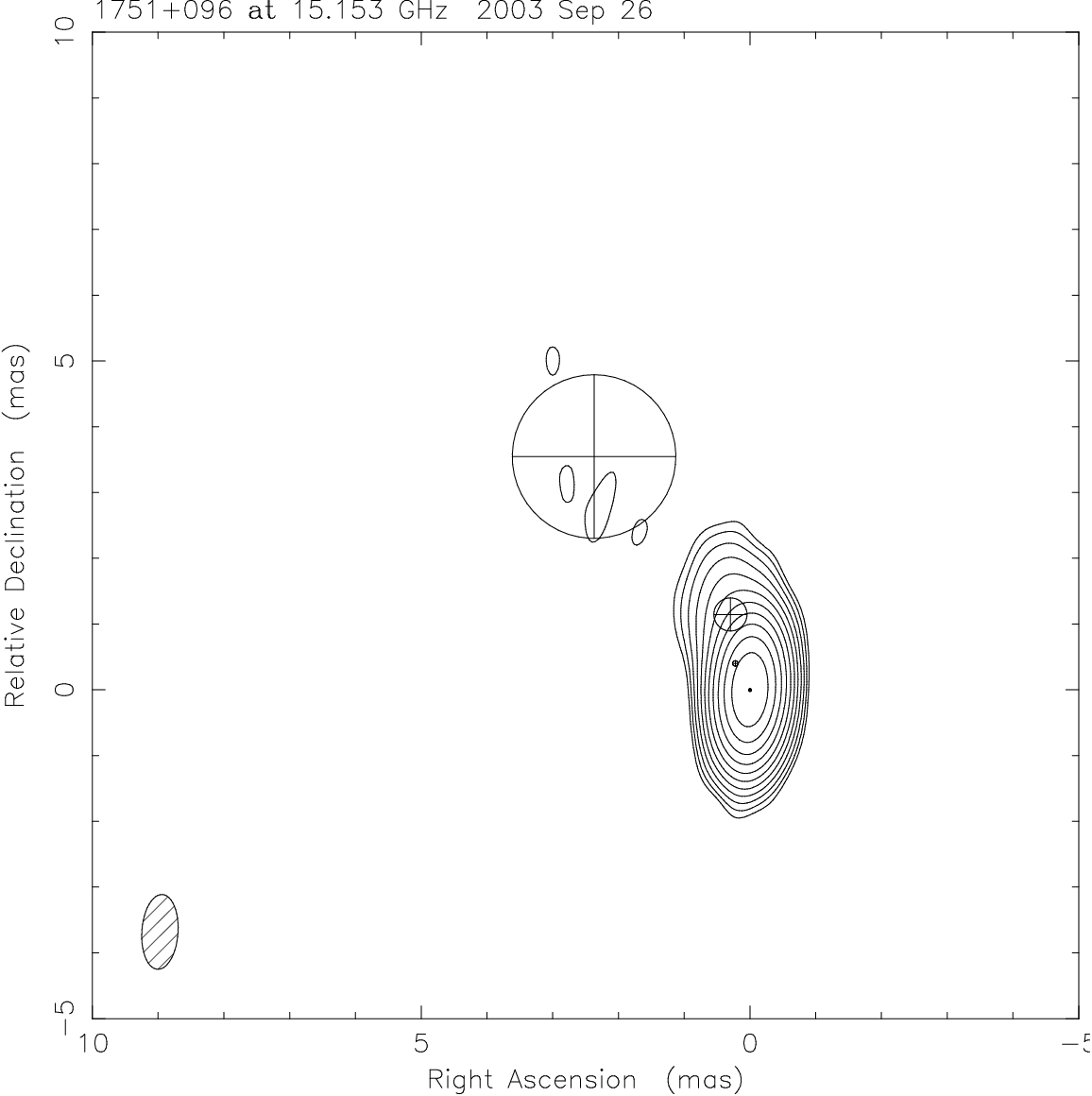}}
 \hspace{0pt}%
 \subfloat[][15\,GHz, 2003.822]{%
 \label{fig:u-37}%
 \includegraphics[width=0.30\textwidth,clip]{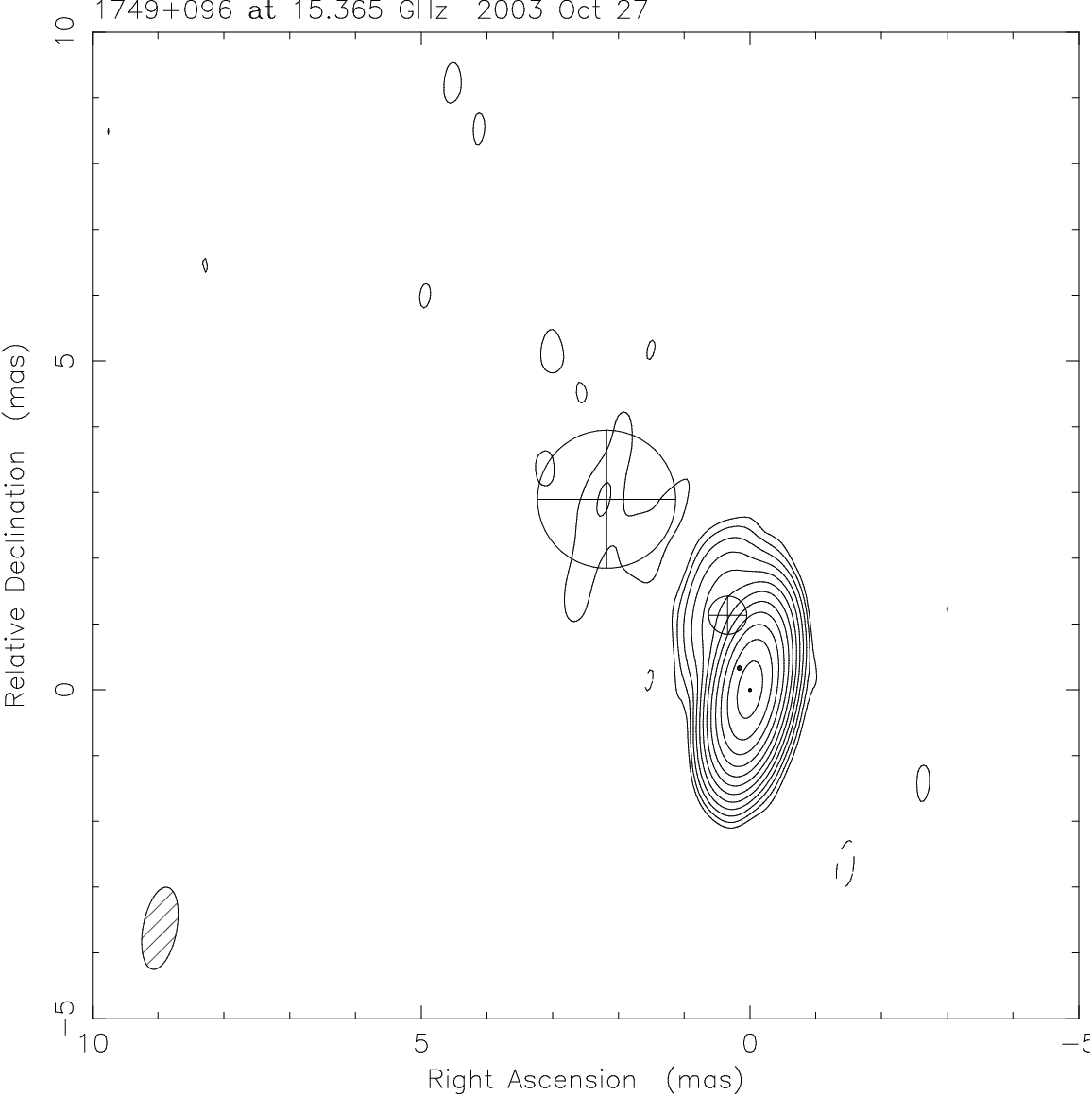}}\\
 \hspace{3pt}%
 \subfloat[][15\,GHz, 2004.224]{%
 \label{fig:u-38}%
 \includegraphics[width=0.30\textwidth,clip]{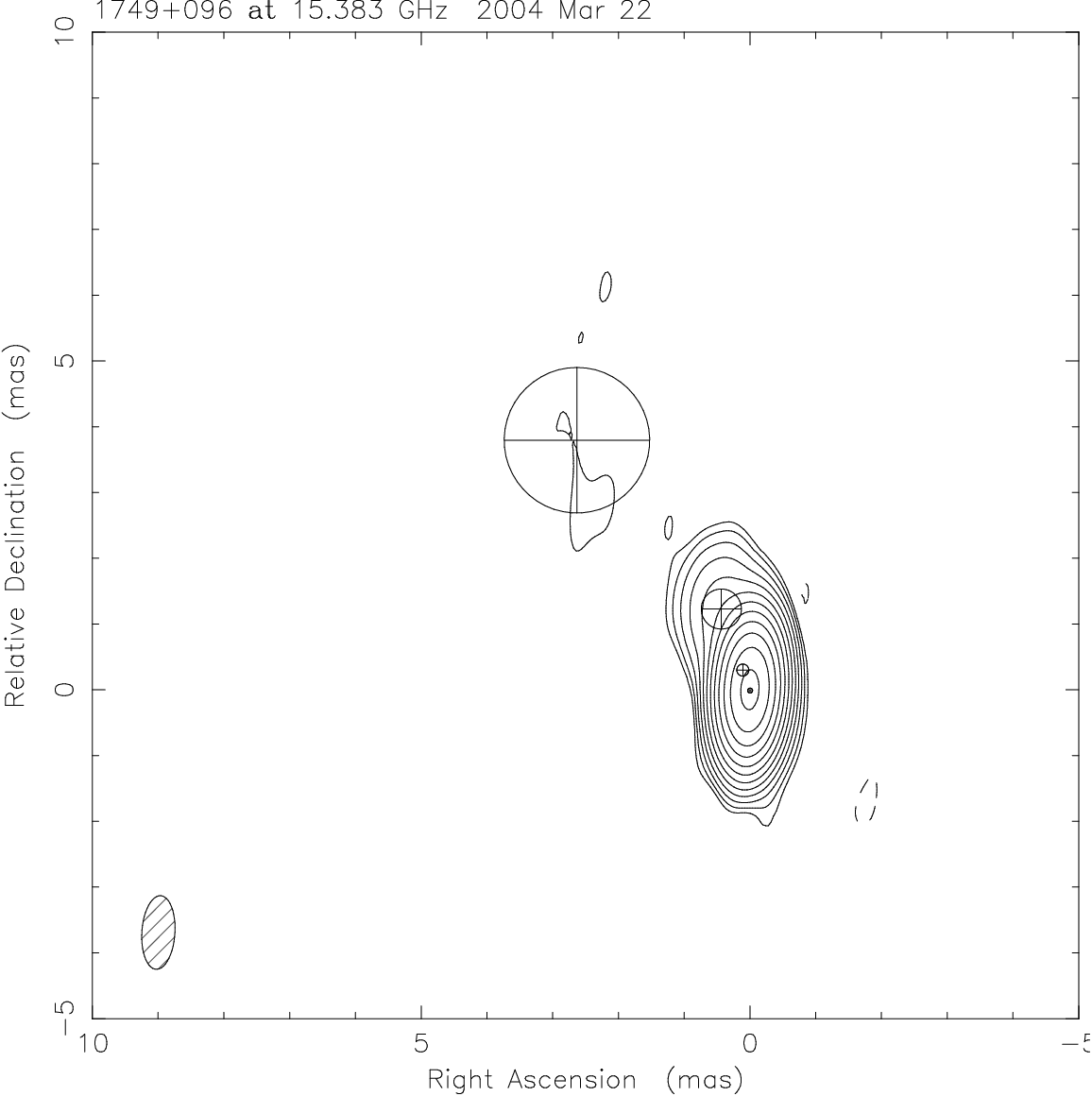}}
 \hspace{3pt}%
 \subfloat[][15\,GHz, 2004.391]{%
 \label{fig:u-39}%
 \includegraphics[width=0.30\textwidth,clip]{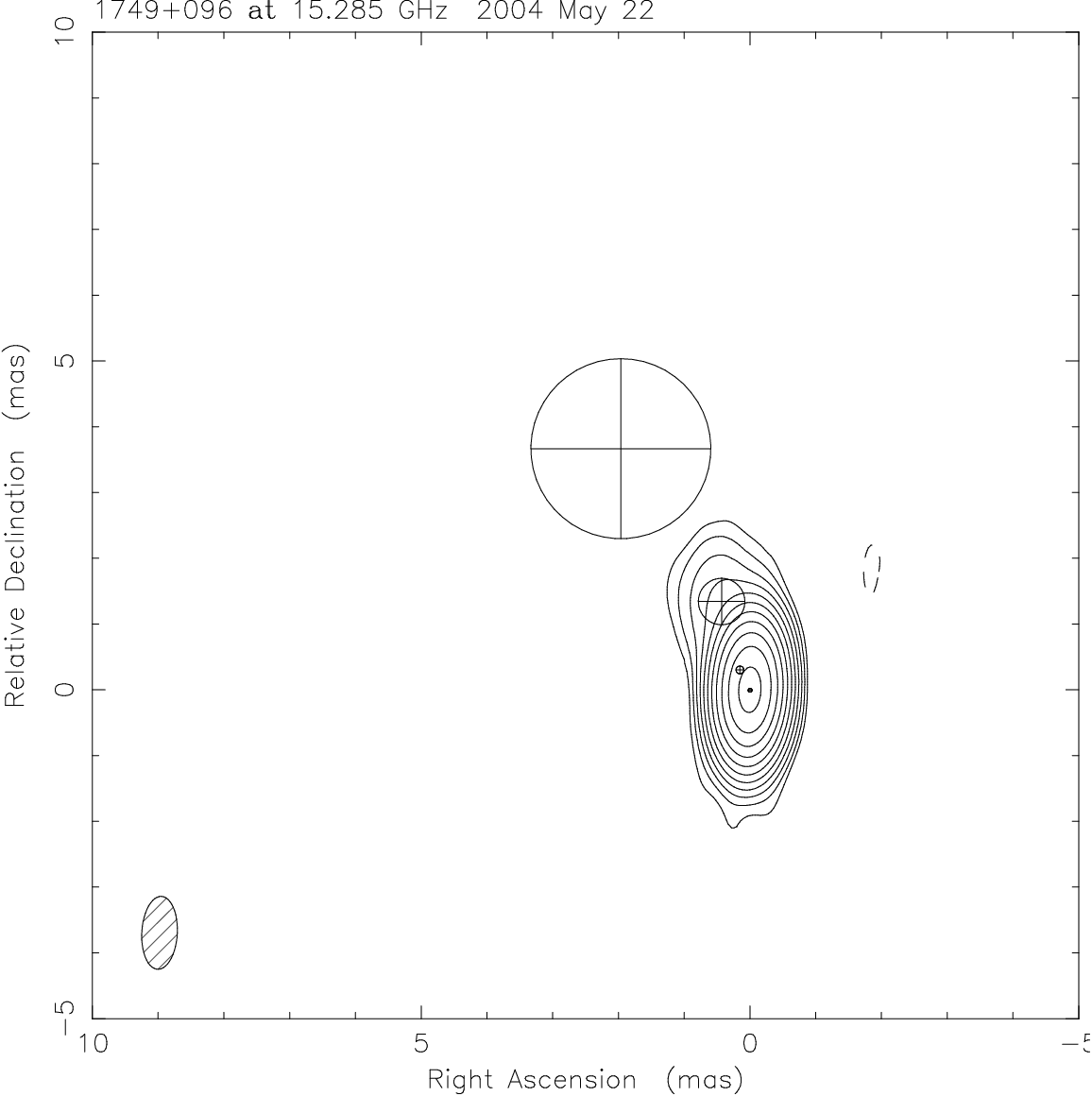}}
 \hspace{3pt}%
 \subfloat[][15\,GHz, 2004.448]{%
 \label{fig:u-40}%
 \includegraphics[width=0.30\textwidth,clip]{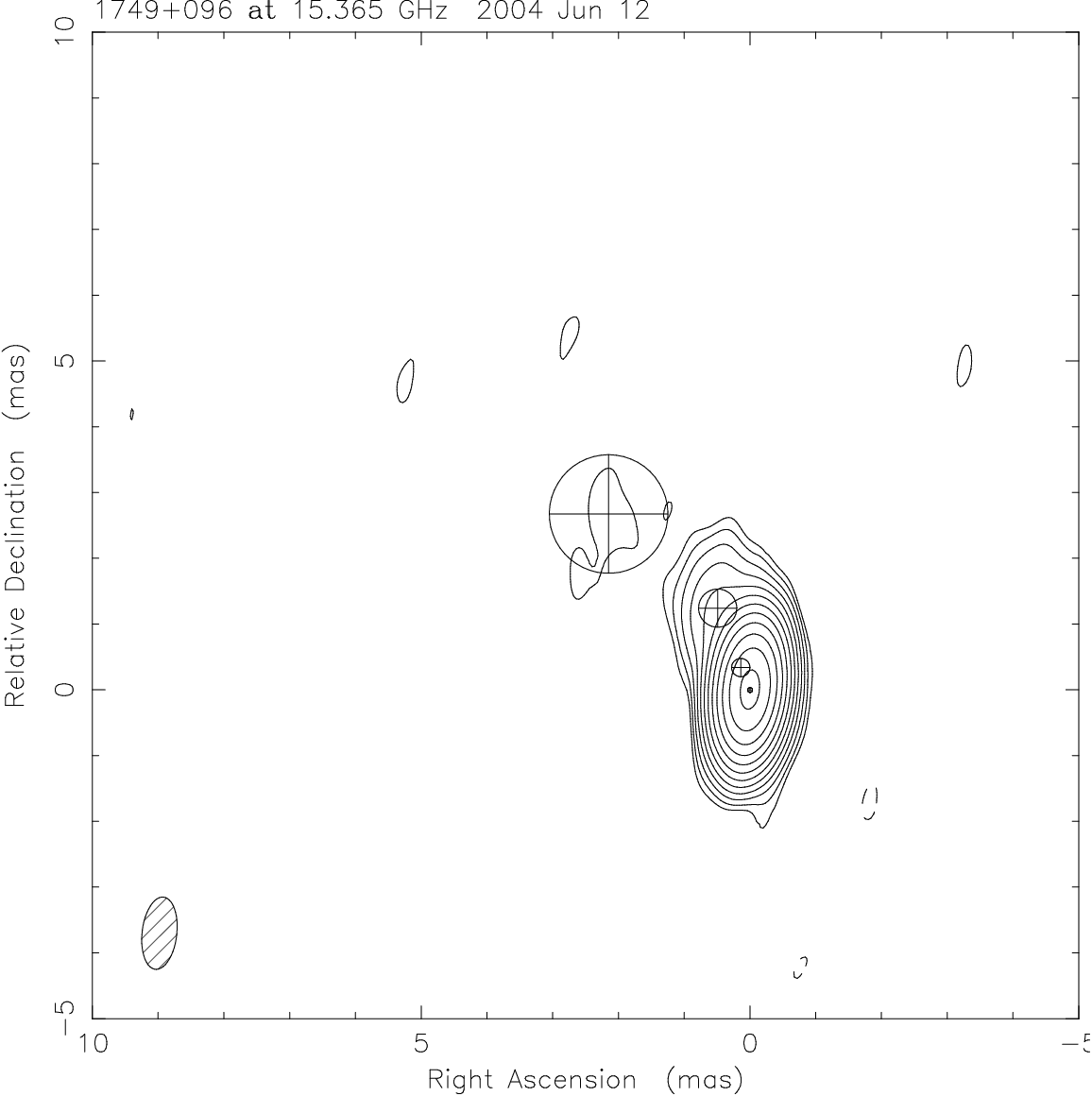}}\\
 \hspace{3pt}%
 \subfloat[][15\,GHz, 2004.691]{%
 \label{fig:u-41}%
 \includegraphics[width=0.30\textwidth,clip]{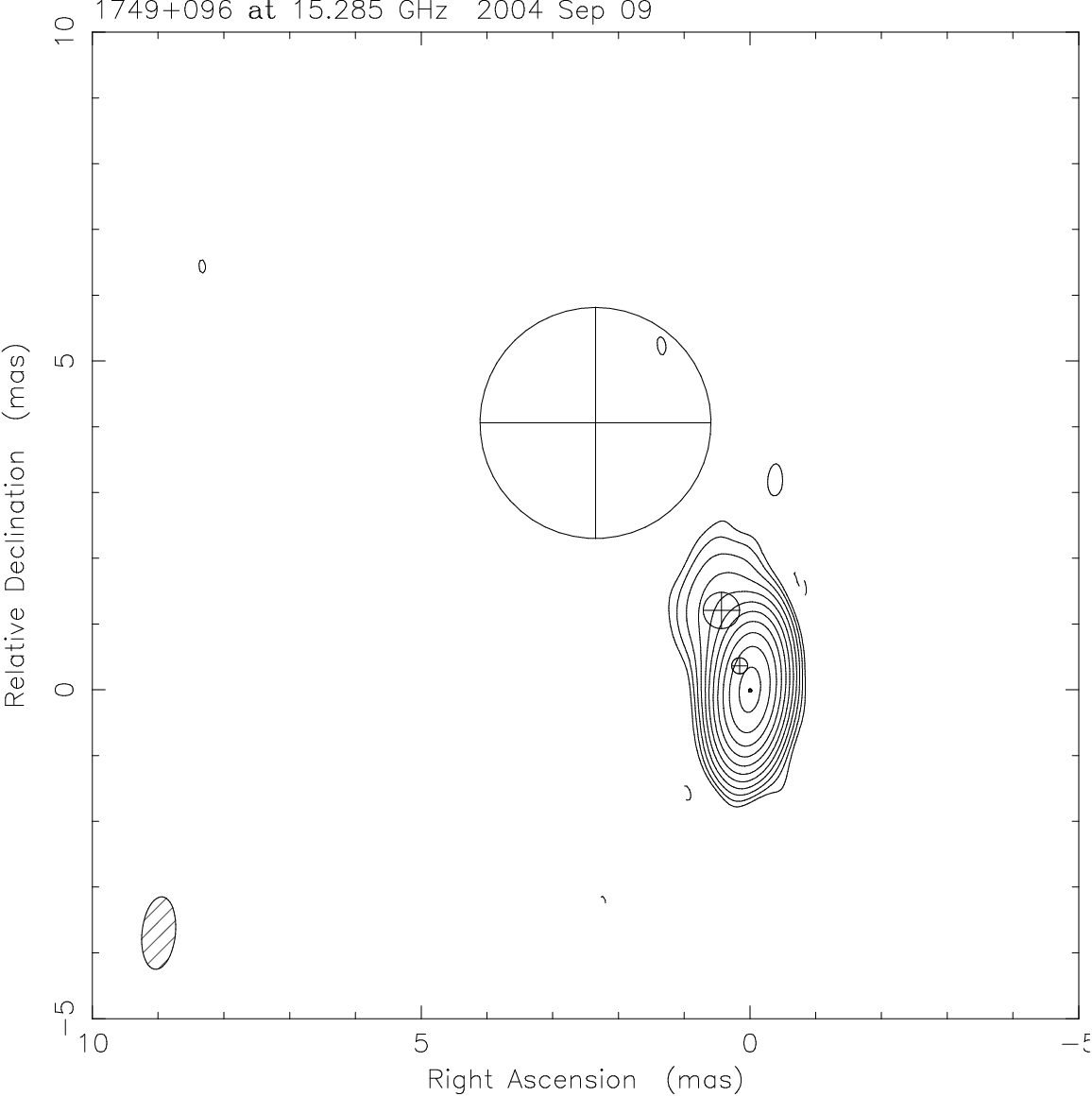}}
 \hspace{3pt}%
 \subfloat[][15\,GHz, 2005.422]{%
 \label{fig:u-42}%
 \includegraphics[width=0.30\textwidth,clip]{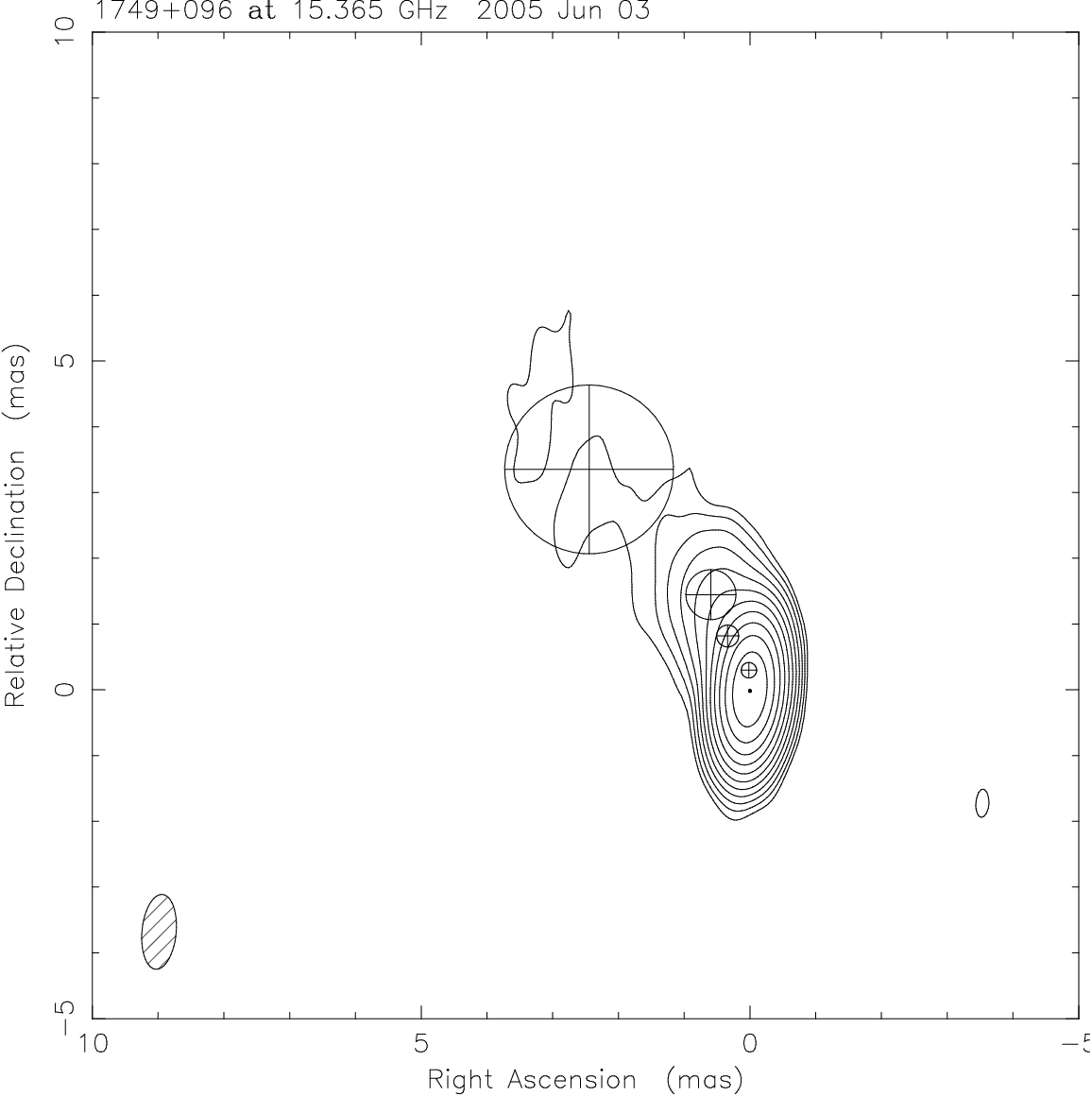}}
\hspace{0pt}%
 \subfloat[][15\,GHz, 2005.534]{%
 \label{fig:u-43}%
 \includegraphics[width=0.30\textwidth,clip]{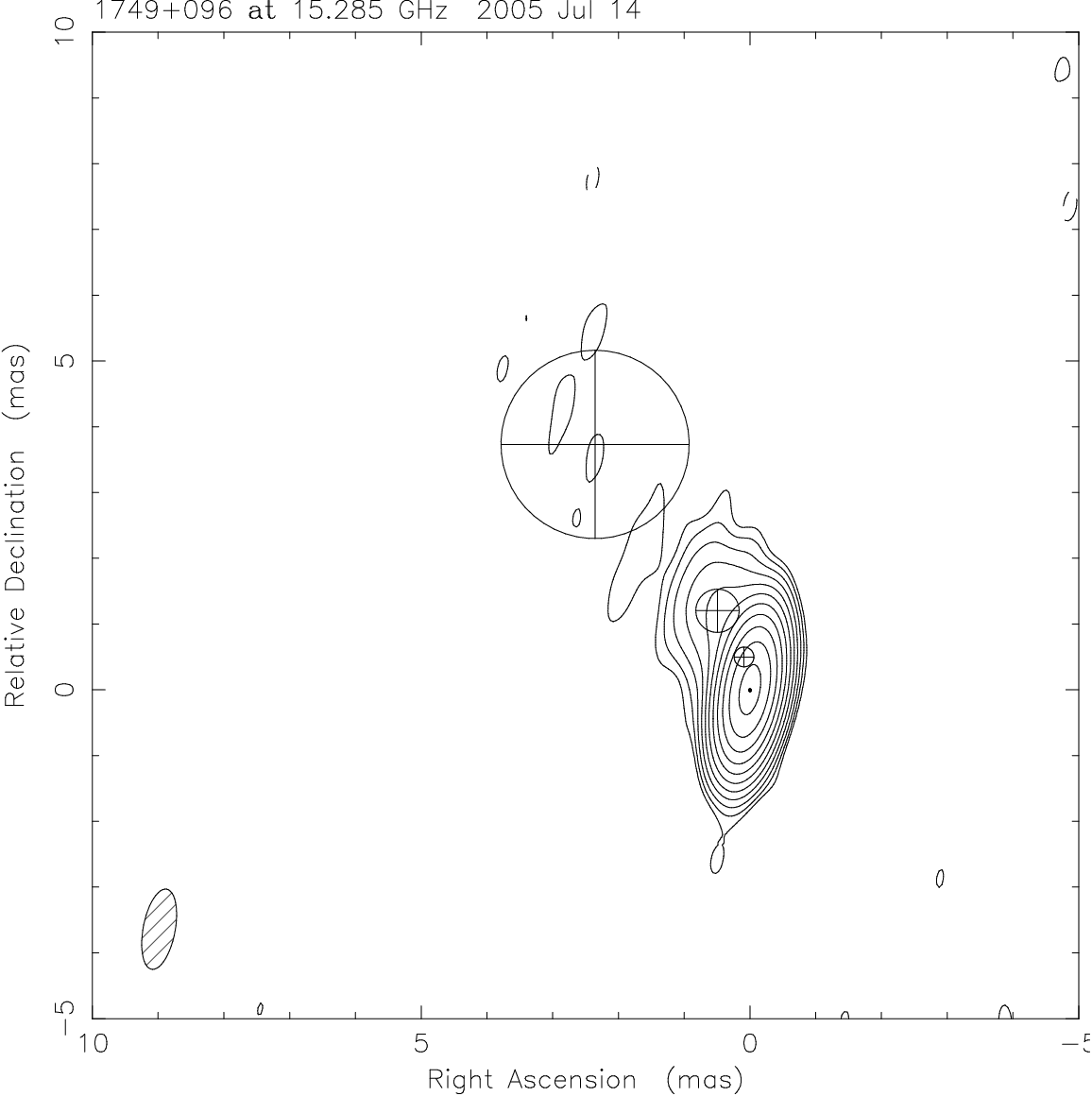}}\\
 \hspace{0pt}%
 \subfloat[][15\,GHz, 2005.668]{%
 \label{fig:u-44}%
 \includegraphics[width=0.30\textwidth,clip]{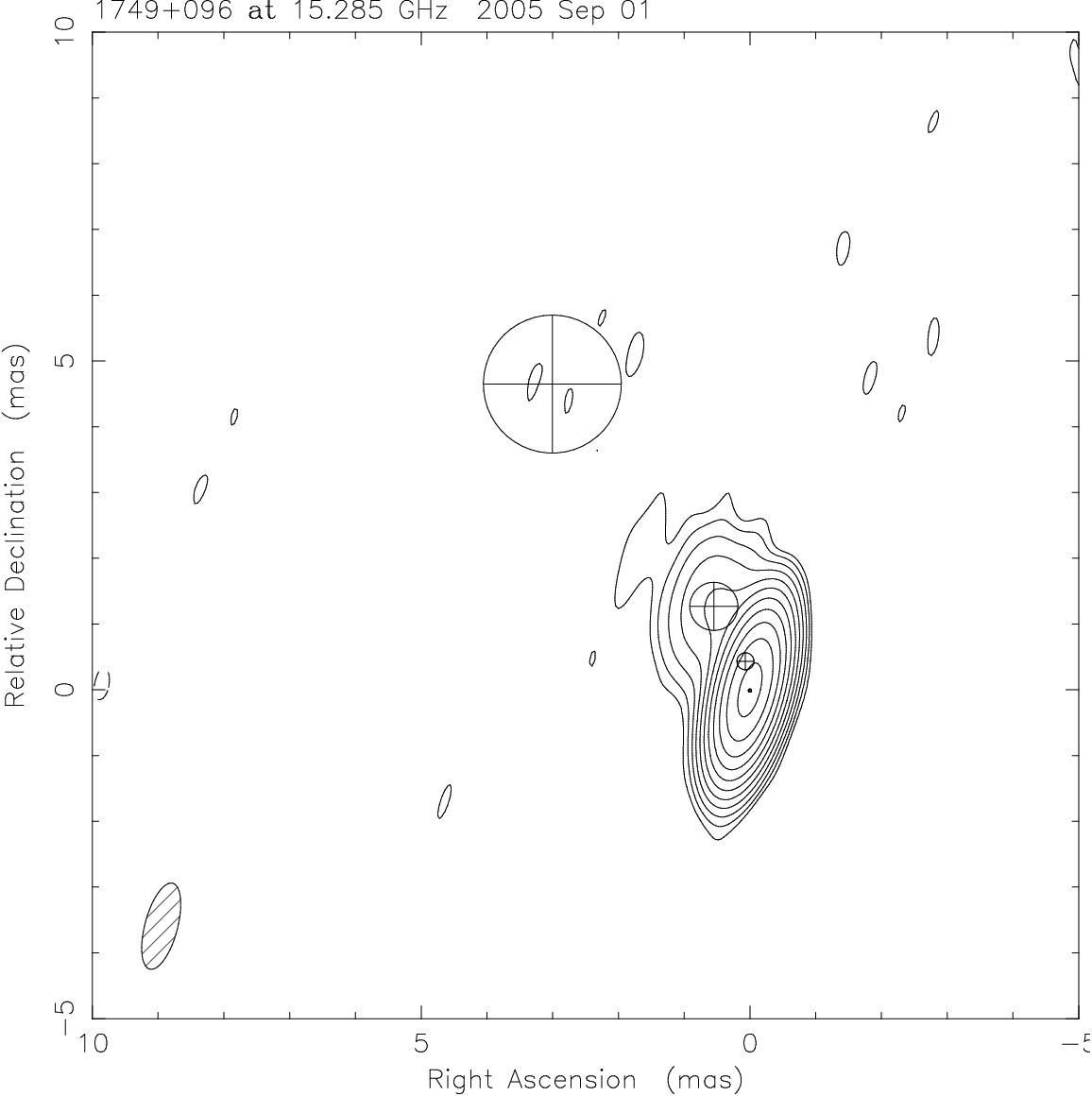}}
 \hspace{3pt}%
 \subfloat[][15\,GHz, 2005.710]{%
 \label{fig:u-45}%
 \includegraphics[width=0.30\textwidth,clip]{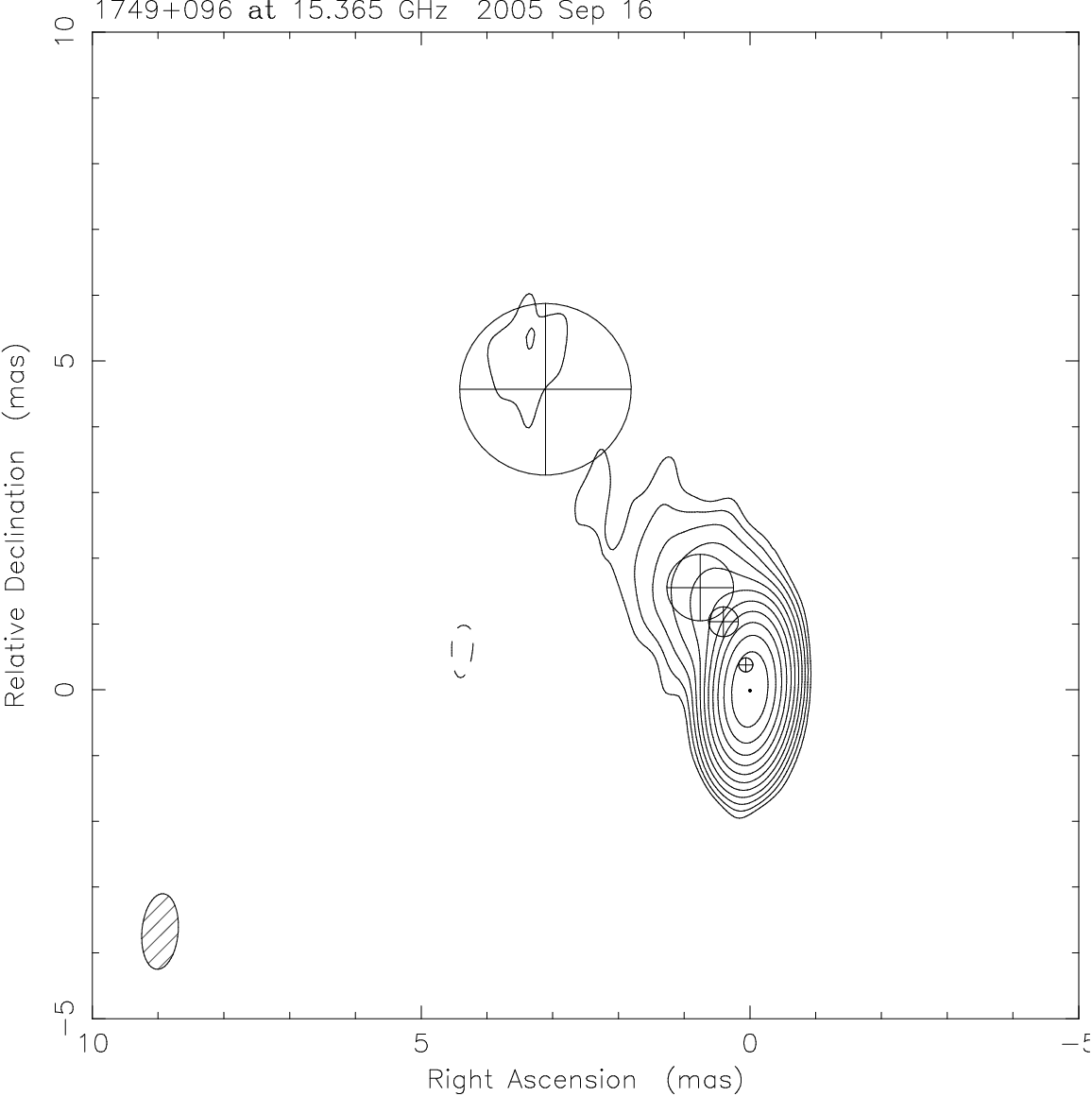}}
  \hspace{3pt}%
  \subfloat[][22\,GHz, 1999.348]{%
  \label{fig:k1}%
  \includegraphics[width=0.30\textwidth,clip]{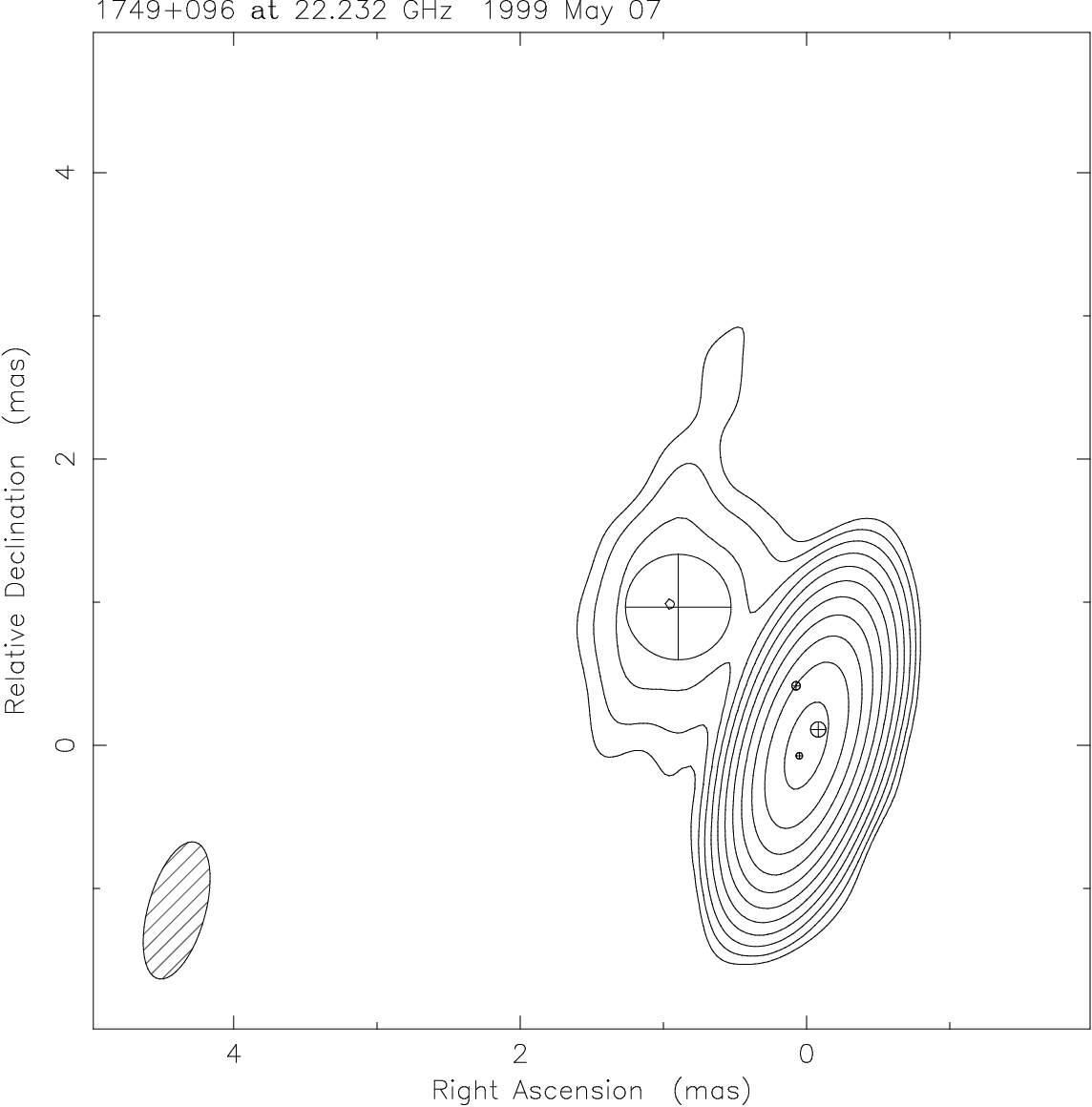}}
  \caption[]{\it{-continued}.}%
  \end{figure*}
  \begin{figure*}%
  \centering
  \ContinuedFloat
  \hspace{0pt}%
  \subfloat[][22\,GHz, 2001.701]{%
  \label{fig:k2}%
  \includegraphics[width=0.30\textwidth,clip]{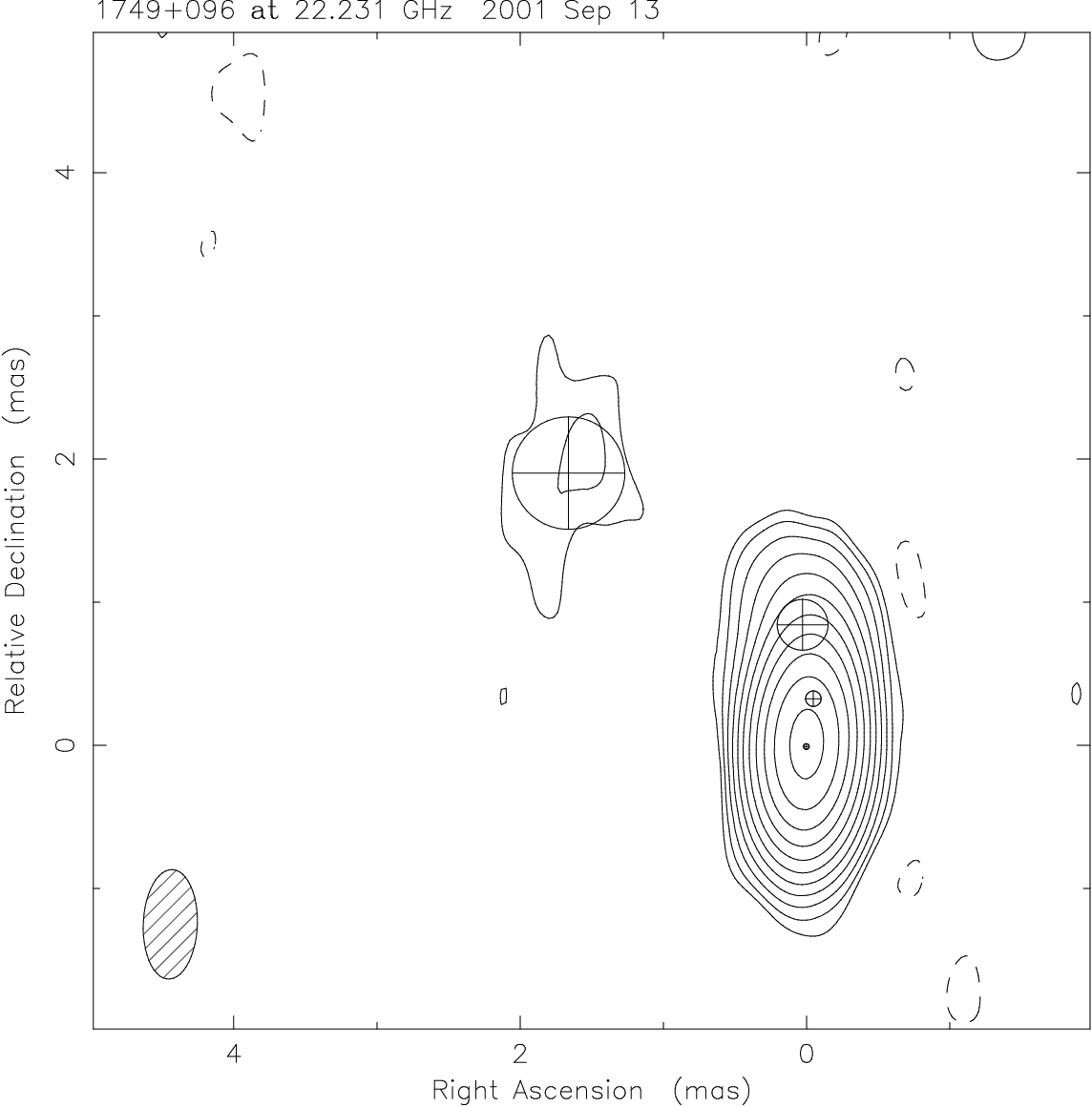}}
  \hspace{3pt}%
  \subfloat[][22\,GHz, 2001.937]{%
  \label{fig:k3}%
  \includegraphics[width=0.30\textwidth,clip]{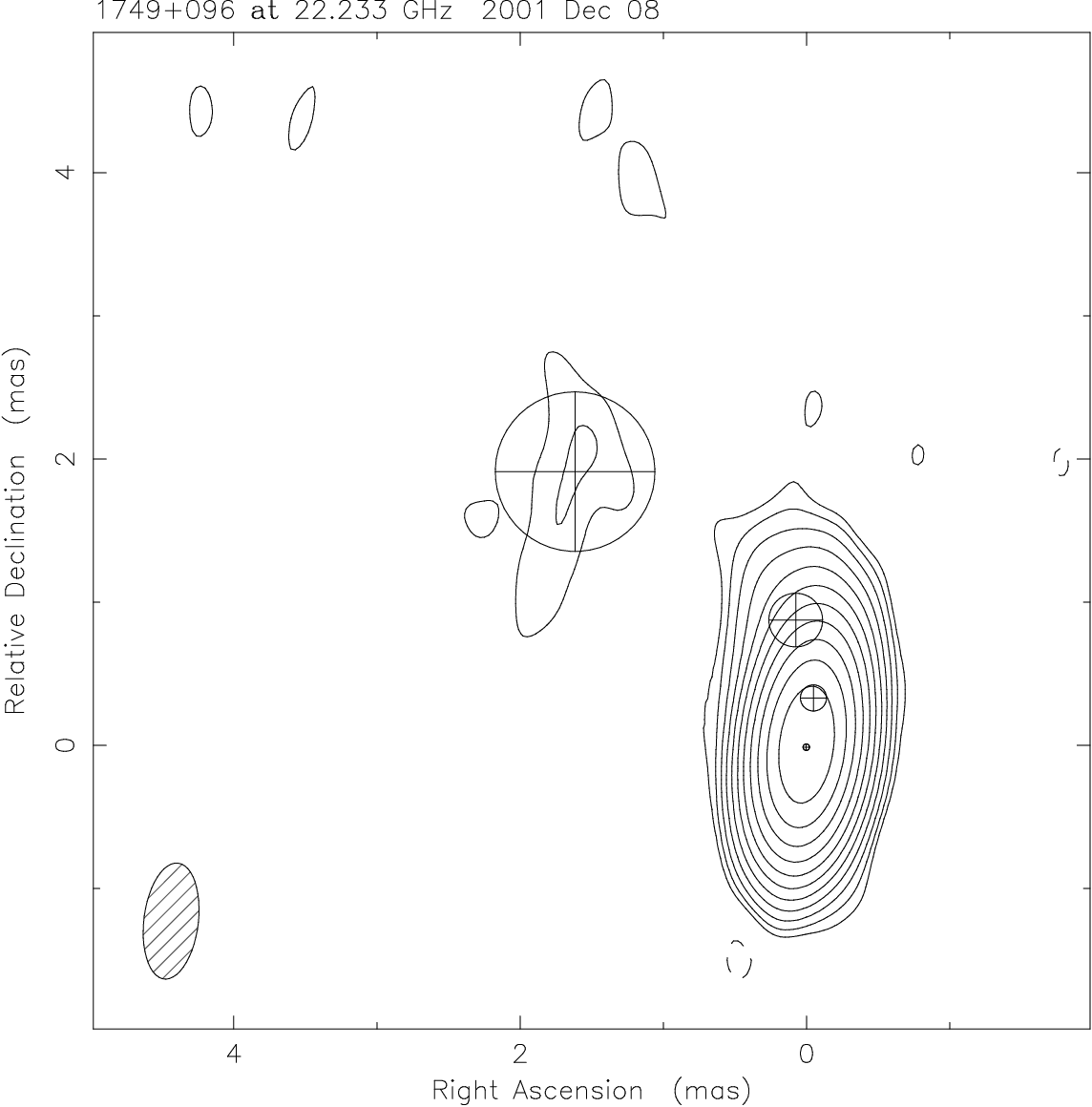}}
  \hspace{3pt}%
  \subfloat[][22\,GHz, 2002.101]{%
  \label{fig:k4}%
  \includegraphics[width=0.30\textwidth,clip]{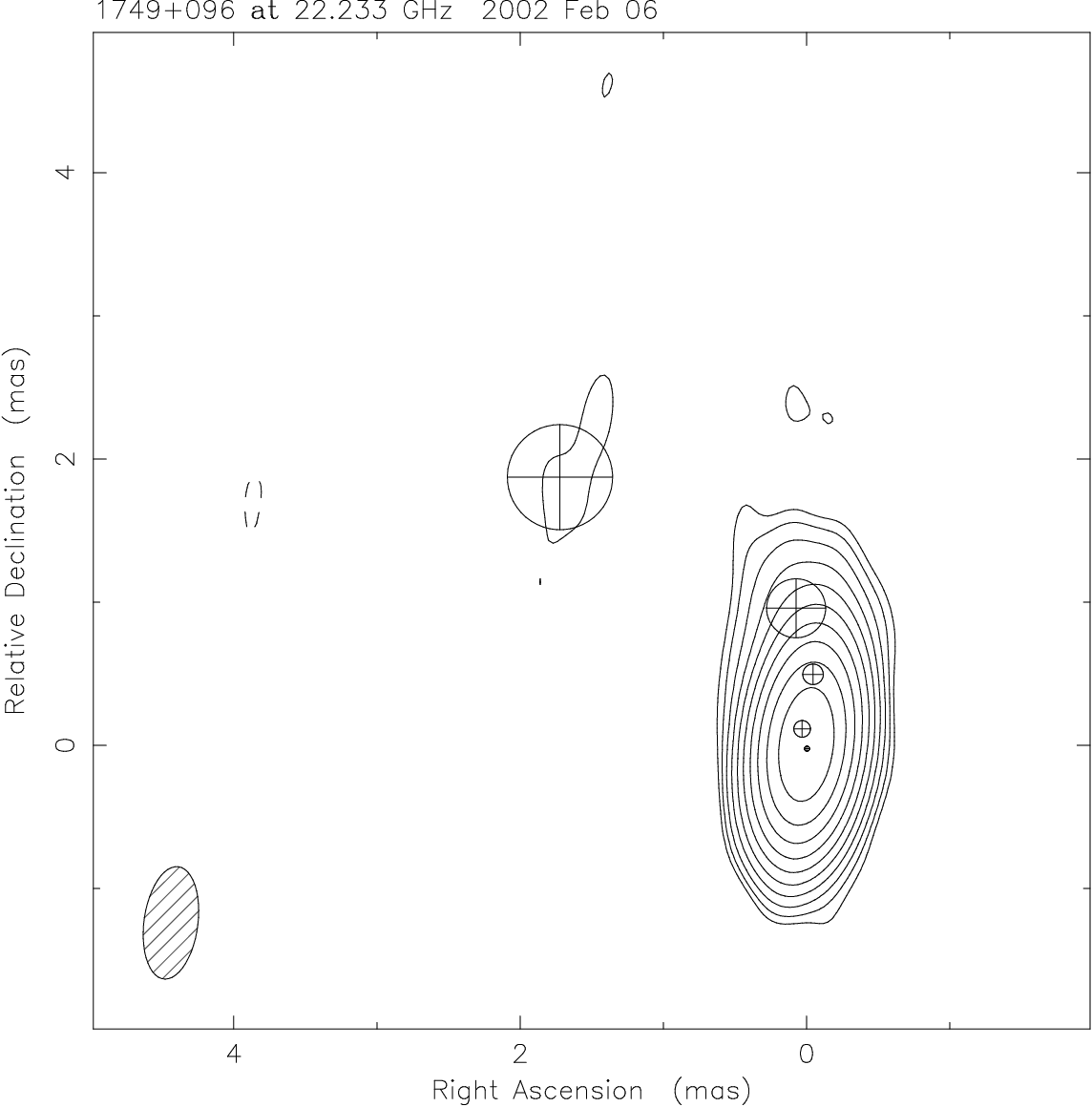}}\\
  \hspace{0pt}%
  \subfloat[][22\,GHz, 2002.449]{%
  \label{fig:k5}%
  \includegraphics[width=0.30\textwidth,clip]{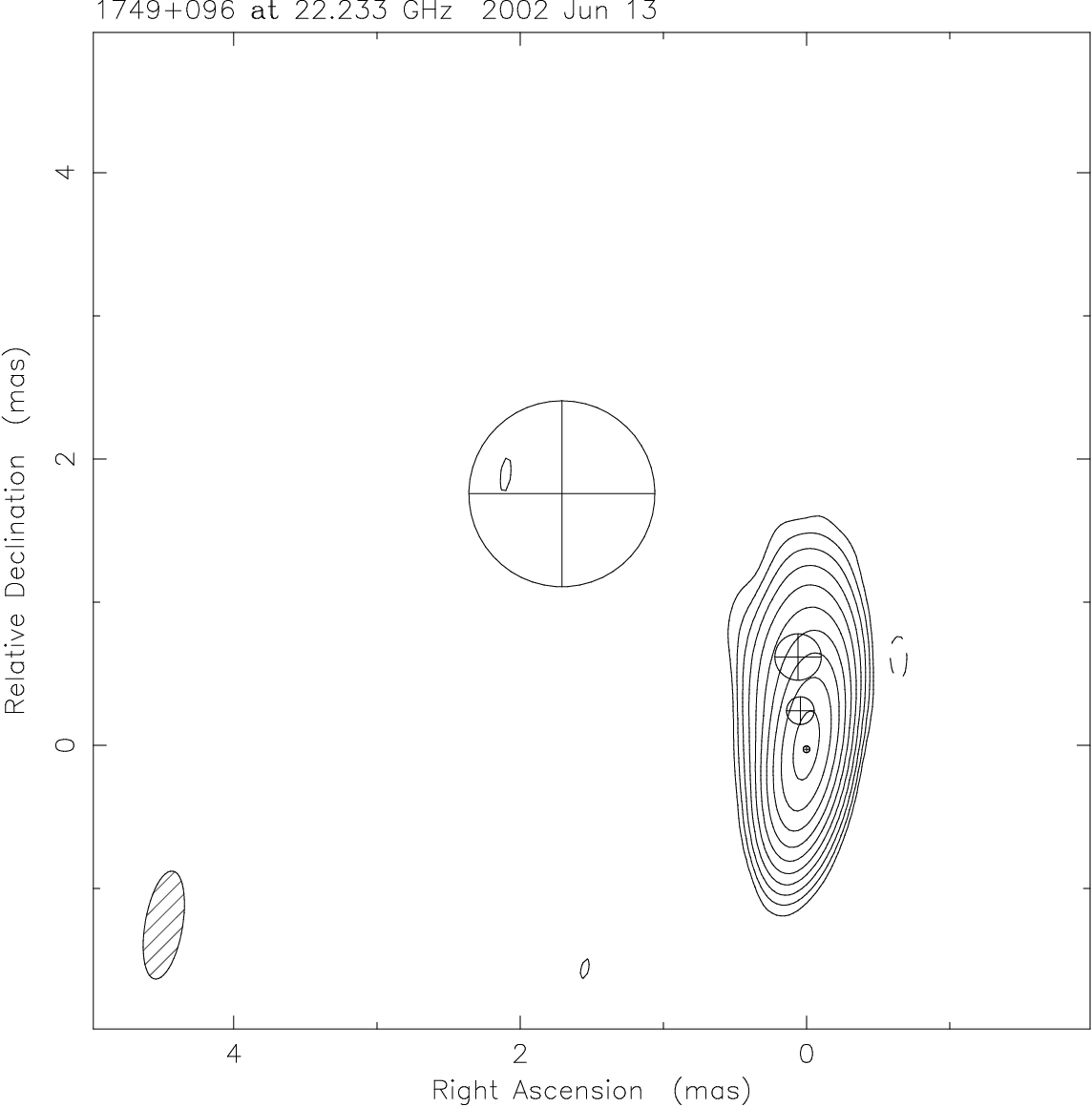}}
  \hspace{3pt}%
  \subfloat[][22\,GHz, 2002.658]{%
  \label{fig:k6}%
  \includegraphics[width=0.30\textwidth,clip]{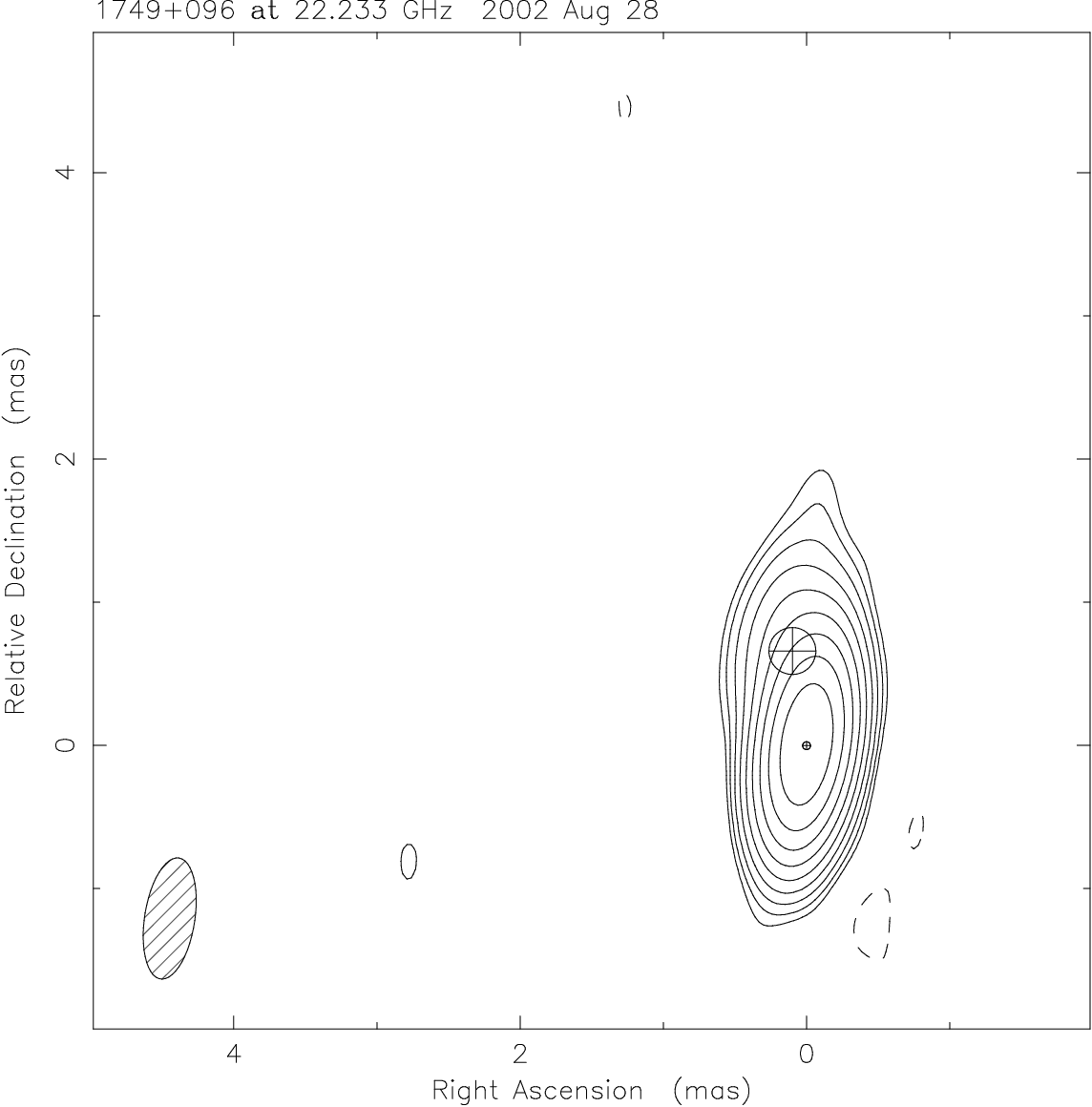}}
  \hspace{3pt}%
  \subfloat[][22\,GHz, 2002.836]{%
  \label{fig:k7}%
  \includegraphics[width=0.30\textwidth,clip]{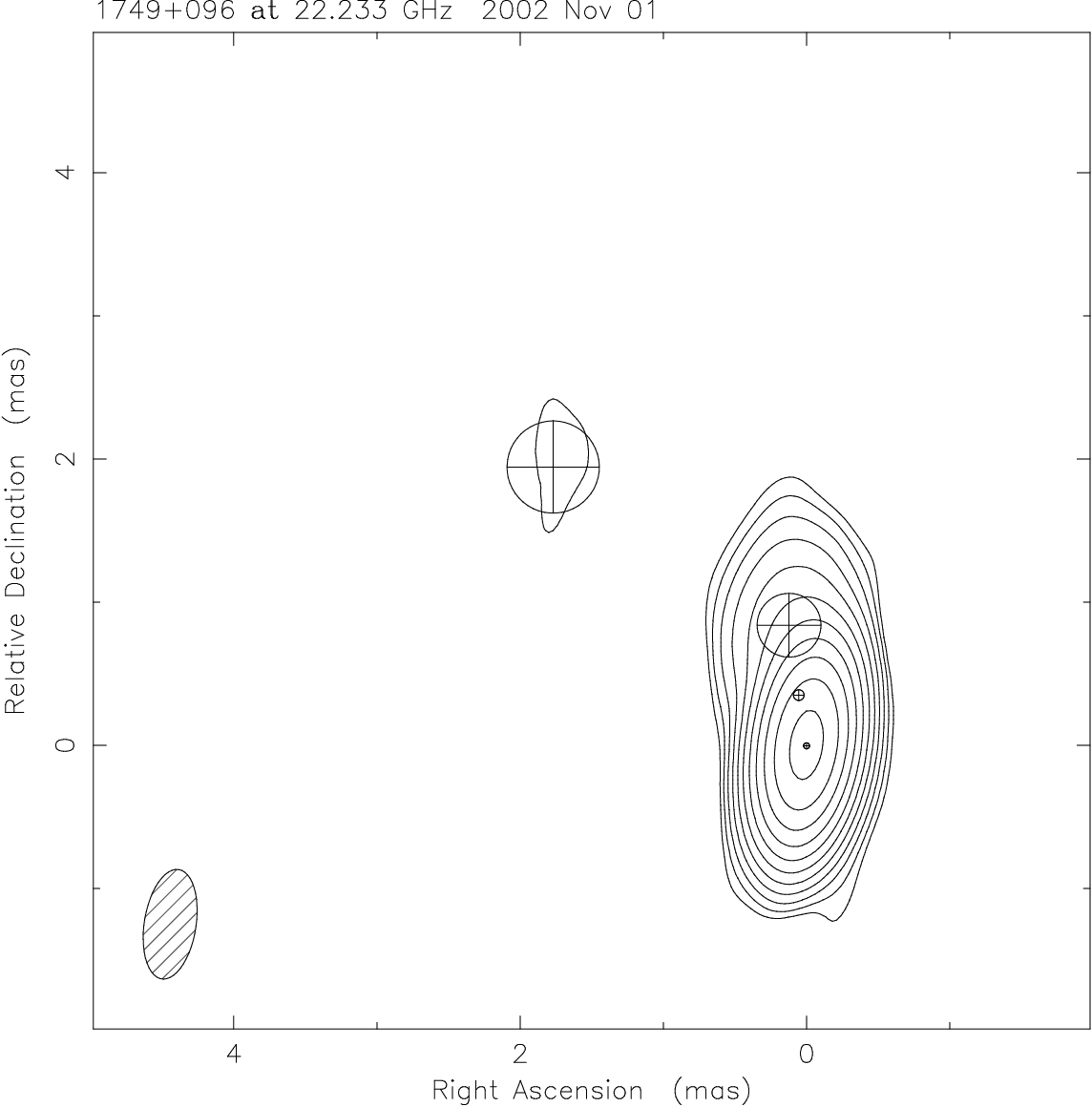}}\\
  \hspace{0pt}%
  \subfloat[][22\,GHz, 2003.008]{%
  \label{fig:k8}%
  \includegraphics[width=0.30\textwidth,clip]{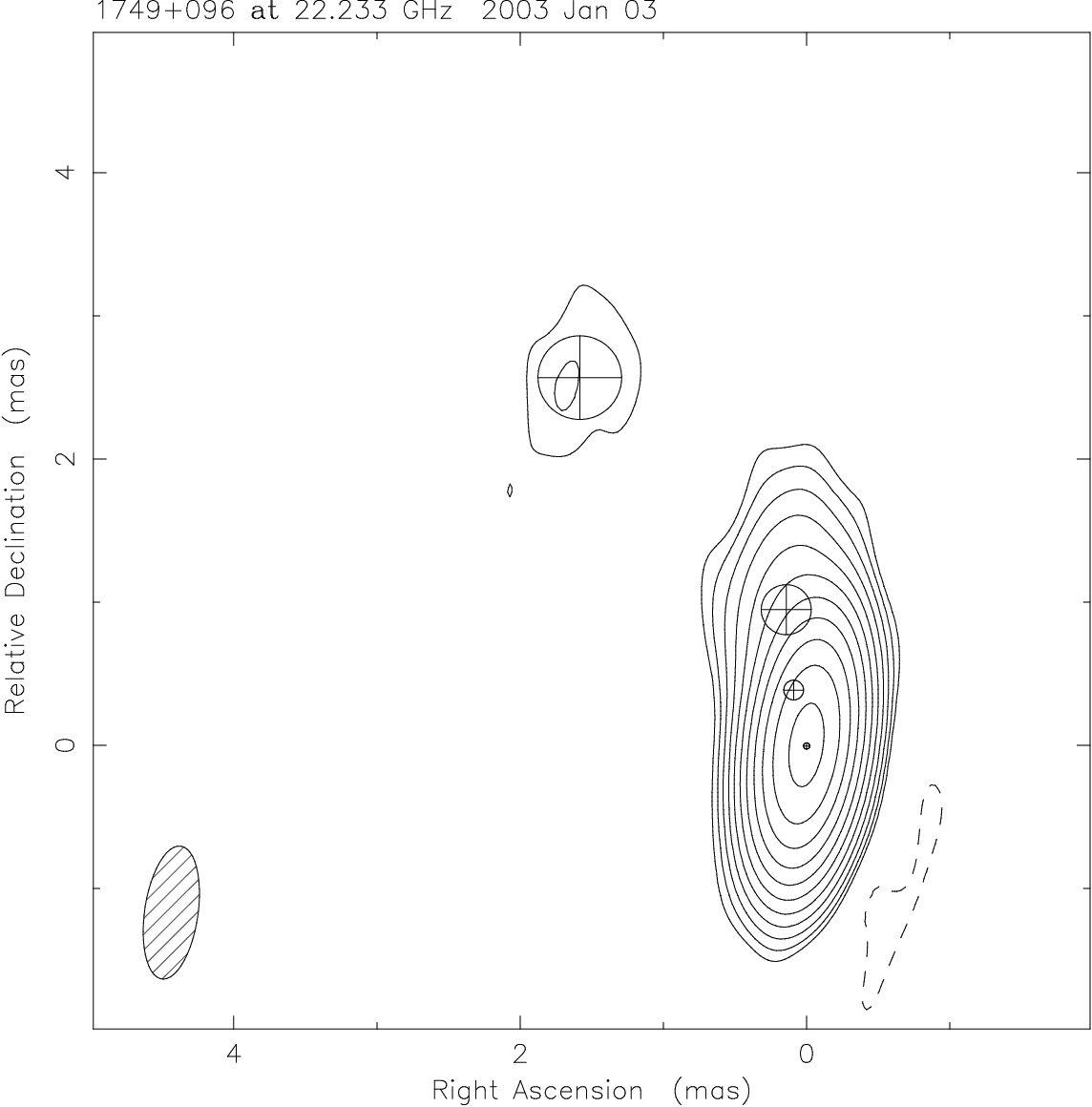}}
  \hspace{3pt}%
  \subfloat[][22\,GHz, 2003.216]{%
  \label{fig:k9}%
  \includegraphics[width=0.30\textwidth,clip]{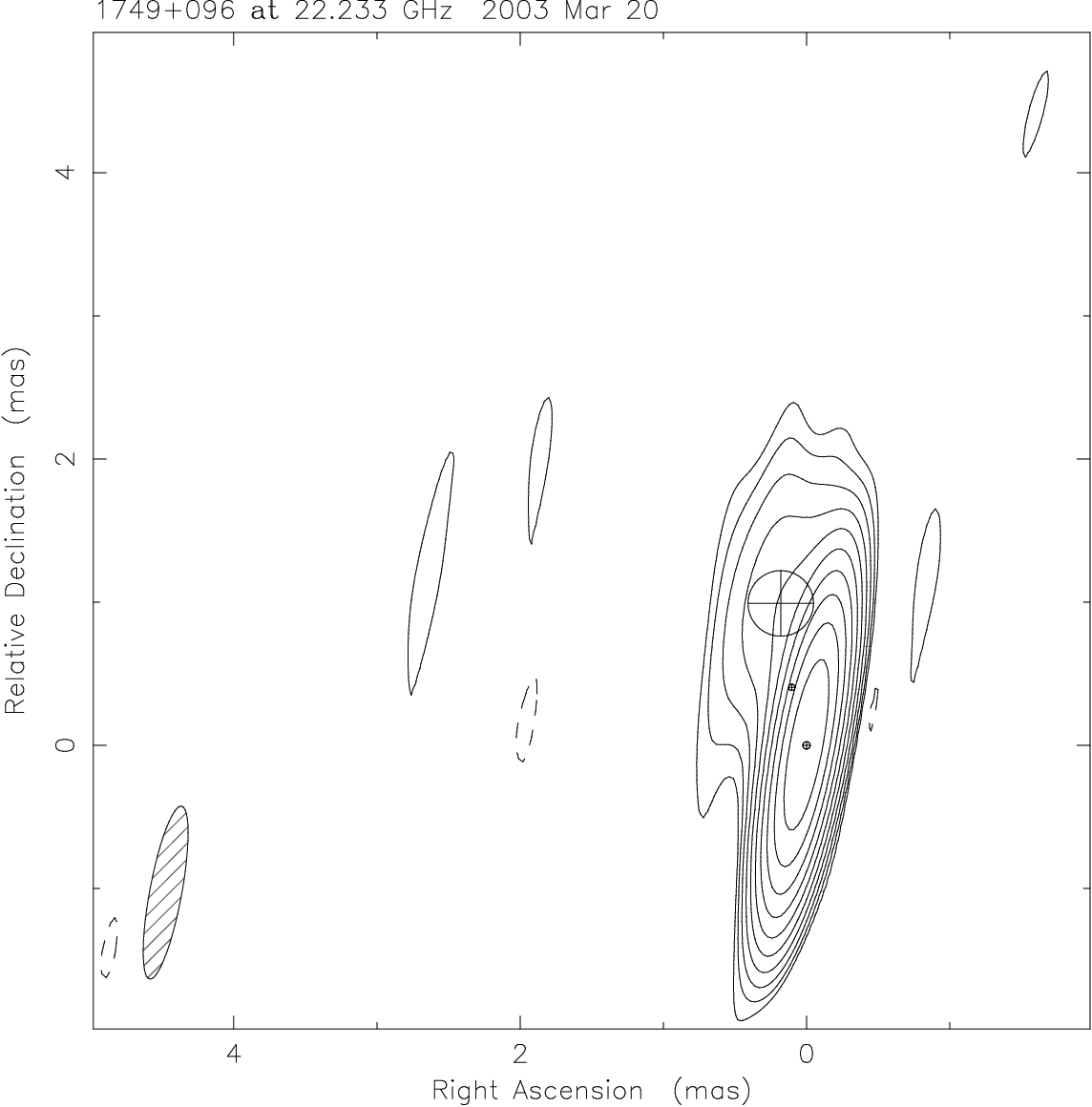}}
  \hspace{3pt}%
  \subfloat[][22\,GHz, 2003.479]{%
  \label{fig:k10}%
  \includegraphics[width=0.30\textwidth,clip]{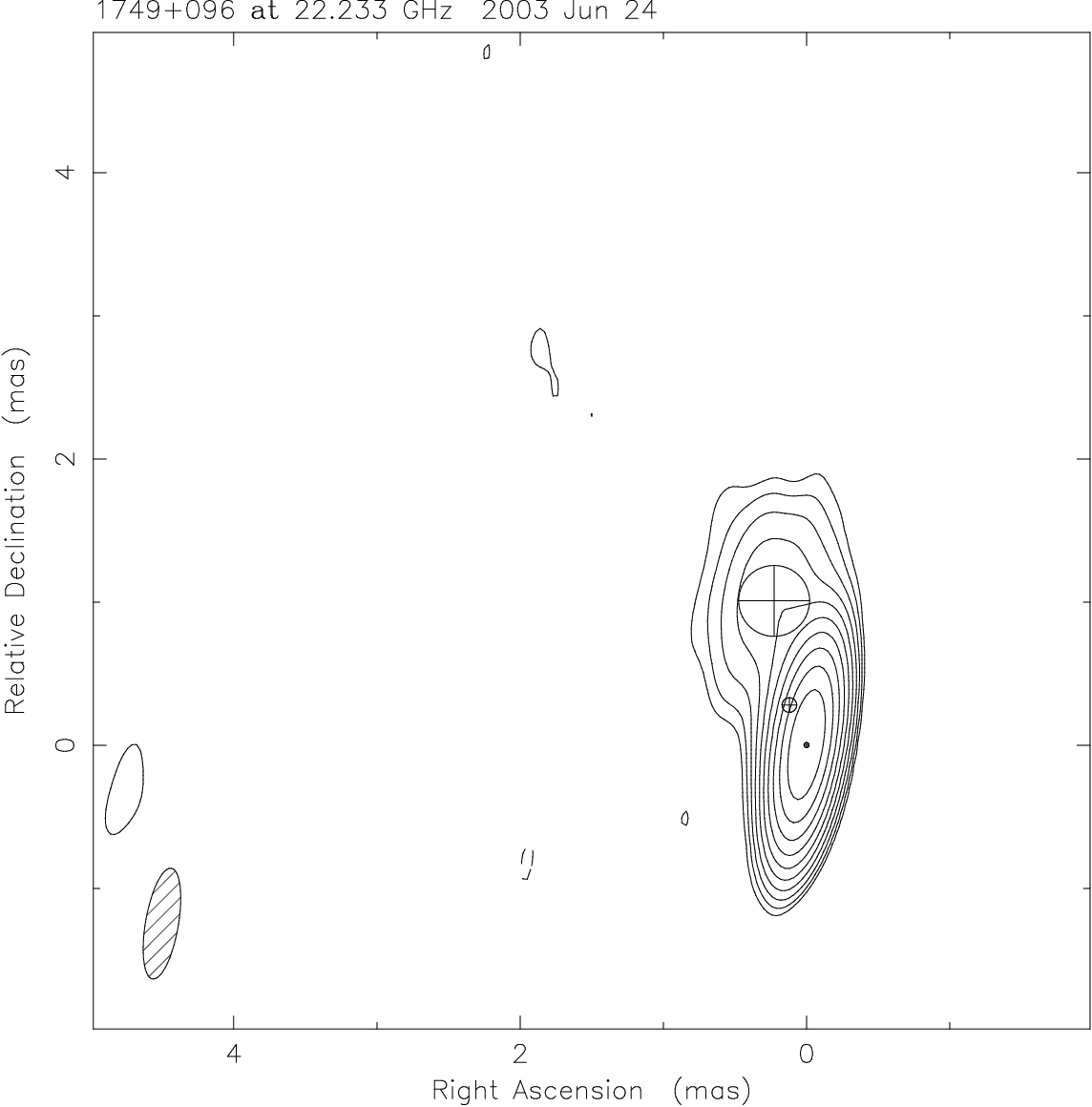}}\\
  \hspace{0pt}%
  \subfloat[][22\,GHz, 2004.153]{%
  \label{fig:k11}%
  \includegraphics[width=0.30\textwidth,clip]{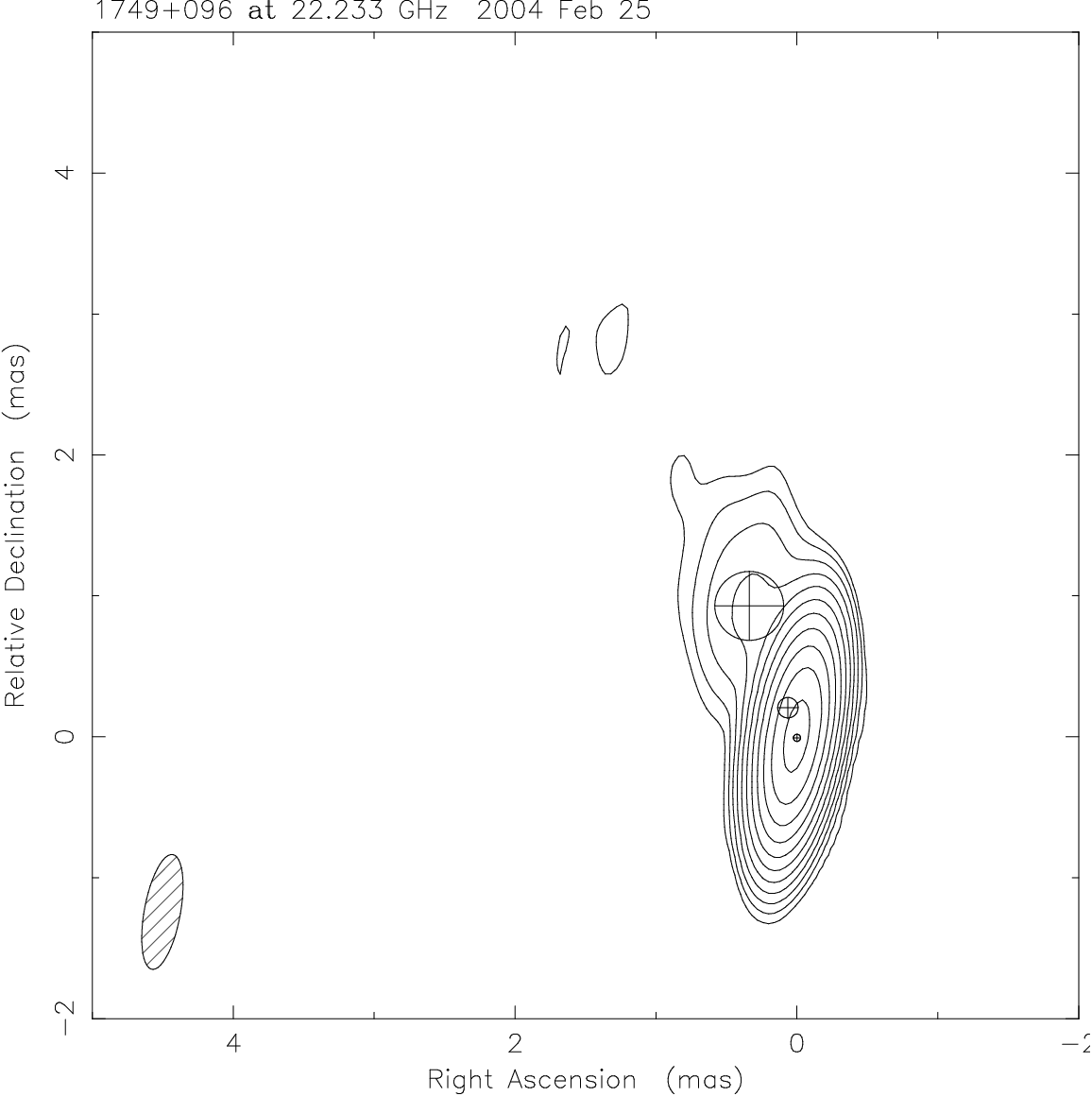}}
  \hspace{3pt}%
  \subfloat[][22\,GHz, 2004.328]{%
  \label{fig:k12}%
  \includegraphics[width=0.30\textwidth,clip]{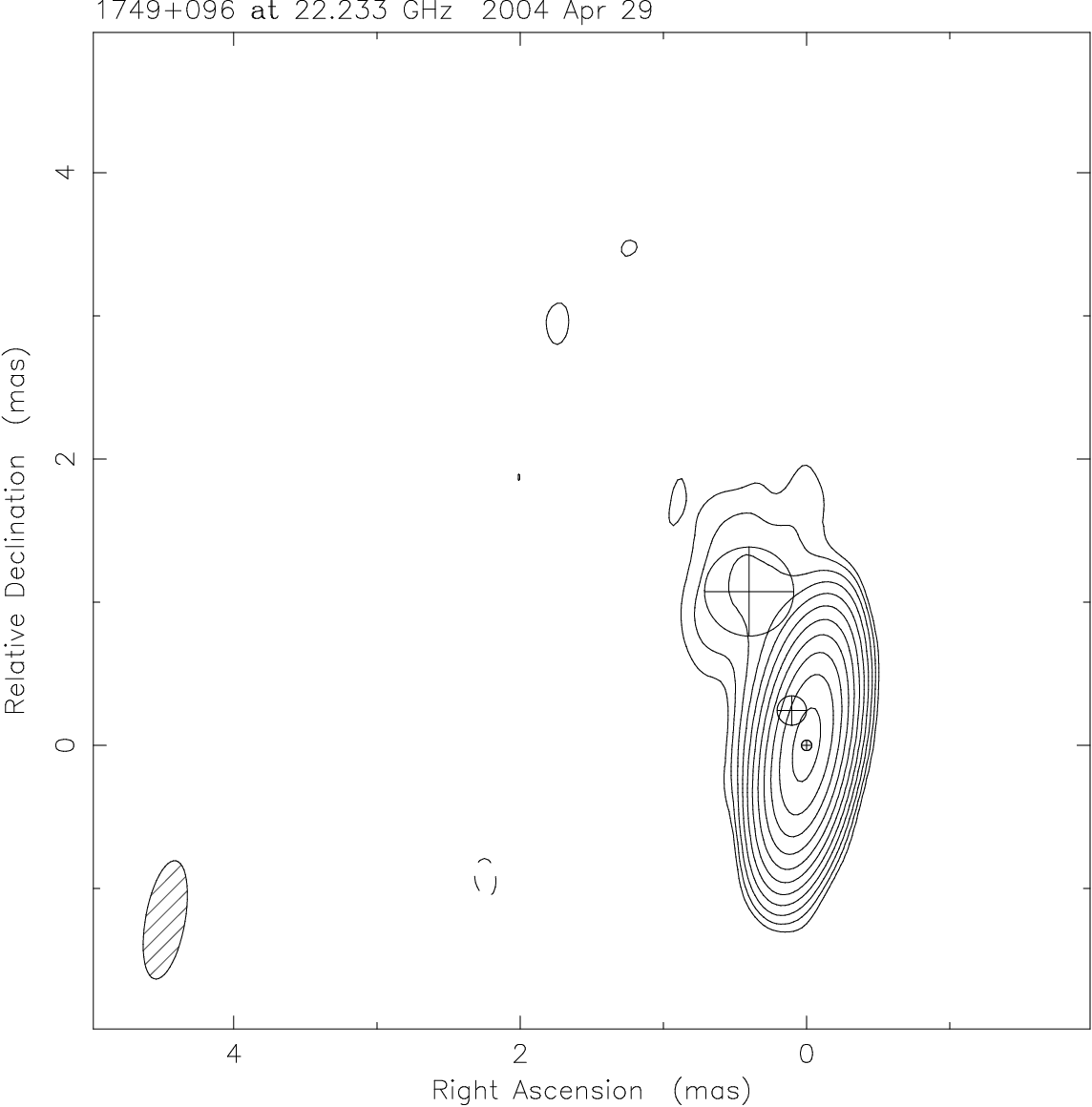}}
  \hspace{3pt}%
  \subfloat[][22\,GHz, 2004.505]{%
  \label{fig:k13}%
  \includegraphics[width=0.30\textwidth,clip]{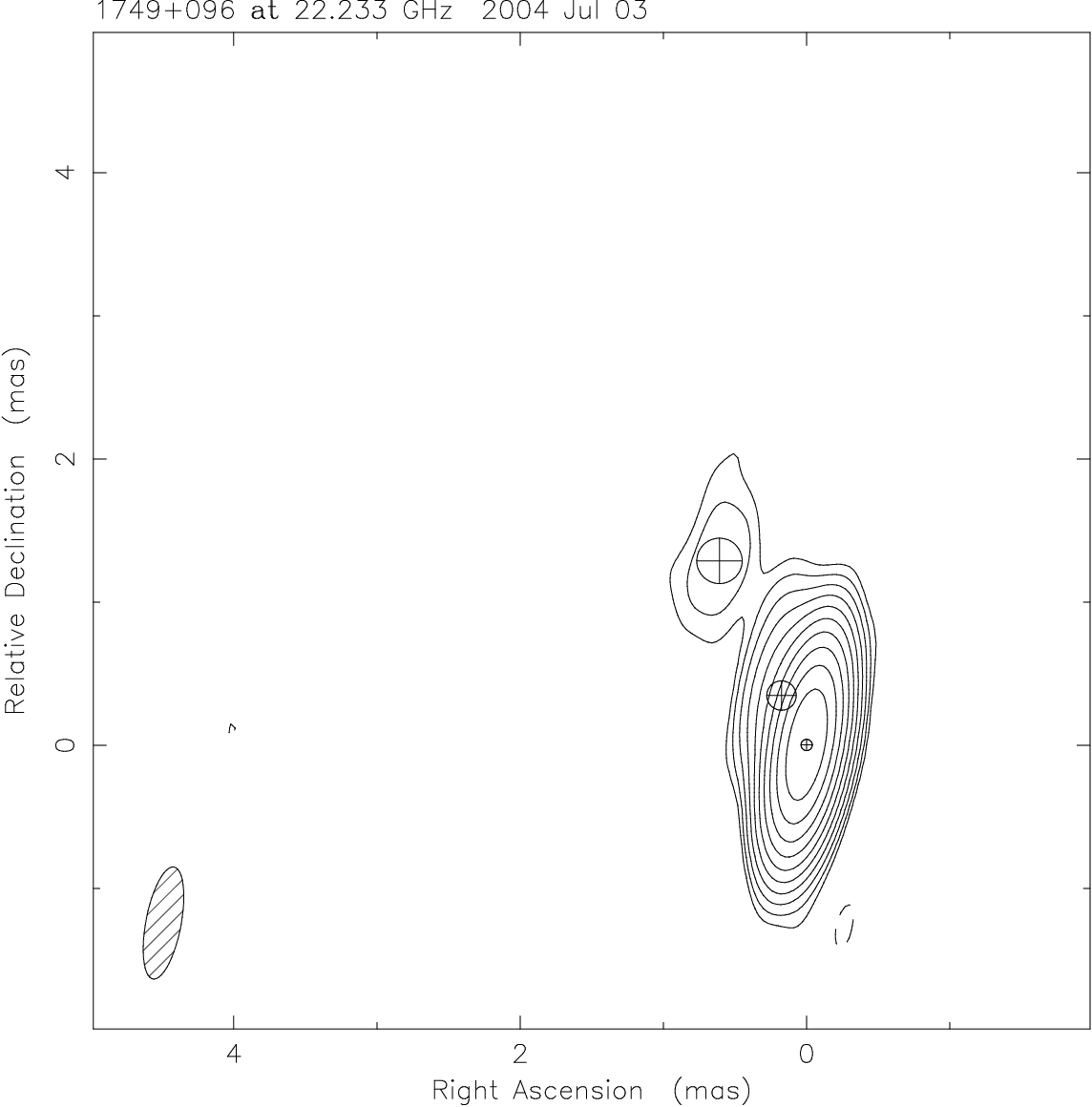}}
  \caption[]{\it{-continued}.}%
  \end{figure*}
  \begin{figure*}%
  \centering
  \ContinuedFloat
  \hspace{0pt}%
  \subfloat[][22\,GHz, 2004.683]{%
  \label{fig:k14}%
  \includegraphics[width=0.30\textwidth,clip]{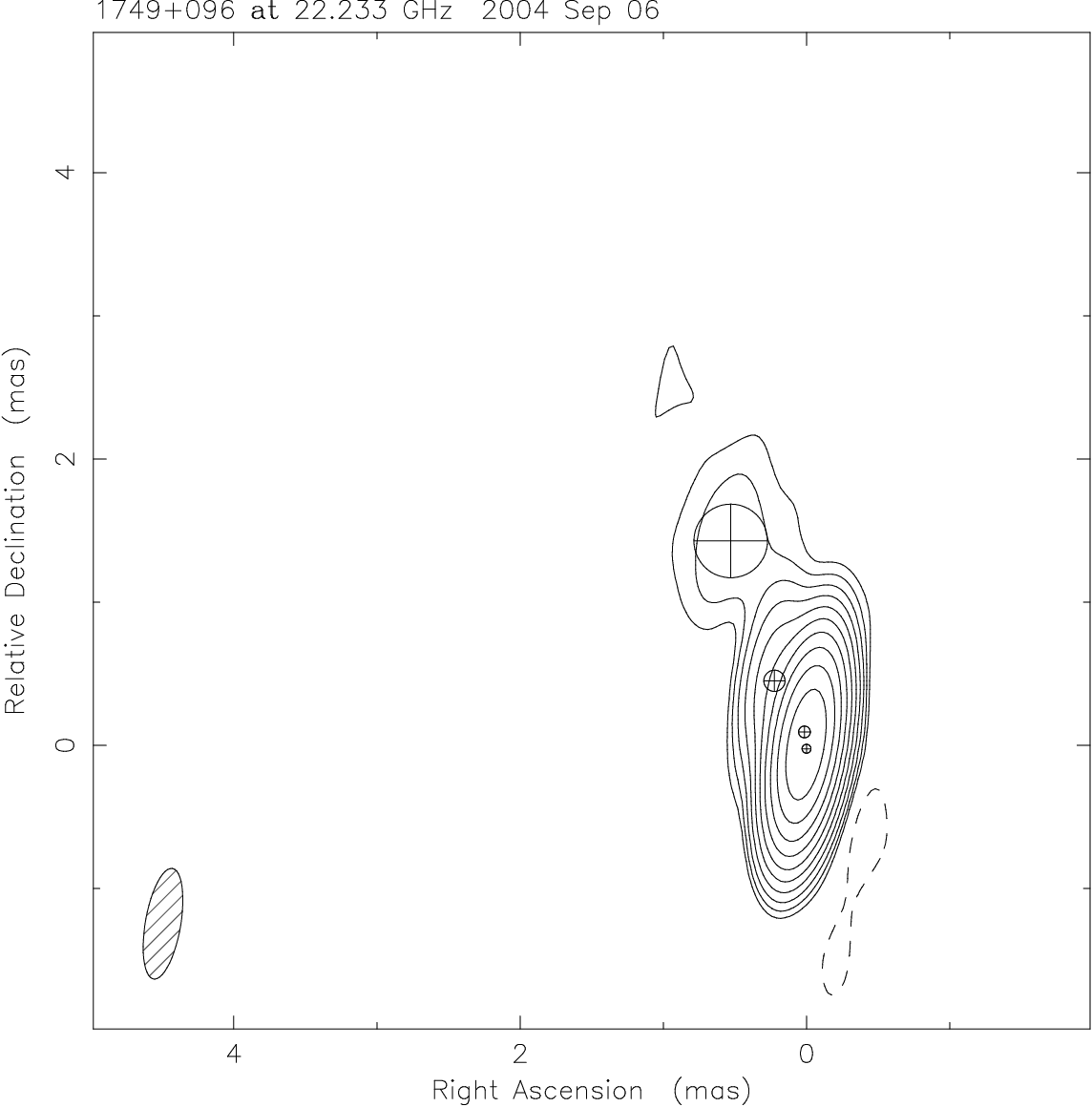}}
  \hspace{3pt}%
  \subfloat[][22\,GHz, 2004.874]{%
  \label{fig:k15}%
  \includegraphics[width=0.30\textwidth,clip]{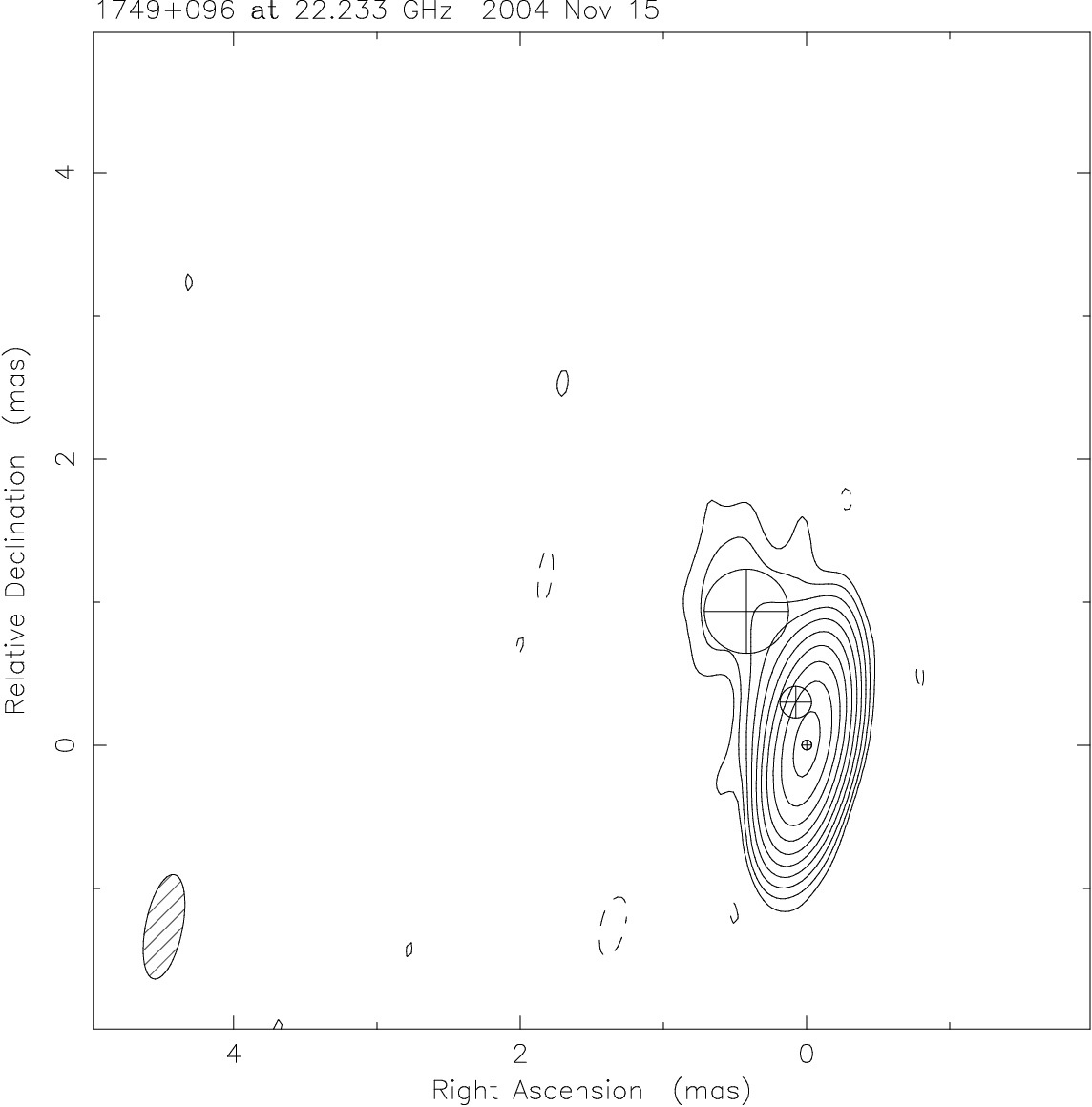}}
  \hspace{3pt}%
  \subfloat[][22\,GHz, 2005.093]{%
  \label{fig:k16}%
  \includegraphics[width=0.30\textwidth,clip]{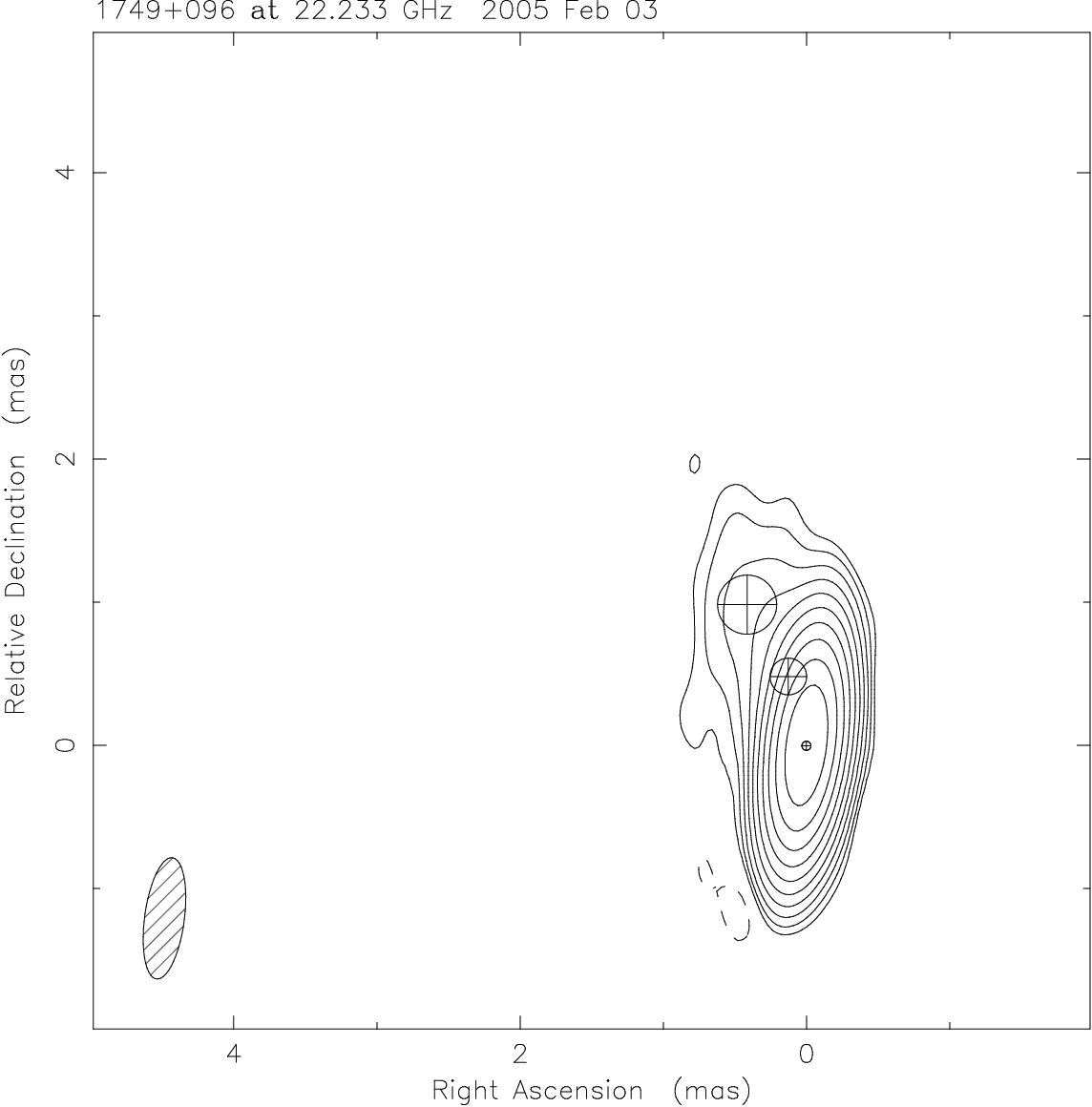}}\\
  \hspace{0pt}%
  \subfloat[][43\,GHz, 1999.348]{%
  \label{fig:q1}%
  \includegraphics[width=0.30\textwidth,clip]{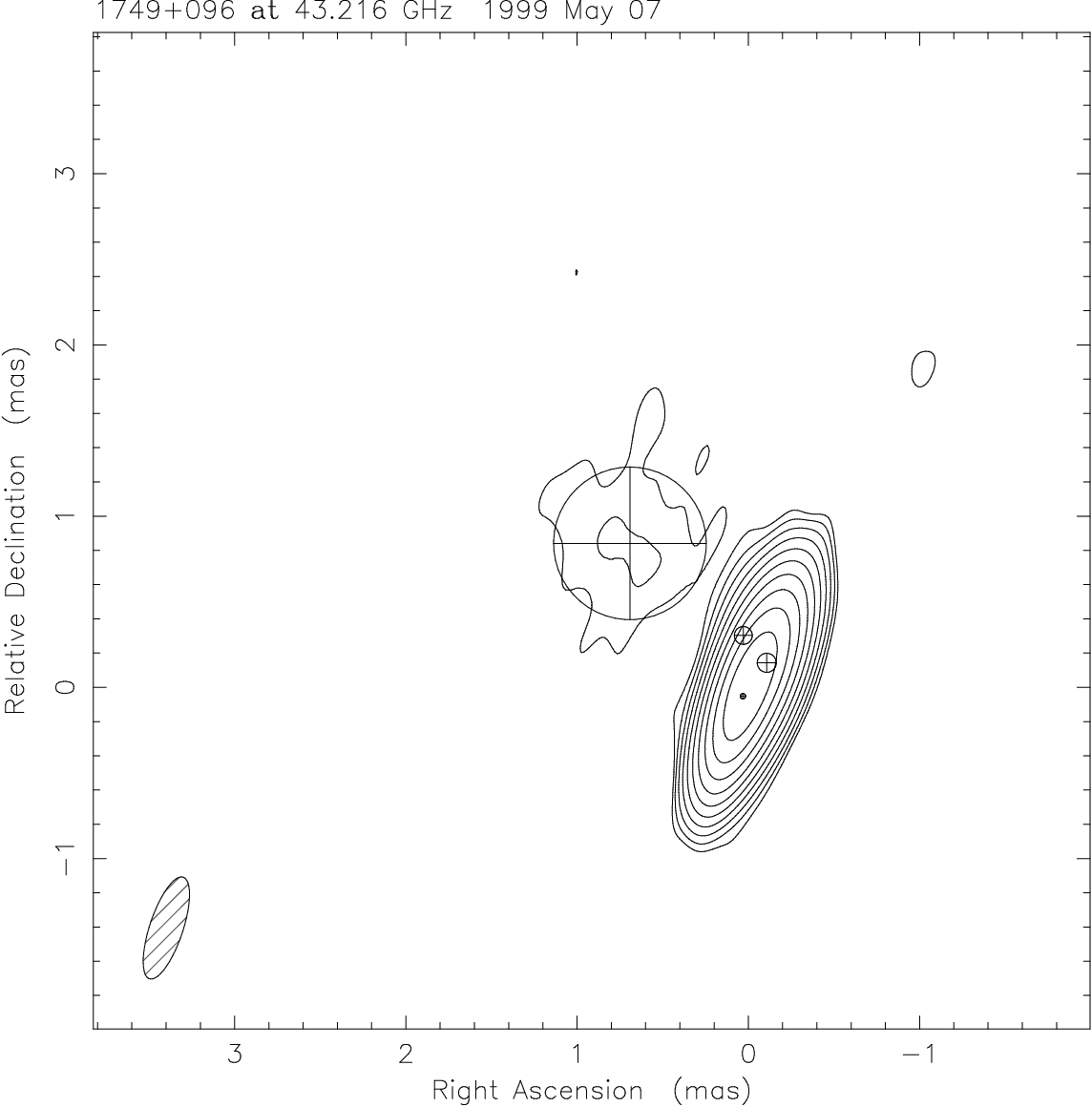}}
  \hspace{0pt}%
  \subfloat[][43\,GHz, 2001.701]{%
  \label{fig:q2}%
  \includegraphics[width=0.30\textwidth,clip]{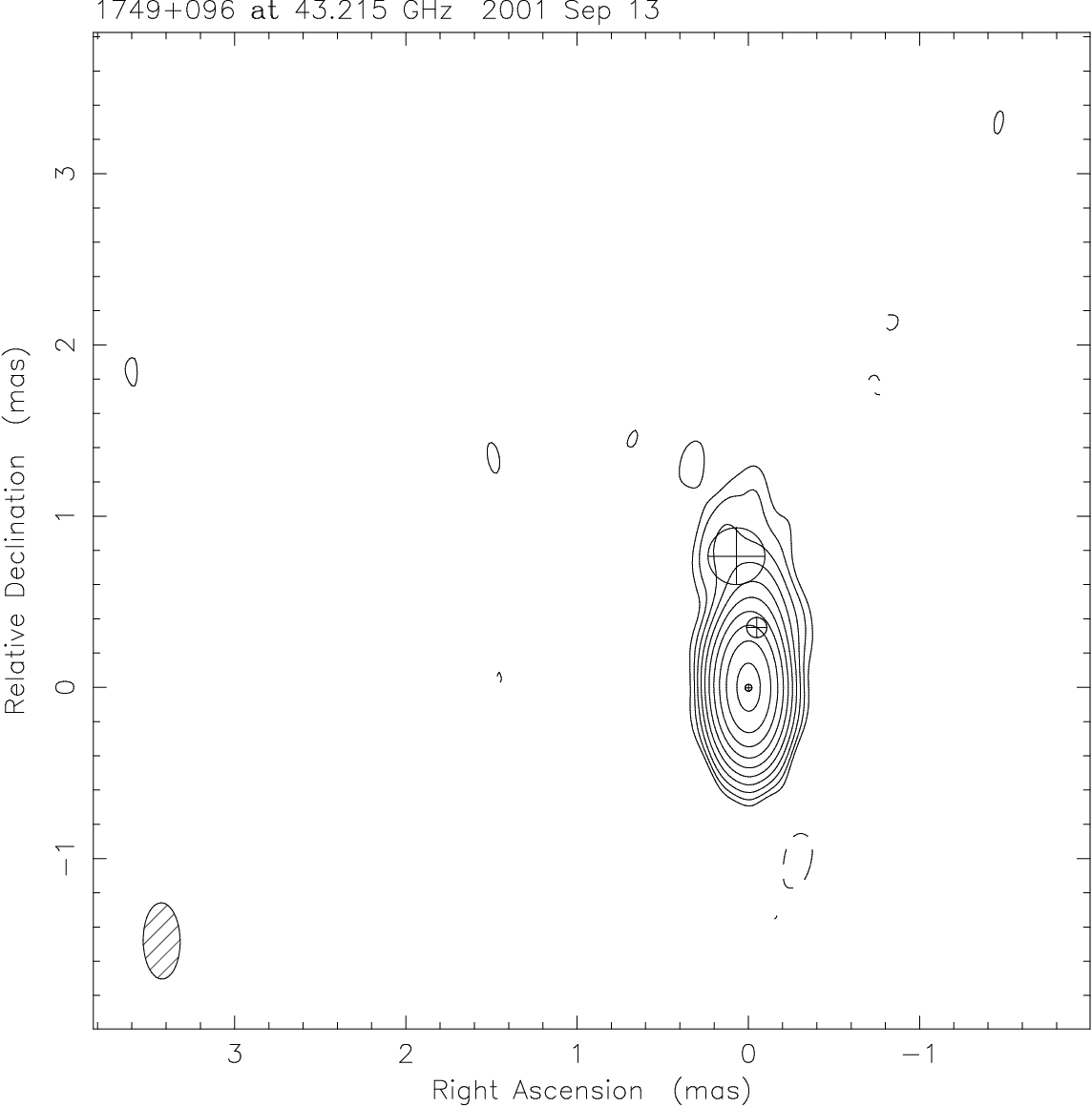}}
  \caption[]{\it{-continued}.}%
  \end{figure*}
 }

\onllongtab{1}{
\begin{longtable}[]{ccccccrcr}
\caption{Description of VLBA images of PKS 1749+096 shown in Fig.~\ref{fig:online_maps}. The columns list the epoch ID, the observing frequency, the peak flux density, the parameters of the restoring elliptical Gaussian beam: the full width at half maximum (FWHM) of the major and minor axes and the PA of the major axis, the rms noise level (3\,$\sigma$) and the contour levels of the image, expressed in percentage of the peak intensity.}\label{tab:online_para}\\
\hline
&&&&\multicolumn{3}{c}{Restoring Beam}&\\[10pt]
\cline{5-7}
ID&$\nu$&Epoch&$S_{\rm peak}$&Major&Minor&PA&$3\times\sigma$&Contours\\
\hline
 &GHz&&Jy/beam&mas&mas&deg&mJy/beam&\\
(1)&(2)&(3)&(4)&(5)&(6)&(7)&(8)&(9)\\
\hline
\endfirsthead
\multicolumn{9}{c}%
{{\bfseries \tablename\ \thetable{} -- continued}} \\
\hline
&&&&\multicolumn{3}{c}{Restoring Beam}&\\
\cline{5-7}
ID&$\nu$&Epoch&$S_{\rm peak}$&Major&Minor&PA&$3\times\sigma$&Contours\\
\hline
&GHz&&Jy/beam&mas&mas&deg&mJy/beam&\\
(1)&(2)&(3)&(4)&(5)&(6)&(7)&(8)&(9)\\
\hline
\endhead
\hline
\endfoot
\hline
\endlastfoot
1&8&1999.348&3.90&2.38&1.06&-14.6&1.7&-0.05, 0.05, 0.1, ..., 51.2\\
2&8&2001.701&2.83&2.20&1.08&0.0&1.3&0.05, 0.1, 0.2, ..., 51.2\\
\hline
3&15&1995.575&5.48&1.10&0.50&11.4&3.5&-0.075, 0.075, 0.15, ..., 76.8\\
4&15&1996.052&2.62&1.10&0.47&3.2&1.9&-0.075, 0.075, 0.15, ..., 76.8\\
5&15&1996.224&0.94&1.08&0.47&-3.2&1.1&-0.15, 0.15, 0.3, ..., 76.8\\
6&15&1996.404&0.74&1.20&0.53&4.3&1.4&-0.15, 0.15, 0.3, ..., 76.8\\
7&15&1996.571&0.78&1.15&0.47&-1.9&1.0&-0.15, 0.15, 0.3, ..., 76.8\\
8&15&1996.740&0.84&1.18&0.48&-8.5&1.2&-0.15, 0.15, 0.3, ..., 76.8\\
9&15&1996.822&1.01&0.93&0.44&-1.6&1.2&-0.1, 0.1, 0.2, ..., 51.2\\
10&15&1996.932&1.82&1.29&0.44&-3.6&1.4&-0.075, 0.075, 0.15, ..., 76.8\\
11&15&1997.110&2.85&1.06&0.52&-0.8&1.5&-0.05, 0.05, 0.10, ..., 51.2\\
12&15&1997.658&3.01&1.04&0.51&3.1&2.2&-0.075, 0.075, 0.15, ..., 76.8\\
13&15&1998.849&4.70&1.21&0.53&-4.3&11.2&-0.25, 0.25, 0.5, ..., 64\\
14&15&1999.348&3.85&1.40&0.59&-15.7&1.5&-0.05, 0.05, 0.1, ..., 51.2\\
15&15&1999.562&3.59&1.67&0.54&-17.8&1.9&-0.05, 0.05, 0.1, ..., 51.2\\
16&15&1999.792&3.20&1.34&0.52&-6.5&4.3&0.15, 0.3, 0.6, ..., 76.8\\
17&15&1999.978&2.51&1.15&0.50&-9.4&2.9&-0.1, 0.1, 0.2, ..., 51.2\\
18&15&1999.989&2.34&1.16&0.56&-8.4&1.7&-0.075, 0.075, 0.15, ..., 76.8\\
19&15&2000.347&2.57&1.19&0.50&-4.9&1.5&0.075, 0.15, 0.3, ..., 76.8\\
20&15&2000.577&3.34&1.15&0.48&-4.1&1.8&-0.05, 0.05, 0.1, ..., 51.2\\
21&15&2000.691&3.20&1.17&0.46&-3.7&1.6&-0.05, 0.05, 0.1, ..., 51.2\\
22&15&2000.740&3.14&1.19&0.51&-3.0&1.6&-0.05, 0.05, 0.1, ..., 51.2\\
23&15&2001.060&4.15&1.21&0.53&-10.0&2.0&-0.05, 0.05, 0.1, ..., 51.2\\
24&15&2001.340&3.62&1.09&0.62&-7.8&1.6&0.05, 0.10, 0.20, ..., 51.2\\
25&15&2001.468&2.57&1.15&0.47&-3.0&1.6&-0.075, 0.075, 0.15, ..., 76.8\\
26&15&2001.496&2.71&1.18&0.48&-6.1&1.5&-0.05, 0.05, 0.1, ..., 51.2\\
27&15&2001.597&2.37&1.21&0.52&-14.1&1.7&0.1, 0.2, 0.4, ..., 51.2\\
28&15&2001.701&3.60&1.22&0.59&0.1&2.0&-0.05, 0.05, 0.1, ..., 51.2\\
29&15&2001.805&4.67&1.09&0.62&-7.2&1.6&-0.04, 0.04, 0.08, ..., 81.9\\
30&15&2001.838&5.68&1.21&0.54&-2.3&2.1&-0.04, 0.04, 0.08, ..., 81.9\\
31&15&2002.019&4.96&1.22&0.54&-3.1&2.9& 0.075, 0.15, ..., 76.8\\
32&15&2002.381&3.13&1.20&0.50&-5.6&2.0&-0.075, 0.075, 0.15, ..., 76.8\\
33&15&2002.416&3.10&1.19&0.55&-8.3&1.0&-0.03, 0.03, 0.06, ..., 61.4\\
34&15&2002.510&2.43&1.10&0.59&-8.3&1.5&-0.075, 0.075, 0.15, ..., 76.8\\
35&15&2002.600&2.79&1.12&0.50&-3.9&1.4&-0.05, 0.05, 0.1, ..., 51.2\\
36&15&2002.973&3.71&1.33&0.57&-12.1&2.0&0.075, 0.15, 0.3, ..., 76.8\\
37&15&2003.099&3.54&1.12&0.52&-3.0&1.2&-0.04, 0.04, 0.08, ..., 81.9\\
38&15&2003.737&4.81&1.14&0.55&-3.8&1.8&0.05, 0.1, 0.2, ..., 51.2\\
39&15&2003.822&5.95&1.27&0.51&-10.4&2.0&-0.035, 0.035, 0.07, ...,71.7\\
40&15&2004.224&4.77&1.12&0.50&-3.7&1.6&-0.04, 0.04, 0.08, ..., 81.9\\
41&15&2004.391&6.26&1.11&0.54&-3.1&3.7&-0.075, 0.075, 0.15, ..., 76.8\\
42&15&2004.448&5.76&1.10&0.53&-6.6&2.1&-0.04, 0.04, 0.08, ..., 81.9\\
43&15&2004.691&3.75&1.11&0.51&-5.9&2.5&-0.075, 0.075, 0.15, ..., 76.8\\
44&15&2005.422&2.44&1.14&0.52&-1.7&1.0&0.05, 0.1, 0.2, ..., 51.2\\
45&15&2005.534&2.76&1.24&0.49&-10.7&2.2&-0.075, 0.075, 0.15, ..., 76.8\\
46&15&2005.668&2.71&1.35&0.50&-15.0&2.1&-0.075, 0.075, 0.15, ..., 76.8\\
47&15&2005.710&2.26&1.15&0.55&-5.1&1.0&-0.05, 0.05, 0.1, ..., 51.2\\
\hline
48&22&1999.348&3.37&0.99&0.40&-15.4&2.0&0.075, 0.15, 0.3, ..., 76.8\\
49&22&2001.701&3.70&0.76&0.38&-2.2&3.1&-0.075, 0.075, 0.15, ..., 76.8\\
50&22&2001.937&4.71&0.81&0.38&-6.9&2.8&-0.05, 0.05, 0.1, ..., 51.2\\
51&22&2002.101&4.07&0.79&0.38&-7.6&3.8&-0.1, 0.1, 0.2, ..., 51.2\\
52&22&2002.449&2.35&0.76&0.26&-9.6&3.3&-0.15, 0.15, 0.3, ..., 76.8\\
53&22&2002.658&3.29&0.85&0.35&-8.7&6.6&-0.2, 0.2, 0.4, ..., 51.2\\
54&22&2002.836&3.85&0.77&0.36&-8.9&2.4&0.075, 0.15, 0.3, ..., 76.8\\
55&22&2003.008&3.79&0.94&0.37&-8.2&2.5&-0.075, 0.075, 0.15, ..., 76.8\\
56&22&2003.216&2.92&1.23&0.22&-10.5&3.7&-0.1, 0.1, 0.2, ..., 51.2\\
57&22&2003.479&2.79&0.78&0.23&-9.7&2.9&-0.1, 0.1, 0.2, ..., 51.2\\
58&22&2004.153&3.35&0.83&0.26&-10.4&2.1&0.075, 0.15, 0.3, ..., 76.8\\
59&22&2004.328&5.51&0.84&0.28&-10.1&3.8&-0.075, 0.075, 0.15, ..., 76.8\\
60&22&2004.505&4.69&0.80&0.25&-10.7&4.5&-0.1, 0.1, 0.2, ..., 51.2\\
61&22&2004.683&5.09&0.78&0.25&-9.7&4.6&-0.1, 0.1, 0.2, ..., 51.2\\
62&22&2004.874&2.75&0.74&0.26&-11.2&3.8&-0.15, 0.15, 0.3, ..., 76.8\\
63&22&2005.093&3.67&0.85&0.28&-7.6&2.9&-0.1, 0.1, 0.2, ..., 51.2\\
\hline
64&43&1999.348&2.50&0.63&0.20&-18.4&2.4&-0.1, 0.1, 0.2,..., 51.2\\
65&43&2001.701&3.69&0.44&0.22&0.8&3.8&-0.15, 0.15, 0.3, ..., 76.8\\
\end{longtable}
}

\onllongtab{2}{
\begin{longtable}[]{cclcrccc}
\caption{Model-fitting results for PKS 1749+096.}\label{tab:online_model} \\
\hline
\multicolumn{1}{c}{Epoch} & \multicolumn{1}{c}{Id.} & \multicolumn{1}{c}{Flux}& \multicolumn{1}{c}{Core Separation}& \multicolumn{1}{c}{PA}& \multicolumn{1}{c}{Size}\\
\multicolumn{1}{c}{}&\multicolumn{1}{c}{}& \multicolumn{1}{c}{[mJy]}&\multicolumn{1}{c}{[mas]}&\multicolumn{1}{c}{[degree]}&\multicolumn{1}{c}{[mas]}\\
\hline
\endfirsthead
\multicolumn{7}{c}%
{{\bfseries \tablename\ \thetable{} -- continued}} \\
\hline
\multicolumn{1}{c}{Epoch} & \multicolumn{1}{c}{Id.} & \multicolumn{1}{c}{Flux}& \multicolumn{1}{c}{Core Separation}& \multicolumn{1}{c}{PA}& \multicolumn{1}{c}{Size}\\
\multicolumn{1}{c}{}&\multicolumn{1}{c}{}& \multicolumn{1}{c}{[mJy]}&\multicolumn{1}{c}{[mas]}&\multicolumn{1}{c}{[degree]}&\multicolumn{1}{c}{[mas]}\\
\hline
\endhead
\hline
\endfoot
\hline 
\endlastfoot
&&\multicolumn{2}{|c|}{(I) $\nu$ = 8\,GHz}&&\\
\hline
1999.348&D&2112.2$\pm$108.0& 0.00$\pm$0.00&  0.0$\pm$0.0&0.05$\pm$0.01\\
        &C6&1775.1$\pm$ 92.0& 0.23$\pm$0.10&-31.2$\pm$3.7&0.03$\pm$0.01\\
        &C5& 120.0$\pm$ 15.2& 1.39$\pm$0.10& 38.5$\pm$4.1&1.02$\pm$0.10\\
      &C1+C2&  26.2$\pm$  7.6& 5.20$\pm$0.41& 29.9$\pm$4.5&3.08$\pm$0.82\\
\hline
2001.701&D&2404.2$\pm$122.1& 0.00$\pm$0.00&  0.0$\pm$0.0&0.04$\pm$0.01\\
    &C7+C6& 488.8$\pm$ 26.2& 0.55$\pm$0.10&-1.1$\pm$10.3&0.30$\pm$0.01\\
        &C5&  66.2$\pm$ 11.9& 2.17$\pm$0.12& 37.0$\pm$3.2&1.55$\pm$0.23\\
    &C1+C2&  10.7$\pm$  4.4& 8.23$\pm$0.81& 30.8$\pm$5.7&4.15$\pm$1.63\\
\hline
&&\multicolumn{2}{|c|}{(II) $\nu$ = 15\,GHz}&&\\
\hline
1995.575&D&3981.4$\pm$212.4& 0.00$\pm$0.00&  0.0$\pm$0.0&0.04$\pm$0.01\\
        &C4&1552.9$\pm$ 95.2& 0.12$\pm$0.05& 34.9$\pm$16.2&0.07$\pm$0.01\\
        &C3&  68.4$\pm$ 11.1& 0.81$\pm$0.05& 28.9$\pm$3.5&0.28$\pm$0.03\\
        &C2&   9.1$\pm$  1.9& 2.46$\pm$0.05& 21.8$\pm$1.2&0.26$\pm$0.04\\
        &C1&  27.2$\pm$ 14.2& 4.34$\pm$0.54& 28.4$\pm$7.1&2.12$\pm$1.08\\
\hline
1996.052&D&1610.9$\pm$ 88.7& 0.00$\pm$0.00&  0.0$\pm$0.0&0.04$\pm$0.01\\
        &C4&1111.8$\pm$ 57.4& 0.17$\pm$0.05& 33.9$\pm$16.3&0.10$\pm$0.01\\
        &C3&  19.5$\pm$  3.1& 0.96$\pm$0.05& 24.8$\pm$3.0&0.53$\pm$0.07\\
        &C2&  16.8$\pm$  3.8& 2.82$\pm$0.07& 32.5$\pm$1.4&0.72$\pm$0.14\\
        &C1&  14.3$\pm$  4.0& 4.57$\pm$0.17& 22.9$\pm$2.1&1.32$\pm$0.33\\
\hline
1996.224&D& 681.1$\pm$ 36.3& 0.00$\pm$0.00&  0.0$\pm$0.0&0.04$\pm$0.01\\
        &C4& 352.3$\pm$ 19.0& 0.26$\pm$0.05& 34.5$\pm$10.8&0.10$\pm$0.01\\
        &C3&  17.0$\pm$  2.3& 0.95$\pm$0.05& 26.3$\pm$3.0&0.51$\pm$0.05\\
        &C2&  20.8$\pm$  5.4& 2.38$\pm$0.14& 27.5$\pm$3.5&1.16$\pm$0.28\\
        &C1&  17.5$\pm$  6.9& 4.74$\pm$0.46& 28.0$\pm$5.6&2.40$\pm$0.92\\
\hline
1996.404&D& 646.5$\pm$ 34.6& 0.00$\pm$0.00&  0.0$\pm$0.0&0.03$\pm$0.01\\
        &C4& 126.6$\pm$  9.6& 0.32$\pm$0.05& 31.2$\pm$7.1&0.08$\pm$0.01\\
        &C3&  14.2$\pm$  2.9& 1.06$\pm$0.05& 23.3$\pm$2.4&0.56$\pm$0.09\\
        &C2&  10.2$\pm$  2.3& 2.70$\pm$0.06& 29.8$\pm$1.2&0.61$\pm$0.11\\
        &C1&  17.1$\pm$  6.7& 5.33$\pm$0.44& 28.6$\pm$4.7&2.31$\pm$0.88\\
\hline
1996.571&D& 749.2$\pm$ 44.5& 0.00$\pm$0.00&  0.0$\pm$0.0&0.05$\pm$0.01\\
        &C4&  53.5$\pm$  6.0& 0.39$\pm$0.05& 26.1$\pm$7.3&0.16$\pm$0.01\\
        &C3&   9.1$\pm$  2.1& 1.32$\pm$0.06& 27.5$\pm$2.7&0.63$\pm$0.12\\
        &C2&  14.2$\pm$  3.3& 2.76$\pm$0.10& 27.8$\pm$2.1&0.95$\pm$0.20\\
        &C1&  13.4$\pm$  4.9& 4.90$\pm$0.38& 27.7$\pm$4.4&2.08$\pm$0.75\\
\hline
1996.740&D& 835.5$\pm$ 42.6& 0.00$\pm$0.00&  0.0$\pm$0.0&0.07$\pm$0.01\\
        &C4&  29.6$\pm$  2.5& 0.46$\pm$0.05& 26.9$\pm$6.2&0.13$\pm$0.01\\
        &C2&  11.3$\pm$  2.1& 2.94$\pm$0.05& 29.1$\pm$1.0&0.70$\pm$0.11\\
        &C1&  17.6$\pm$  5.5& 4.48$\pm$0.29& 27.9$\pm$3.6&1.95$\pm$0.58\\
\hline
1996.822&D&1012.2$\pm$ 61.2& 0.00$\pm$0.00&  0.0$\pm$0.0&0.07$\pm$0.01\\
        &C4&  30.6$\pm$  5.7& 0.42$\pm$0.05& 25.2$\pm$6.8&0.32$\pm$0.04\\
        &C3&   6.1$\pm$  1.9& 1.27$\pm$0.05& 26.8$\pm$2.0&0.37$\pm$0.09\\
        &C2&  13.7$\pm$  4.1& 2.90$\pm$0.15& 26.2$\pm$2.9&1.07$\pm$0.30\\
        &C1&  17.0$\pm$  9.5& 5.03$\pm$0.80& 26.9$\pm$9.0&2.88$\pm$1.61\\
\hline
1996.932&D&1826.1$\pm$ 92.4& 0.00$\pm$0.00&  0.0$\pm$0.0&0.05$\pm$0.01\\
        &C4&  15.2$\pm$  3.1& 0.47$\pm$0.05& 29.4$\pm$3.0&0.05$\pm$0.01\\
        &C3&   3.3$\pm$  1.9& 1.40$\pm$0.05& 34.3$\pm$2.0&0.14$\pm$0.06\\
        &C2&   3.8$\pm$  1.9& 3.15$\pm$0.24& 33.2$\pm$4.3&1.12$\pm$0.49\\
        &C1&  19.5$\pm$  6.3& 5.04$\pm$0.34& 26.5$\pm$3.8&2.20$\pm$0.69\\
\hline
1997.110&D&1852.0$\pm$ 95.4& 0.00$\pm$0.00&  0.0$\pm$0.0&0.05$\pm$0.01\\
        &C5&1027.2$\pm$ 53.6& 0.09$\pm$0.05& 34.1$\pm$23.8&0.08$\pm$0.01\\
        &C4&  22.0$\pm$  3.8& 0.68$\pm$0.05& 29.0$\pm$4.2&0.27$\pm$0.03\\
      &C1+C2&  22.9$\pm$  7.5& 4.19$\pm$0.35& 28.5$\pm$4.9&2.27$\pm$0.71\\
\hline
1997.658&D&2636.1$\pm$135.0& 0.00$\pm$0.00&  0.0$\pm$0.0&0.04$\pm$0.01\\
        &C5& 507.0$\pm$ 28.8& 0.24$\pm$0.05& 42.9$\pm$11.7&0.18$\pm$0.01\\
        &C4&  62.4$\pm$  5.5& 0.61$\pm$0.05& 31.5$\pm$4.7&0.32$\pm$0.02\\
      &C1+C2&  21.9$\pm$  9.9& 4.29$\pm$0.49& 26.5$\pm$6.5&2.24$\pm$0.98\\
\hline
1998.849&D&4850.0$\pm$248.5& 0.00$\pm$0.00&  0.0$\pm$0.0&0.12$\pm$0.01\\
        &C5& 159.1$\pm$ 35.6& 0.97$\pm$0.06& 44.7$\pm$3.5&0.65$\pm$0.12\\
\hline
1999.348&D&2121.2$\pm$108.4& 0.00$\pm$0.00&  0.0$\pm$0.0&0.05$\pm$0.01\\
        &C6&1845.1$\pm$ 94.3& 0.22$\pm$0.05&-33.1$\pm$12.7&0.10$\pm$0.01\\
        &C5&  70.2$\pm$  7.0& 1.50$\pm$0.05& 38.1$\pm$1.9&0.71$\pm$0.05\\
      &C1+C2&  16.5$\pm$  6.5& 5.21$\pm$0.51& 28.9$\pm$5.7&2.72$\pm$1.02\\
\hline
1999.562&D&2564.7$\pm$128.6& 0.00$\pm$0.00&  0.0$\pm$0.0&0.04$\pm$0.01\\
        &C6&1118.3$\pm$ 57.1& 0.28$\pm$0.05&-17.1$\pm$10.1&0.15$\pm$0.01\\
        &C5&  66.5$\pm$  9.9& 1.36$\pm$0.05& 36.4$\pm$2.3&0.87$\pm$0.11\\
      &C1+C2&  15.8$\pm$  4.7& 5.07$\pm$0.21& 31.4$\pm$2.4&1.50$\pm$0.41\\
\hline
1999.792&D&2616.9$\pm$136.4& 0.00$\pm$0.00&  0.0$\pm$0.0&0.05$\pm$0.01\\
        &C6& 687.5$\pm$ 49.3& 0.25$\pm$0.05&-11.7$\pm$11.3&0.20$\pm$0.01\\
        &C5&  64.2$\pm$ 20.9& 1.51$\pm$0.19& 35.4$\pm$7.4&1.30$\pm$0.39\\
\hline
1999.978&D&2039.3$\pm$108.0& 0.00$\pm$0.00&  0.0$\pm$0.0&0.03$\pm$0.01\\
        &C6& 594.5$\pm$ 35.8& 0.32$\pm$0.05& -8.3$\pm$8.8&0.20$\pm$0.01\\
        &C5&  58.9$\pm$ 15.4& 1.44$\pm$0.15& 37.3$\pm$6.2&1.25$\pm$0.30\\
      &C1+C2&  15.9$\pm$  7.1& 5.81$\pm$0.47& 29.2$\pm$4.7&2.19$\pm$0.94\\
\hline
1999.989&D&1870.6$\pm$110.2& 0.00$\pm$0.00&  0.0$\pm$0.0&0.04$\pm$0.01\\
        &C6& 597.2$\pm$ 66.2& 0.32$\pm$0.05& -7.4$\pm$8.8&0.17$\pm$0.01\\
        &C5&  52.3$\pm$ 18.7& 1.54$\pm$0.20& 35.2$\pm$7.7&1.19$\pm$0.40\\
      &C1+C2&   9.8$\pm$  5.4& 5.66$\pm$0.29& 32.6$\pm$3.0&1.15$\pm$0.59\\
\hline
2000.347&D&2107.0$\pm$107.3& 0.00$\pm$0.00&  0.0$\pm$0.0&0.03$\pm$0.01\\
        &C6a& 504.3$\pm$ 25.5& 0.31$\pm$0.05&-15.1$\pm$9.1&0.11$\pm$0.01\\
        &C6b& 100.8$\pm$  7.3& 0.63$\pm$0.05& -0.8$\pm$4.5&0.24$\pm$0.01\\
        &C5&  43.4$\pm$ 10.2& 1.81$\pm$0.13& 38.5$\pm$4.3&1.24$\pm$0.27\\
      &C1+C2&  12.1$\pm$  5.0& 5.88$\pm$0.30& 28.7$\pm$2.9&1.53$\pm$0.60\\
\hline
2000.577&D&2825.0$\pm$144.4& 0.00$\pm$0.00&  0.0$\pm$0.0&0.05$\pm$0.01\\
        &C6a& 531.5$\pm$ 27.0& 0.27$\pm$0.05&-19.0$\pm$5.3&0.05$\pm$0.01\\
        &C6b& 160.4$\pm$ 16.1& 0.63$\pm$0.05& -0.6$\pm$4.5&0.18$\pm$0.01\\
        &C5&  38.6$\pm$ 11.5& 1.99$\pm$0.19& 37.3$\pm$5.6&1.39$\pm$0.39\\
      &C1+C2&   9.9$\pm$  4.7& 5.28$\pm$0.27& 30.2$\pm$2.9&1.21$\pm$0.53\\
\hline
2000.691&D&2601.6$\pm$131.5& 0.00$\pm$0.00&  0.0$\pm$0.0&0.07$\pm$0.01\\
        &C6a& 731.9$\pm$ 38.0& 0.31$\pm$0.05&-15.7$\pm$9.1&0.12$\pm$0.01\\
        &C6b&  99.1$\pm$  6.8& 0.73$\pm$0.05&  4.3$\pm$3.9&0.26$\pm$0.01\\
        &C5&  28.4$\pm$  7.4& 2.07$\pm$0.11& 39.9$\pm$3.0&0.90$\pm$0.21\\
      &C1+C2&  10.9$\pm$  6.0& 5.39$\pm$0.62& 31.2$\pm$6.6&2.34$\pm$1.25\\
\hline
2000.740&D&2684.0$\pm$134.9& 0.00$\pm$0.00&  0.0$\pm$0.0&0.04$\pm$0.01\\
        &C6a& 480.6$\pm$ 25.5& 0.33$\pm$0.05&-20.4$\pm$6.0&0.07$\pm$0.01\\
        &C6b& 171.0$\pm$ 10.2& 0.62$\pm$0.05&  1.0$\pm$4.6&0.19$\pm$0.01\\
        &C5&  40.5$\pm$ 12.4& 1.89$\pm$0.19& 36.7$\pm$5.9&1.35$\pm$0.38\\
      &C1+C2&   7.5$\pm$  3.7& 5.76$\pm$0.39& 30.1$\pm$3.9&1.68$\pm$0.79\\
\hline
2001.060&D&3677.3$\pm$184.8& 0.00$\pm$0.00&  0.0$\pm$0.0&0.05$\pm$0.01\\
        &C6a& 368.8$\pm$ 20.1& 0.25$\pm$0.05&-23.5$\pm$9.0&0.08$\pm$0.01\\
        &C6b& 310.3$\pm$ 19.1& 0.57$\pm$0.05& -4.3$\pm$5.0&0.25$\pm$0.01\\
      &C1+C2&  38.5$\pm$ 11.1& 1.94$\pm$0.19& 38.4$\pm$5.7&1.40$\pm$0.38\\
\hline
2001.340&D&2887.8$\pm$146.7& 0.00$\pm$0.00&  0.0$\pm$0.0&0.03$\pm$0.01\\
        &C7& 741.6$\pm$ 39.0& 0.21$\pm$0.05&-13.8$\pm$13.3&0.12$\pm$0.01\\
        &C6& 152.4$\pm$ 10.5& 0.66$\pm$0.05&  0.1$\pm$4.3&0.28$\pm$0.01\\
        &C5&  31.9$\pm$  9.4& 1.97$\pm$0.19& 37.8$\pm$5.7&1.42$\pm$0.39\\
      &C1+C2&  10.2$\pm$  4.3& 5.15$\pm$0.50& 32.7$\pm$5.6&2.44$\pm$1.00\\
\hline
2001.468&D&2211.5$\pm$118.0& 0.00$\pm$0.00&  0.0$\pm$0.0&0.03$\pm$0.01\\
        &C7& 479.0$\pm$ 28.7& 0.36$\pm$0.05& -4.3$\pm$7.9&0.19$\pm$0.01\\
        &C6&  51.2$\pm$  5.2& 0.98$\pm$0.05&  6.2$\pm$2.9&0.37$\pm$0.03\\
        &C5&  16.3$\pm$  3.3& 2.41$\pm$0.05& 42.2$\pm$1.1&0.56$\pm$0.09\\
      &C1+C2&  10.8$\pm$  4.0& 4.71$\pm$0.21& 32.5$\pm$2.5&1.19$\pm$0.41\\
\hline
2001.496&D&2140.1$\pm$108.8& 0.00$\pm$0.00&  0.0$\pm$0.0&0.03$\pm$0.01\\
        &C7& 609.3$\pm$ 32.6& 0.27$\pm$0.05& -7.7$\pm$10.4&0.14$\pm$0.01\\
        &C6& 132.8$\pm$  9.4& 0.74$\pm$0.05&  1.1$\pm$3.8&0.35$\pm$0.01\\
        &C5&  36.2$\pm$ 11.2& 2.07$\pm$0.20& 38.3$\pm$5.8&1.40$\pm$0.41\\
      &C1+C2&   8.0$\pm$  3.4& 6.06$\pm$0.35& 27.5$\pm$3.3&1.70$\pm$0.70\\
\hline
2001.597&D&1862.2$\pm$ 94.5& 0.00$\pm$0.00&  0.0$\pm$0.0&0.03$\pm$0.01\\
        &C7& 536.2$\pm$ 29.2& 0.29$\pm$0.05& -7.1$\pm$9.7&0.11$\pm$0.01\\
        &C6& 158.5$\pm$ 10.0& 0.72$\pm$0.05&  1.3$\pm$4.0&0.38$\pm$0.01\\
        &C5&  34.1$\pm$  7.6& 2.39$\pm$0.16& 36.0$\pm$3.9&1.55$\pm$0.32\\
\hline
2001.701&D&3286.4$\pm$166.3& 0.00$\pm$0.00&  0.0$\pm$0.0&0.05$\pm$0.01\\
        &C7& 378.4$\pm$ 22.6& 0.37$\pm$0.05& -5.8$\pm$7.7&0.13$\pm$0.01\\
        &C6&  86.4$\pm$ 11.4& 0.92$\pm$0.05&  4.1$\pm$3.1&0.41$\pm$0.04\\
        &C5&  26.1$\pm$  9.4& 2.36$\pm$0.20& 41.6$\pm$5.0&1.27$\pm$0.41\\
\hline
2001.805&D&4414.6$\pm$222.1& 0.00$\pm$0.00&  0.0$\pm$0.0&0.05$\pm$0.01\\
        &C7& 329.5$\pm$ 19.8& 0.36$\pm$0.05& -6.4$\pm$7.9&0.16$\pm$0.01\\
        &C6&  87.8$\pm$  7.3& 0.87$\pm$0.05&  5.9$\pm$3.3&0.38$\pm$0.02\\
        &C5&  29.2$\pm$  6.5& 2.30$\pm$0.12& 40.4$\pm$3.0&1.20$\pm$0.24\\
\hline
2001.838&D&5424.8$\pm$274.7& 0.00$\pm$0.00&  0.0$\pm$0.0&0.04$\pm$0.01\\
        &C7& 347.0$\pm$ 22.0& 0.38$\pm$0.05& -5.2$\pm$7.5&0.17$\pm$0.01\\
        &C6&  61.3$\pm$  9.4& 0.99$\pm$0.05&  6.4$\pm$2.9&0.35$\pm$0.04\\
        &C5&  31.4$\pm$  9.7& 2.45$\pm$0.21& 39.5$\pm$5.0&1.49$\pm$0.43\\
\hline
2002.019&D&4742.5$\pm$244.0& 0.00$\pm$0.00&  0.0$\pm$0.0&0.05$\pm$0.01\\
        &C7& 376.4$\pm$ 24.8& 0.50$\pm$0.05& -6.0$\pm$5.7&0.22$\pm$0.01\\
        &C6&  44.5$\pm$ 10.5& 1.33$\pm$0.06& 15.1$\pm$2.5&0.60$\pm$0.12\\
        &C5&  15.3$\pm$  6.9& 2.73$\pm$0.21& 50.0$\pm$4.3&1.05$\pm$0.41\\
\hline
2002.381&D&2798.2$\pm$149.2& 0.00$\pm$0.00&  0.0$\pm$0.0&0.05$\pm$0.01\\
        &C7& 556.0$\pm$ 31.4& 0.47$\pm$0.05&  5.5$\pm$6.0&0.20$\pm$0.01\\
        &x&  19.0$\pm$  2.9& 1.80$\pm$0.05& 10.0$\pm$1.6&0.12$\pm$0.01\\
        &C5&  18.7$\pm$  7.8& 3.20$\pm$0.25& 44.3$\pm$4.6&1.32$\pm$0.51\\
\hline
2002.416&D&2644.3$\pm$133.5& 0.00$\pm$0.00&  0.0$\pm$0.0&0.04$\pm$0.01\\
        &C7a& 622.2$\pm$ 33.2& 0.40$\pm$0.05&  6.6$\pm$7.1&0.22$\pm$0.01\\
        &C7b&  99.5$\pm$  7.6& 0.84$\pm$0.05& -1.1$\pm$3.4&0.34$\pm$0.02\\
        &C6&  20.2$\pm$  3.6& 1.39$\pm$0.05& 18.2$\pm$2.0&0.63$\pm$0.09\\
        &C5&  19.2$\pm$  3.5& 2.62$\pm$0.09& 43.1$\pm$2.0&1.11$\pm$0.18\\
      &C1+C2&   6.5$\pm$  2.6& 5.58$\pm$0.37& 28.2$\pm$3.9&1.96$\pm$0.75\\
\hline
2002.510&D&2251.1$\pm$118.4& 0.00$\pm$0.00&  0.0$\pm$0.0&0.04$\pm$0.01\\
        &C7& 411.2$\pm$ 27.6& 0.57$\pm$0.05&  5.8$\pm$5.0&0.26$\pm$0.01\\
        &C6&  43.8$\pm$  8.0& 1.23$\pm$0.07&  8.4$\pm$3.2&0.86$\pm$0.13\\
        &C5&  12.5$\pm$  2.9& 2.70$\pm$0.10& 43.8$\pm$2.1&0.96$\pm$0.19\\
      &C1+C2&  10.3$\pm$  4.0& 5.05$\pm$0.29& 34.0$\pm$3.4&1.61$\pm$0.59\\
\hline
2002.600&D&2658.4$\pm$134.9& 0.00$\pm$0.00&  0.0$\pm$0.0&0.03$\pm$0.01\\
        &C7& 329.4$\pm$ 25.7& 0.61$\pm$0.05&  7.6$\pm$4.7&0.26$\pm$0.01\\
        &C6&  52.7$\pm$  8.3& 1.16$\pm$0.05&  6.7$\pm$2.5&0.56$\pm$0.07\\
        &C5&  18.7$\pm$  5.1& 2.84$\pm$0.17& 48.6$\pm$3.4&1.31$\pm$0.33\\
      &C1+C2&  10.0$\pm$  4.2& 5.13$\pm$0.29& 25.7$\pm$3.3&1.45$\pm$0.58\\
\hline
2002.973&D&3642.3$\pm$186.9& 0.00$\pm$0.00&  0.0$\pm$0.0&0.03$\pm$0.01\\
        &C7& 182.8$\pm$ 11.6& 0.65$\pm$0.05&  8.9$\pm$4.4&0.24$\pm$0.01\\
        &C6&  82.0$\pm$  6.5& 1.18$\pm$0.05& 10.9$\pm$2.4&0.64$\pm$0.03\\
      &C5+C1&  17.5$\pm$  6.0& 4.53$\pm$0.32& 34.6$\pm$4.0&2.01$\pm$0.64\\
\hline
2003.099&D&3341.7$\pm$179.6& 0.00$\pm$0.00&  0.0$\pm$0.0&0.03$\pm$0.01\\
      &C8+C9& 219.0$\pm$ 15.0& 0.28$\pm$0.05& 17.1$\pm$8.1&0.08$\pm$0.01\\
        &C7& 166.7$\pm$ 11.2& 0.92$\pm$0.05& 10.0$\pm$3.1&0.40$\pm$0.01\\
        &C6&  19.8$\pm$  4.2& 1.56$\pm$0.07& 19.6$\pm$2.6&0.77$\pm$0.14\\
      &C5+C1&  16.2$\pm$  7.4& 4.45$\pm$0.63& 35.3$\pm$8.1&2.82$\pm$1.26\\
\hline
2003.737& D&4786.3$\pm$244.2& 0.00$\pm$0.00&  0.0$\pm$0.0&0.04$\pm$0.01\\
        &C9&  70.8$\pm$ 18.4& 0.46$\pm$0.05& 28.9$\pm$4.9&0.08$\pm$0.01\\
        &C7& 101.1$\pm$ 11.0& 1.19$\pm$0.05& 14.6$\pm$2.4&0.50$\pm$0.04\\
      &C5+C1&  20.9$\pm$  8.6& 4.27$\pm$0.49& 33.8$\pm$6.6&2.49$\pm$0.98\\
\hline
2003.822&  D&5929.2$\pm$300.8& 0.00$\pm$0.00&  0.0$\pm$0.0&0.04$\pm$0.01\\
        & C9&  75.0$\pm$ 12.5& 0.37$\pm$0.05& 25.8$\pm$4.6&0.06$\pm$0.01\\
        & C7& 103.0$\pm$ 11.9& 1.19$\pm$0.05& 16.6$\pm$2.4&0.58$\pm$0.05\\
      &C5+C1&  24.3$\pm$  8.7& 3.63$\pm$0.35&36.9$\pm$5.5&2.10$\pm$0.70\\
\hline
2004.224&  D&4650.5$\pm$235.4& 0.00$\pm$0.00&  0.0$\pm$0.0&0.07$\pm$0.01\\
        &C10& 260.7$\pm$ 15.9& 0.33$\pm$0.05& 19.7$\pm$8.6&0.18$\pm$0.01\\
        & C7&  77.6$\pm$  7.8& 1.31$\pm$0.05& 19.5$\pm$2.2&0.61$\pm$0.05\\
       &C5+C1&  17.3$\pm$  7.5& 4.63$\pm$0.46&34.7$\pm$5.7&2.21$\pm$0.92\\
\hline
2004.391&  D&6142.5$\pm$335.0& 0.00$\pm$0.00&  0.0$\pm$0.0&0.06$\pm$0.01\\
        &C10& 242.8$\pm$ 25.9& 0.35$\pm$0.05& 26.6$\pm$8.1&0.12$\pm$0.01\\
        & C7&  79.1$\pm$ 19.5& 1.42$\pm$0.08& 17.8$\pm$3.1&0.71$\pm$0.15\\
      &C5+C1&  19.3$\pm$ 12.3& 4.17$\pm$0.83&28.1$\pm$11.3&2.74$\pm$1.66\\
\hline
2004.448&  D&5680.2$\pm$286.0& 0.00$\pm$0.00&  0.0$\pm$0.0&0.07$\pm$0.01\\
        &C10& 244.9$\pm$ 16.8& 0.37$\pm$0.05& 22.5$\pm$7.7&0.28$\pm$0.01\\
        & C7&  65.7$\pm$  8.5& 1.34$\pm$0.05& 21.6$\pm$2.1&0.58$\pm$0.06\\
      &C5+C1&  17.2$\pm$  6.4& 3.43$\pm$0.31& 38.8$\pm$5.2&1.80$\pm$0.62\\
\hline
2004.691&  D&3655.4$\pm$187.9& 0.00$\pm$0.00&  0.0$\pm$0.0&0.05$\pm$0.01\\
        &C10& 220.5$\pm$ 20.2& 0.41$\pm$0.05& 22.9$\pm$6.9&0.24$\pm$0.01\\
        & C7&  64.3$\pm$ 10.8& 1.29$\pm$0.05& 19.7$\pm$2.2&0.55$\pm$0.07\\
      &C5+C1&  20.6$\pm$  9.2& 4.70$\pm$0.76& 30.0$\pm$9.2&3.51$\pm$1.51\\
\hline
2005.422&  D&2284.4$\pm$120.7& 0.00$\pm$0.00&  0.0$\pm$0.0&0.04$\pm$0.01\\
        &C12& 199.2$\pm$ 17.6& 0.31$\pm$0.05&  2.9$\pm$9.1&0.23$\pm$0.01\\
        &C11&  60.8$\pm$  4.9& 0.90$\pm$0.05& 22.1$\pm$3.2&0.33$\pm$0.02\\
        &C10&  39.1$\pm$  4.8& 1.58$\pm$0.05& 22.1$\pm$1.8&0.76$\pm$0.08\\
      &C5+C1&  19.8$\pm$  9.2& 4.16$\pm$0.58& 36.0$\pm$8.0&2.57$\pm$1.17\\
\hline
2005.534&  D&2726.6$\pm$138.6& 0.00$\pm$0.00&  0.0$\pm$0.0&0.04$\pm$0.01\\
        &C12&  97.1$\pm$ 10.8& 0.51$\pm$0.05& 10.1$\pm$5.6&0.30$\pm$0.02\\
    &C10+C11&  71.2$\pm$ 13.4& 1.31$\pm$0.05& 22.2$\pm$2.3&0.66$\pm$0.10\\
      &C5+C1&  18.3$\pm$  9.5& 4.42$\pm$0.72& 32.2$\pm$9.2&2.86$\pm$1.43\\
\hline
2005.668&  D&2657.6$\pm$137.0& 0.00$\pm$0.00&  0.0$\pm$0.0&0.05$\pm$0.01\\
        &C12& 110.1$\pm$ 11.1& 0.45$\pm$0.05&  9.4$\pm$6.3&0.25$\pm$0.02\\
    &C10+C11&  72.3$\pm$ 14.0& 1.39$\pm$0.06& 23.2$\pm$2.5&0.74$\pm$0.12\\
      &C5+C1&  12.8$\pm$  5.6& 5.55$\pm$0.43& 32.7$\pm$4.4&2.03$\pm$0.85\\
\hline
2005.710&  D&2179.6$\pm$119.0& 0.00$\pm$0.00&  0.0$\pm$0.0&0.03$\pm$0.01\\
        &C12& 129.0$\pm$ 14.2& 0.40$\pm$0.05&  9.3$\pm$7.1&0.21$\pm$0.02\\
        &C11&  34.1$\pm$ 10.5& 1.12$\pm$0.05& 21.1$\pm$2.8&0.45$\pm$0.11\\
        &C10&  39.9$\pm$  7.9& 1.74$\pm$0.09& 25.8$\pm$2.9&1.01$\pm$0.17\\
      &C5+C1&  12.8$\pm$  4.7& 5.54$\pm$0.46& 34.2$\pm$4.8&2.61$\pm$0.92\\
\hline
&&\multicolumn{2}{|c|}{(I) $\nu$ = 22\,GHz}&&\\
\hline
1999.348&D&2139.9$\pm$115.7& 0.00$\pm$0.00&  0.0$\pm$0.0&0.05$\pm$0.01\\
        &C6&1455.4$\pm$ 76.9& 0.23$\pm$0.05&-35.7$\pm$12.2&0.11$\pm$0.01\\
        &x&  31.1$\pm$  8.9& 0.49$\pm$0.05&  2.7$\pm$3.5&0.06$\pm$0.01\\
        &C5&  54.2$\pm$ 11.1& 1.34$\pm$0.07& 39.1$\pm$2.9&0.74$\pm$0.13\\
\hline
2001.701&D&3543.5$\pm$179.7& 0.00$\pm$0.00&  0.0$\pm$0.0&0.04$\pm$0.01\\
        &C7& 287.4$\pm$ 27.7& 0.34$\pm$0.05& -8.4$\pm$8.3&0.11$\pm$0.01\\
        &C6&  93.7$\pm$ 17.5& 0.85$\pm$0.05&  1.8$\pm$3.3&0.36$\pm$0.05\\
        &C5&  21.1$\pm$  7.4& 2.53$\pm$0.12& 41.0$\pm$2.8&0.79$\pm$0.25\\
\hline
2001.937&D&4532.1$\pm$231.8& 0.00$\pm$0.00&  0.0$\pm$0.0&0.04$\pm$0.01\\
        &C7& 339.7$\pm$ 20.5& 0.35$\pm$0.05& -8.3$\pm$8.1&0.18$\pm$0.01\\
        &C6&  67.3$\pm$  9.0& 0.89$\pm$0.05&  4.7$\pm$3.2&0.38$\pm$0.04\\
        &C5&  20.2$\pm$  7.9& 2.51$\pm$0.20& 40.0$\pm$4.6&1.12$\pm$0.40\\
\hline
2002.101&D&3456.7$\pm$183.6& 0.00$\pm$0.00&  0.0$\pm$0.0&0.03$\pm$0.01\\
        &C8& 687.1$\pm$ 48.7& 0.14$\pm$0.05& 13.9$\pm$19.6&0.12$\pm$0.01\\
        &C7& 121.1$\pm$ 14.6& 0.52$\pm$0.05& -4.7$\pm$5.5&0.14$\pm$0.01\\
        &C6&  63.1$\pm$ 11.3& 0.98$\pm$0.05&  4.4$\pm$2.9&0.41$\pm$0.06\\
        &C5&  11.8$\pm$  6.1& 2.56$\pm$0.15& 42.3$\pm$3.5&0.73$\pm$0.31\\
\hline
2002.449&D&2187.3$\pm$114.5& 0.00$\pm$0.00&  0.0$\pm$0.0&0.05$\pm$0.01\\
        &C8& 347.7$\pm$ 34.2& 0.27$\pm$0.05&  9.1$\pm$10.4&0.19$\pm$0.01\\
        &C7& 225.0$\pm$ 35.1& 0.65$\pm$0.05&  5.2$\pm$4.4&0.32$\pm$0.04\\
        &C5&  10.9$\pm$  6.2& 2.47$\pm$0.33& 43.7$\pm$7.7&1.30$\pm$0.67\\
\hline
2002.658&D&3312.1$\pm$238.4& 0.00$\pm$0.00&  0.0$\pm$0.0&0.05$\pm$0.01\\
        &C7& 233.7$\pm$ 37.5& 0.67$\pm$0.05&  8.5$\pm$4.2&0.33$\pm$0.04\\
\hline
2002.836&D&3817.8$\pm$201.5& 0.00$\pm$0.00&  0.0$\pm$0.0&0.04$\pm$0.01\\
        &C8& 112.8$\pm$  8.2& 0.36$\pm$0.05&  8.6$\pm$5.5&0.07$\pm$0.01\\
        &C7& 146.7$\pm$ 20.7& 0.85$\pm$0.05&  8.2$\pm$3.3&0.45$\pm$0.05\\
        &C5&  10.9$\pm$  4.2& 2.63$\pm$0.11& 42.3$\pm$2.5&0.64$\pm$0.23\\
\hline
2003.008&D&3760.3$\pm$214.0& 0.00$\pm$0.00&  0.0$\pm$0.0&0.04$\pm$0.01\\
        &C8& 116.5$\pm$ 14.4& 0.40$\pm$0.05& 13.0$\pm$7.1&0.14$\pm$0.01\\
        &C7&  98.7$\pm$  8.5& 0.96$\pm$0.05&  8.5$\pm$3.0&0.35$\pm$0.02\\
        &C5&  12.6$\pm$  4.1& 3.02$\pm$0.08& 31.6$\pm$1.6&0.58$\pm$0.16\\
\hline
2003.216&D&2985.2$\pm$154.8& 0.00$\pm$0.00&  0.0$\pm$0.0&0.05$\pm$0.01\\
        &C8&  58.2$\pm$ 13.8& 0.42$\pm$0.05& 14.2$\pm$2.7&0.04$\pm$0.01\\
        &C7& 105.5$\pm$ 15.6& 1.01$\pm$0.05& 10.3$\pm$2.8&0.46$\pm$0.06\\
\hline
2003.479&D&2797.8$\pm$164.1& 0.00$\pm$0.00&  0.0$\pm$0.0&0.03$\pm$0.01\\
        &C9&  94.5$\pm$ 13.7& 0.30$\pm$0.05& 23.2$\pm$9.4&0.10$\pm$0.01\\
        &C7& 113.7$\pm$ 17.5& 1.03$\pm$0.05& 12.7$\pm$2.8&0.50$\pm$0.07\\
\hline
2004.153&D&3253.4$\pm$191.2& 0.00$\pm$0.00&  0.0$\pm$0.0&0.05$\pm$0.01\\
        &C10& 311.0$\pm$ 30.8& 0.22$\pm$0.05&16.8$\pm$12.7&0.14$\pm$0.01\\
        &C9&  58.2$\pm$  8.0& 1.00$\pm$0.05& 19.9$\pm$2.8&0.49$\pm$0.06\\
\hline
2004.328&D&5616.1$\pm$382.1& 0.00$\pm$0.00&  0.0$\pm$0.0&0.07$\pm$0.01\\
        &C10& 233.0$\pm$ 42.9& 0.27$\pm$0.05&23.0$\pm$10.4&0.20$\pm$0.03\\
        &C9&  59.3$\pm$ 14.9& 1.15$\pm$0.07& 20.5$\pm$3.5&0.62$\pm$0.14\\
\hline
2004.505&D&4925.6$\pm$254.9& 0.00$\pm$0.00&  0.0$\pm$0.0&0.08$\pm$0.01\\
        &C10& 169.6$\pm$ 25.5& 0.39$\pm$0.05& 26.9$\pm$7.3&0.21$\pm$0.02\\
        &C9&  31.3$\pm$  8.7& 1.42$\pm$0.05& 25.3$\pm$2.0&0.32$\pm$0.08\\
\hline
2004.683&D&4005.9$\pm$230.8& 0.00$\pm$0.00&  0.0$\pm$0.0&0.06$\pm$0.01\\
        &C11&1386.8$\pm$ 98.0& 0.12$\pm$0.05& 7.0$\pm$18.3&0.08$\pm$0.01\\
        &C10& 161.2$\pm$ 30.4& 0.52$\pm$0.05& 25.3$\pm$5.5&0.15$\pm$0.02\\
        &C9&  51.6$\pm$ 11.7& 1.54$\pm$0.05& 20.0$\pm$1.9&0.51$\pm$0.10\\
\hline
2004.874&D&2789.7$\pm$160.9& 0.00$\pm$0.00&  0.0$\pm$0.0&0.07$\pm$0.01\\
        &C11& 184.0$\pm$ 21.9& 0.31$\pm$0.05& 15.1$\pm$9.1&0.22$\pm$0.02\\
        &C10&  47.6$\pm$ 17.3& 1.04$\pm$0.10& 24.2$\pm$5.3&0.58$\pm$0.19\\
\hline
2005.093&D&3758.5$\pm$198.4& 0.00$\pm$0.00&  0.0$\pm$0.0&0.06$\pm$0.01\\
        &C11&  99.7$\pm$ 11.3& 0.50$\pm$0.05& 14.6$\pm$5.7&0.26$\pm$0.02\\
        &C10&  38.3$\pm$  7.8& 1.07$\pm$0.05& 22.8$\pm$2.7&0.41$\pm$0.07\\
\hline
&&\multicolumn{2}{|c|}{(I) $\nu$ = 43\,GHz}&&\\
\hline
1999.348&D&1916.8$\pm$193.4& 0.00$\pm$0.00&  0.0$\pm$0.0&0.03$\pm$0.01\\
        &C6&1121.5$\pm$113.3& 0.24$\pm$0.03&-35.5$\pm$7.1&0.11$\pm$0.01\\
        &x&  45.8$\pm$  6.5& 0.36$\pm$0.03& -0.4$\pm$4.7&0.10$\pm$0.01\\
        &C5&  39.2$\pm$ 10.7& 1.11$\pm$0.11& 36.5$\pm$5.8&0.89$\pm$0.22\\
\hline
2001.701&D&3747.4$\pm$376.6& 0.00$\pm$0.00&  0.0$\pm$0.0&0.04$\pm$0.01\\
        &C7& 160.8$\pm$ 19.3& 0.36$\pm$0.03& -7.8$\pm$4.7&0.12$\pm$0.01\\
        &C6&  60.6$\pm$ 18.6& 0.77$\pm$0.04&  5.2$\pm$3.1&0.33$\pm$0.08\\
\end{longtable}
}

\begin{acknowledgements}
We thank the referee for the useful comments and suggestions. The Very Long Baseline Array is a facility of the National Radio Astronomy Observatory, USA, operated by Associated Universities Inc., under cooperative agreement with the National Science Foundation. This research has made use of data from the MOJAVE database that is maintained by the MOJAVE team (Lister et al., 2009, AJ, 137, 3718). The authors thank M. Aller for providing UMRAO data, partly prior to publication. UMRAO is supported by the National Science Foundation under a series of grants, most recently AST-0607523, and by funds from the University of Michigan. Special thanks are due to T. Savolainen for providing data that were used for spectral analysis. This work is also based on observations with the 100-m telescope of the MPIfR (Max-Planck-Institut f\"ur Radioastronomie) at Effelsberg. ZQS acknowledged the support by China Ministry of Science and Technology under 
State Key Development Program for Basic Research (2012CB821800), 
the National Natural Science Foundation of China (grants 10625314, 10821302 and 11173046),
and the CAS/SAFEA International Partnership Program for Creative Research Teams.

\end{acknowledgements}
\bibliographystyle{aa}
\bibliography{OT081}
\end{document}